\documentclass[aps,twocolumn,footinbib, notitlepage,superscriptaddress, longbibliography]{revtex4-1}
\usepackage{graphicx}
 \usepackage{amssymb}
 \usepackage{amsmath}
 \usepackage{amsfonts}
\usepackage{overpic}
\usepackage{enumerate}
\usepackage{color}
\usepackage[dvipsnames]{xcolor}
\usepackage{xspace}
\usepackage[normalsize]{subfigure}
\usepackage{hyperref}
\usepackage{bm}
\usepackage{empheq}
\usepackage[capitalize]{cleveref}

\crefname{section}{Sec.}{Secs.}

\DeclareFontFamily{OT1}{pzc}{}
\DeclareFontShape{OT1}{pzc}{m}{it}{<-> s * [1.10] pzcmi7t}{}
\DeclareMathAlphabet{\mathpzc}{OT1}{pzc}{m}{it}

\DeclareMathAccent{\ring}{\mathalpha}{operators}{"17}
\providecommand{\st}[1]{_{\text{#1}}}
\providecommand{\sfrac}[2]{#1/#2}
\providecommand{\ut}[1]{^{\text{#1}}}

\def\onehalf{\frac{1}{2}}
\def\onequarter{\frac{1}{4}}

\def\bra{\ensuremath{\langle}}
\def\ket{\ensuremath{\rangle}}

\def\const{\mathrm{const}}

\def\pd{\partial}
\def\tr{\mathrm{tr}}

\def\im{\mathrm{i}}
\def\kv{\bv{k}}

\def\xv{\bv{x}}
\def\yv{\bv{y}}

\def\pv{\bv{p}}
\def\rv{\bv{r}}
\def\rvp{\bv{r}_\parallel}
\def\sv{\bv{s}}

\def\b0{\bv{0}}

\def\ra{\rightarrow}

\def\fcal{\mathpzc{f}}
\def\fcalc{\ring\fcal}
\def\Fcal{\mathcal{F}}
\def\Fcalc{\ring{\Fcal}}
\def\Gcal{G}
\def\Gcalc{\ring{G}}
\def\Hcal{\mathcal{H}}
\def\Hc2{\Hcal^{(2)}}
\def\Ccal{C}
\def\Ccalc{\ring{C}}
\def\Dcal{\mathcal{D}}
\def\Kcal{\mathcal{K}}
\def\Lcal{\mathcal{L}}

\def\Ocal{\mathcal{O}}
\def\Scal{\mathcal{S}}

\def\Vcal{\mathcal{V}}
\def\Zcal{\mathcal{Z}}
\def\Zcalc{\ring{\Zcal}}
\def\Zcalgc{\Zcal}

\def\dHdMc{\frac{\delta \Hcal}{\delta \Mc}}
\def\dHgcdMc{\frac{\delta \Hcal(h;[\Mc])}{\delta \Mc}}
\def\dHdMcsf{\sfrac{\delta \Hcal}{\delta \Mc}}

\def\hyp13{{_1 F_3}}
\def\sqtau{\sqrt{\tau}}

\def\reg{\text{reg}}

\def\mtau{\hat\tau}

\def\Mc{\psi}
\def\intOP{\Phi}
\def\constrOP{\Sigma}
\def\avOP{\varphi}
\def\eigenf{\varsigma}
\def\wt{\mathpzc{w}}
\def\res{\st{res}}
\def\resR{\st{\textit{R},res}}

\def\br0{{(0)}}

\def\bcs{boundary conditions\xspace}
\def\pbc{\ut{(p)}}
\def\Dbc{\ut{(D)}}
\def\Nbc{\ut{(N)}}
\def\pNbc{\ut{(p,N)}}
\def\CCF{CCF\xspace}
\def\CCFs{CCFs\xspace}
\def\gfunc{\mathpzc{g}}
\def\tren{t}
\def\tscal{x}
\def\hscal{\mathpzc{h}}
\def\phiscal{\mathpzc{m}}
\def\sqtscal{\sqrt{\tscal}}
\def\mtscal{\hat \tscal}

\def\amplPhimu{\phi_h^\br0}

\def\amplXip{\xi_+^\br0}
\def\amplXiPhi{\xi_\avOP^\br0}
\def\amplXimu{\xi_{h}^\br0}

\def\lagparam{\mu}
\def\CCF{CCF\xspace}
\def\CCFs{CCFs\xspace}
\def\resFE{residual finite-size free energy\xspace}
\def\d{\mathrm{d}}
\def\modesum{\Scal}
\def\msumBare{\mathrm{S}}

\newcommand{\bitem}{\begin{itemize}}
\newcommand{\eitem}{\end{itemize}}
\newcommand{\benum}{\begin{enumerate}}
\newcommand{\eenum}{\end{enumerate}}
\newcommand{\btab}[1]{\begin{tabular}{#1}}
\newcommand{\etab}{\end{tabular}}
\newcommand{\beq}{\begin{equation}}
\newcommand{\eeq}{\end{equation}}
\newcommand{\beqn}{\begin{equation*}}
\newcommand{\eeqn}{\end{equation*}}

\newcommand{\bv}[1]{\mathbf{#1}}

\begin{document}
\title{Statistical field theory with constraints:\\application to critical Casimir forces in the canonical ensemble}
\author{Markus Gross}
\email{gross@is.mpg.de}
\affiliation{Max-Planck-Institut f\"{u}r Intelligente Systeme, Heisenbergstra{\ss}e 3, 70569 Stuttgart, Germany}
\affiliation{IV.\ Institut f\"{u}r Theoretische Physik, Universit\"{a}t Stuttgart, Pfaffenwaldring 57, 70569 Stuttgart, Germany}
\author{Andrea Gambassi}
\affiliation{SISSA -- International School for Advanced Studies and INFN, via Bonomea 265, 34136 Trieste, Italy}
\author{S. Dietrich}
\affiliation{Max-Planck-Institut f\"{u}r Intelligente Systeme, Heisenbergstra{\ss}e 3, 70569 Stuttgart, Germany}
\affiliation{IV.\ Institut f\"{u}r Theoretische Physik, Universit\"{a}t Stuttgart, Pfaffenwaldring 57, 70569 Stuttgart, Germany}
\date{\today}

\begin{abstract}
The effect of imposing a constraint on a fluctuating scalar order parameter field in a system of finite volume is studied within statistical field theory. 
The canonical ensemble, corresponding to a fixed total integrated order parameter (e.g., the total number of particles), is obtained as a special case of the theory.
A perturbative expansion is developed which allows one to systematically determine the constraint-induced finite-volume corrections to the free energy and to correlation functions.
In particular, we focus on the Landau-Ginzburg model in a film geometry (i.e., in a rectangular parallelepiped with a small aspect ratio) with periodic, Dirichlet, or Neumann \bcs in the transverse direction and periodic \bcs in the remaining, lateral directions. 
Within the expansion in terms of $\epsilon=4-d$, where $d$ is the spatial dimension of the bulk, the finite-size contribution to the free energy of the confined system and the associated critical Casimir force are calculated to leading order in $\epsilon$ and are compared to the corresponding expressions for an unconstrained (grand canonical) system.
The constraint restricts the fluctuations within the system and it accordingly modifies the residual finite-size free energy.
The resulting critical Casimir force is shown to depend on whether it is defined by assuming a fixed transverse area or a fixed total volume.
In the former case, the constraint is typically found to significantly enhance the attractive character of the force as compared to the grand canonical case.
In contrast to the grand canonical Casimir force, which, for supercritical temperatures, vanishes in the limit of thick films, in the canonical case with fixed transverse area the critical Casimir force attains for thick films a negative value for all \bcs studied here. Typically, the dependence of the critical Casimir force both on the temperature- and on the field-like scaling variables is different in the two ensembles.
\end{abstract}


\maketitle

\section{Introduction}
\label{sec_intro}
In general, statistical ensembles of systems of finite size are not equivalent \cite{lebowitz_ensemble_1967, allen_computer_1989,roman_ensemble_2008}. 
The primary reason is that imposing a constraint on an extensive thermodynamic variable restricts the fluctuation spectrum of that quantity.
For instance, for a fluid the total number of particles is fixed in the canonical ensemble, whereas it fluctuates in the grand canonical one.
While liquids are typically studied in the grand canonical ensemble \cite{hansen_theory_2006}, there is a number of cases in which the difference between the canonical and the grand canonical ensemble becomes significant: 
most notably, these are systems composed of relatively few particles, such as fluids confined to nanoscale pores or capillaries \cite{gonzalez_density_1997, gonzalez_how_1998}.
This issue has prompted the development of canonical density functional methods \cite{white_density-functional_2000, white_ornstein-zernike_2001, white_extended_2002, de_las_heras_full_2014} which explicitly take fluctuation corrections into account.
Recently, static and dynamic critical phenomena have been investigated also within molecular dynamics \cite{puri_dynamics_1993, das_critical_2006, das_static_2006, roy_transport_2011,das_finite-size_2012,roy_structure_2016,puosi_direct_2016} or lattice Boltzmann simulations \cite{gross_simulation_2012, gross_critical_2012}.
These simulation methods typically operate in the canonical ensemble and require finite-size corrections in order to extract physical properties of bulk systems \cite{lebowitz_long-range_1961, salacuse_finite-size_1996, roman_fluctuations_1997,roman_ensemble_2008}.
Ensemble differences have also been studied extensively in the context of Bose-Einstein condensation (see, e.g., Refs.\ \cite{ziff_ideal_1977, gajda_fluctuations_1997, glaum_condensation_2007}).

In the present study, we consider statistical field theory for an order parameter (OP) field $\phi(\rv)$, which represents, for instance, the deviation of the density of a one-component fluid from its critical value or the deviation of the local concentration from the critical composition of a binary liquid mixture. 
For simplicity, henceforth we adopt the notation pertaining to a one-component fluid.
While the field theory discussed here is rather general, explicit results for the \resFE and the critical Casimir force (CCF) are obtained for the so-called $\phi^4$-Landau-Ginzburg model in a film geometry.
We use the notion \emph{film} for a finite system of volume $V$ with an aspect ratio smaller than unity, while the \emph{thin-film limit} refers to the limit of a vanishing aspect ratio.
The volume integral
\beq \intOP=\int_V \d^d r\, \phi(\rv)
\label{eq_totMass}\eeq 
represents the ``total mass'' in the system, which can fluctuate in the grand canonical ensemble but is fixed to a certain value in the canonical ensemble. 
This constraint is mirrored by the fluctuations within the system and, as shown here, it turns out to typically enhance the attractive character of the CCF.
For a general introduction to the topic of CCFs, we refer to Refs.\ \cite{krech_casimir_1994,brankov_theory_2000,gambassi_casimir_2009}. 
There are relatively few theoretical studies which focus on the effect of an OP constraint on critical phenomena under confinement \cite{eisenriegler_helmholtz_1987, brankov_probabilistic_1989,blote_three-dimensional_2000, caracciolo_finite-size_2001,pleimling_crossing_2001, gulminelli_transient_2003, deng_constrained_2005}.
Constraining a \emph{non}-ordering degree of freedom which is coupled to the OP gives rise to the so-called Fisher renormalization of critical exponents and amplitudes \cite{fisher_renormalization_1968, imry_theory_1973, achiam_phase_1975,anisimov_general_1995, krech_critical_1999,mryglod_corrections_2001, izmailian_universal_2014}. 
A discussion of ensemble differences for critical fluid films within mean field theory (MFT) is presented in Ref.\ \cite{gross_critical_2016} for so-called $(++)$ and $(+-)$ \bcs, where $\pm$ denotes surface fields of strength $h_1=\pm\infty$, which express the preference of the confining walls for one or the other coexisting liquid phase.

In the present study, we investigate the effect of the OP constraint on the OP \emph{fluctuations}, focusing on systems of finite volume with periodic, Dirichlet, or Neumann \bcs. 
Within the framework of boundary critical phenomena, the latter two realize the so-called ordinary and special surface universality class, respectively \cite{diehl_field-theoretical_1986}.
In the case of Dirichlet \bcs, we focus on the case of zero total mass $\intOP=0$ [\cref{eq_totMass}], while, for periodic and Neumann \bcs, we consider also nonzero values of $\intOP$.
In Ref.\ \cite{gross_critical_2016}, it has been shown that an OP constraint can induce drastic qualitative changes in the CCF, affecting, \emph{inter alia}, its sign and its decay behavior upon increasing the film thickness or the associated scaling variables. 
These changes occur already within MFT, i.e., in the absence of fluctuations. 
Here it is useful to recall that, within MFT and under the same thermodynamic conditions \cite{gross_critical_2016}, the film pressures are identical in both ensembles.
Accordingly, in this situation, the differences in the CCF are due to the differences in the bulk pressures.
In turn, they arise because in the two ensembles film and bulk are coupled differently: in the grand canonical ensemble, film and bulk experience the same chemical potential, whereas, in the canonical ensemble, it is natural to require that film and bulk have the same density. 
As it will be shown in the present study, fluctuations induce a further change of the CCF in addition to this mean field effect, since the OP constraint explicitly affects the film pressure itself, rather than only the coupling between film and bulk.

The present study is organized as follows: In \cref{sec_fieldtheory}, the statistical field theory which accounts for an OP constraint is presented and the construction of the associated perturbation theory is described. In \cref{sec_spec}, this field theory is specialized to the Landau-Ginzburg model in a finite volume, and various \bcs are investigated. In particular, perturbative expressions of the residual finite-size contribution to the free energy are derived. In \cref{sec_scaling}, these results are cast into scaling form, and the corresponding scaling functions for the finite-size free energy and the CCFs are obtained. 
Our main results are discussed in \cref{sec_discussion} and summarized in \cref{sec_summary}.
Important details of calculations are presented in Appendices \ref{app_lattice}--\ref{app_CCF_vol}. A glossary of the most frequently used quantities is provided in Table \ref{tab_glossary}.

\begin{table*}[t!]
  \begin{center}
    \begin{tabular}{c c c}
	\hline\hline
	quantity$^\dag$ & description & definition in \\
	\hline
	$\phi$ & order parameter (OP) field & \cref{sec_intro} \\
	$\intOP$ & total OP (``total mass'') in the system & \cref{eq_totMass} \\
	$\avOP$ & mean OP, $\avOP=\intOP/V$ & \cref{eq_meanOP} \\
	$\wt$ & weight function$^\ddag$ & \cref{eq_constr,eq_weightfnc_unity}\\
	$\constrOP_\wt$ & constrained value of the weighted total OP $\intOP$ & \cref{eq_constr}\\
	$\constrOP$ & constrained value of the total OP, $\constrOP\equiv \constrOP_1$ & \cref{eq_constr_spec}\\
	$\Zcalc$ & constrained (canonical) partition function$^\ddag$ & \cref{eq_Z_constr} \\
	$\Zcal$ & unconstrained (grand canonical) partition function & \cref{eq_Z_unconstr} \\
	$\Hcal$ & effective Hamiltonian & \cref{eq_Hamiltonian,eq_Z_unconstr} \\
	$\Lcal$ & effective free energy functional & \cref{eq_Hamiltonian,eq_landau_fe}\\
	$h$ & bulk field & \cref{eq_Z_unconstr,eq_landau_fe_gc}\\
	$\lagparam$ & Lagrange multiplier associated with the constraint & \cref{eq_meanfield,eq_meanfield_spec}\\
	$\Mc$ & mean OP field & \cref{eq_phi_Mc}\\
	$\sigma$ & fluctuation part of the OP field & \cref{eq_phi_Mc}\\
	$\Gcal$ & Green function & \cref{eq_Greens_op_inv,eq_Greens_func_inhom_DE}\\
	$\Gcalc$ & constraint-induced Green function & \cref{eq_G_redef}\\
	$\Fcal$ & unconstrained (grand canonical) film free energy & \cref{eq_freeE_Gauss_gc,eq_F_gc_pert}\\
	$\Fcalc$ & constrained (canonical) film free energy & \cref{eq_freeE_Gauss,eq_freeE_inhom,eq_F_can_pert}\\
	$d$ & spatial dimension of the film & \cref{sec_spec,fig_setup}\\
	$L$ & film thickness  &  \cref{sec_spec,fig_setup} \\
	$A$ & transverse area & \cref{sec_spec,fig_setup} \\
	$V$ & film volume, $V=AL$ & \cref{sec_spec,fig_setup} \\
	$z$ & coordinate along the transverse direction & \cref{sec_spec,fig_setup}\\
	$\rv_\parallel$ & coordinates along the lateral directions & \cref{sec_spec,fig_setup}\\
	$\tau$ & temperature parameter & \cref{eq_landau_fe}\\
	$g$ & quartic coupling constant & \cref{eq_landau_fe} \\
	$\tren$ & reduced (renormalized) temperature & \cref{eq_t_red,eq_Fres_c_scal_L} \\
	$\mtau$ & effective temperature parameter & \cref{eq_mtau}\\
	$\eigenf(z)$ & eigenfunctions & \cref{eq_sigma_eigeneq} \\
	$\rho$ & aspect ratio & \cref{eq_aspectratio}\\
	$\fcal\res,\fcalc\res$ & residual finite-size free energy per volume & \cref{eq_Fres_def,eq_Fres_gc_def}\\
	$\modesum$ & scaling function of the regularized mode sum & \cref{eq_Sreg_coll,eq_modesum_per}\\
	$u^*$ & fixed point value of the renormalized quartic coupling constant & \cref{eq_u_fixpt}\\
	$r$ & numerical constant & \cref{eq_u_numfact}\\
	$\tscal$ & finite-size scaling variable associated with $t$ & \cref{eq_Fres_c_scal_L}\\
	$\mtscal$ & scaled effective temperature parameter & \cref{eq_tscal_mod}\\
	$\phiscal$ & scaled OP & \cref{eq_Fres_c_scal_L} \\ 
	$\hscal$ & scaled bulk field & \cref{eq_Fres_gc_scal_L}\\
	$\amplXip$, $\amplXiPhi$, $\amplXimu$ & correlation length amplitudes associated with $t$, $\avOP$, and $h$ & \cref{eq_ampl_def1,eq_defgc_ximu} \\
	$\Kcal,\ring\Kcal$ & critical Casimir force (CCF) & \cref{eq_CCF_def_Fres} \\
	$\Theta, \ring\Theta$ & scaling functions of the residual free energy & \cref{eq_Fres_c_scal_L,eq_Fres_gc_scal_L}\\
	$\Xi,\ring\Xi$ & scaling functions of the CCF & \cref{eq_Casi_force_c,eq_Casi_force_gc}\\
	\hline\hline
    \end{tabular}
\end{center}
\caption{Glossary of quantities frequently used in the present study. $^\dag$A subscript $R$ on a quantity indicates its renormalized counterpart (see \cref{sec_scaling}). Periodic, Dirichlet, and Neumann \bcs are indicated by the superscripts (p), (D), and (N), respectively. $^\ddag$The canonical ensemble corresponds to the special case $\wt=1$.}
\label{tab_glossary}
\end{table*}

\section{Statistical field theory with a global constraint}
\label{sec_fieldtheory}

\subsection{Notation and conventions}
In order to simplify the presentation of the analytical calculations carried out in the present study, we introduce the shorthand notation 
\beq \int_\rv \equiv \int_V \d^d r
\eeq 
for the integration over a finite, $d$-dimensional volume $V$.
Following Ref.\ \cite{rudnick_constraint_1981}, we define, for two arbitrary scalar functions $u(\rv)$ and $v(\rv)$ as well as for a function $G(\rv,\rv')$ which is symmetric with respect to its two arguments, the shorthand notations
\beq (u,v) \equiv \int_\rv  u(\rv) v(\rv),
\eeq 
\beq (G,v)_\rv \equiv \int_{\rv'}  G(\rv,\rv') v(\rv') = \int_{\rv'}  G(\rv',\rv) v(\rv'),
\eeq 
and
\beq (u,G,v) \equiv \int_\rv \int_{\rv'}\, u(\rv)G(\rv,\rv') v(\rv').
\label{eq_uGv_notation}\eeq 
In particular, we have $(G,1)_\rv \equiv\int_{\rv'} G(\rv,\rv')$.
A ring ($\ring\:$) above a quantity indicates that it refers to a constrained system. 

\subsection{General framework}
\label{sec_general_theory}
A method to cope with an OP constraint within a statistical field theory has been described in Ref.\ \cite{rudnick_constraint_1981} and is recalled briefly here. Building upon this approach, we study the free energy and correlation functions, focusing on the corrections induced by the constraint, and develop a systematic perturbation theory in the canonical ensemble. 
We consider in this section a finite $d$-dimensional volume $V$ with no additional restriction on its geometry. In \cref{sec_spec}, the theory developed here will be applied to more specific systems.
The fluctuating OP field $\phi(\rv)$ is required to satisfy a constraint of the form
\beq  (\wt,\phi) \equiv \int_\rv \wt(\rv) \phi(\rv)  = \constrOP_{\wt},
\label{eq_constr}\eeq 
where $\constrOP_{\wt}$ is a constant and $\wt(\rv)$ is a given weight function.
The case of total mass conservation corresponds to $\wt=1$.
In fact, our expressions generally represent approximations of the true free energy of a constrained system. (An exception is the Gaussian model, for which exact results can be obtained.)
The linear nature of \cref{eq_constr} is sufficiently flexible to encompass constraints which fix the value of $\phi$ or its derivative at a certain point $\bv s$ in space, corresponding to the choices $\wt(\rv)=\delta(\rv-\bv s)$ and $\wt(\rv) = \delta'(\rv-\bv s)$, respectively. In addition, the present framework can be straightforwardly extended to encompass more than a single constraint.

Under the effect of the constraint in \cref{eq_constr}, the statistics of the field $\phi$ is governed by the \emph{constrained probability distribution}
\beq \mathcal{\ring P}([\phi],\constrOP_\wt) \equiv\frac{1}{\Zcalc} \exp(-\Hcal[\phi]) \delta\left((\wt,\phi)-\constrOP_{\wt}\right), \label{eq_P_constr}
\eeq 
where
\beq \Hcal[\phi] \equiv \int_\rv  \Lcal(\rv; [\phi])
\label{eq_Hamiltonian}\eeq 
is the effective Hamiltonian which controls the statistics of the fluctuations of $\phi$ in the absence of the constraint and $\Lcal$ is its density.
Accordingly, the \emph{constrained partition function} $\Zcalc$ is given by 
\begin{widetext}
\beq\begin{split}
\Zcalc(\constrOP_{\wt}) &\equiv \int \Dcal \phi\, \exp(-\Hcal[\phi]) \delta\left((\wt,\phi)-\constrOP_{\wt}\right) =  \int_{-\infty}^\infty \frac{\d J}{2\pi a^{-1-d/2}} \int\Dcal\phi\, \exp\left(-\Hcal[\phi] + \im  J(\wt,\phi) - \im  J \constrOP_{\wt}\right), 
\end{split}\label{eq_Z_constr}\eeq 
\end{widetext}
where in the last equation we have made use of the Fourier representation of the $\delta$-function. 
As usual, the functional integration in \cref{eq_Z_constr} is defined as the limit $N\to\infty$ of the multiple integrals over a field $\phi_i=\phi(\rv_i)$, $i=1,\ldots,N$ defined on a lattice of size $N$ \cite{zinn-justin_quantum_2002}, i.e., 
\beq \int\Dcal \phi\, \,\widehat =\, \prod_{i=1}^N \int_{-\infty}^\infty \frac{\d\phi_1}{a^{1-d/2}} \int_{-\infty}^\infty \frac{\d \phi_2}{a^{1-d/2}} \cdots \int_{-\infty}^\infty \frac{\d\phi_N}{a^{1-d/2}},
\label{eq_func_int}\eeq
where the quantity $a$ represents the lattice constant, the presence of which in \cref{eq_Z_constr,eq_func_int} renders the partition function dimensionless.
However, in order to simplify the notation and because $a$ formally vanishes in the continuum limit, we shall henceforth not indicate it; $a$ can be re-instantiated straightforwardly into the various expressions on the basis of dimensional analysis and of \cref{eq_Z_constr,eq_func_int}.
As a consequence, certain logarithms will seemingly have dimensionful arguments, while, in fact, in the corresponding lattice field theory, these arguments are multiplied by suitable powers of $a$ which renders them dimensionless \footnote{Alternatively, this issue can be dealt with by considering $\ring{\Zcal}/\ring{\Zcal}\st{ref}$, where $\Zcalc\st{ref}$ is a chosen reference partition function \cite{zinn-justin_quantum_2002,eisenriegler_helmholtz_1987}. However, such a definition induces a shift of the associated free energy, which is undesired for the present purposes \cite{krech_casimir_1994}.}.
Concerning an example, we refer to the explicit calculations within a lattice field theory presented in \cref{app_lattice}.
We shall occasionally comment on this issue further [see, e.g., \cref{eq_freeE_inhom} below].
Returning to \cref{eq_Z_constr}, we remark that, although $\Hcal[\phi]$ can in principle depend on external fields, this dependence does not affect the construction of the constrained partition function $\Zcalc$ and therefore it will not be considered henceforth.
The specific expression of $\Lcal$ is not relevant for the general discussion in this section, which will be put in practice for the Landau-Ginzburg model in Sec.~\ref{sec_spec}.

The \emph{grand canonical} partition function $\Zcalgc(h)$ in the presence of a (spatially uniform) external field $h$ is given by 
\begin{multline} \Zcalgc(h) \equiv \int \Dcal \phi\, \exp\left(-\Hcal(h;[\phi])\right),\\ \text{with}\quad \Hcal(h;[\phi]) \equiv \Hcal[\phi] - h \int_\rv \phi.
\label{eq_Z_unconstr}\end{multline}
It immediately follows from the first equation in \cref{eq_Z_constr} that, for $\wt=1$, $\Zcalgc(h)$ is related to the canonical partition function $\Zcalc(\Sigma)$ at a fixed order parameter $\constrOP\equiv \constrOP_1$ via 
\beq \Zcalgc(h) = \int_{-\infty}^\infty \d\constrOP\, e^{h \constrOP} \Zcalc(\constrOP).
\label{eq_Z_gc_c_rel}\eeq 
This equation forms the basis of many finite-size studies of the grand canonical free energy and of the CCF \cite{rudnick_finite-size_1985, brezin_finite_1985, schloms_minimal_1989, dohm_diversity_2008, dohm_critical_2011}. In contrast to the perturbative approach developed below, in the grand canonical ensemble \cref{eq_Z_gc_c_rel} treats fluctuations of the total OP non-perturbatively. This allows one to overcome the well-known artifacts related to the presence of a so-called zero mode. We will return to this aspect in \cref{sec_F_gc}.

Following standard approaches \cite{domb_field_1976, rudnick_constraint_1981}, the partition functions in \cref{eq_Z_constr,eq_Z_unconstr} are evaluated by means of a saddle-point approximation.
To this end, the OP field $\phi(\rv)$ is split into its mean part $\Mc(\rv)\equiv\bra \phi(\rv)\ket$ and a fluctuation $\sigma(\rv)$,
\beq \phi(\rv) = \Mc(\rv) + \sigma(\rv).
\label{eq_phi_Mc}\eeq
Accordingly, the integration measure $\int\Dcal\phi$ in \cref{eq_Z_constr,eq_Z_unconstr} turns into $\int \Dcal\sigma$ and \cref{eq_phi_Mc} implies that
\beq  \bra \sigma(\rv)\ket = 0,
\label{eq_sigma_avg}\eeq
where the average $\bra\ldots\ket = \int\mathcal{D}\phi \ldots \mathcal{\ring P}([\phi],\constrOP_w)$ is performed over the probability distribution given in \cref{eq_P_constr}. 
The mean OP $\Mc$ is left unspecified at this point, but at a later stage it will be determined self-consistently from \cref{eq_sigma_avg}, which in fact reduces to the equation of state relating $\Mc$ and $h$ in the grand canonical and $\Mc$ and $\constrOP$ in the canonical ensemble, respectively [see \cref{eq_meanfield} below].
In the following, we focus on developing a perturbation theory in the presence of a constraint; we simply state the corresponding and well-known \cite{domb_field_1976,zinn-justin_quantum_2002} results in the absence of it.

Inserting \cref{eq_phi_Mc} into the constraint in \cref{eq_constr} yields, after averaging, 
\begin{multline} \bra \constrOP_\wt \ket =  \constrOP_{\wt} = \int_\rv\, \wt(\rv) \left[ \Mc(\rv) + \bra \sigma(\rv)\ket \right]\\ = \int_\rv\, \wt(\rv) \Mc(\rv) \equiv (\wt,\Mc),
\label{eq_Mc_constr}\end{multline}
i.e., the constant value $\constrOP_\wt$ of the constraint is entirely determined by the nonfluctuating part $\psi$ of the OP alone.
As an immediate consequence of Eqs.~\eqref{eq_Mc_constr} and \eqref{eq_constr} one finds that the weighted volume integral of the fluctuations must vanish:
\beq  \int_\rv \wt(\rv) \sigma(\rv) = 0.
\label{eq_fluct_constr}\eeq 
Returning to the calculation of $\Zcalc$, we expand the action $\Hcal$ in terms of $\sigma$ as \cite{domb_field_1976} 
\begin{widetext}
\beq  \Hcal[\Mc+\sigma] = \Hcal[\Mc] + \int_{\rv_1} \frac{\delta \Hcal[\Mc]}{\delta \Mc(\rv_1)} \sigma(\rv_1) + \frac{1}{2!} \int_{\rv_1}\int_{\rv_2} \frac{\delta^2 \Hcal[\Mc]}{\delta \Mc(\rv_1) \delta \Mc(\rv_2)} \sigma(\rv_1)\sigma(\rv_2) + \int_\rv \Vcal(\rv; [\Mc,\sigma]),
\label{eq_Hamilt_expansion}\eeq
where, extending the analysis presented in Ref.\ \cite{rudnick_constraint_1981}, we account also for non-Gaussian contributions in the action via the potential $\Vcal$:
\beq \begin{split} \Vcal(\rv; [\Mc,\sigma]) &\equiv \frac{1}{3!} \int_{\rv_1} \int_{\rv_2} \frac{\delta^3 \Hcal[\Mc]}{\delta \Mc(\rv_1)\delta \Mc(\rv_2)\delta \Mc(\rv)} \sigma(\rv_1)\sigma(\rv_2)\sigma(\rv) + \ldots.
\end{split}\label{eq_potential}
\eeq  
In order to facilitate the calculation of correlation functions, we add to $\Hcal$ a source term $K(\rv)$ which couples to the fluctuation $\sigma(\rv)$, i.e., in the generating functional in \cref{eq_Z_constr} we replace $\Hcal[\phi]$ according to
\beq \Hcal[\phi] \ra \Hcal[\phi] - \int_\rv K(\rv)\sigma(\rv).
\eeq 
Denoting the quadratic part of the action by
\beq \Hc2(\rv_1,\rv_2; [\Mc]) \equiv \frac{\delta^2 \Hcal[\Mc]}{\delta \Mc(\rv_1) \delta \Mc(\rv_2)}\,,
\label{eq_Greens_op_def}\eeq 
the Green function $\Gcal(\rv_1,\rv_2)$ is defined as the inverse of $\Hc2$: \beq \int_{\rv_2} \Gcal(\rv_1,\rv_2) \Hc2(\rv_2,\rv_3) = \int_{\rv_2} \Hc2(\rv_1,\rv_2)\Gcal(\rv_2,\rv_3) = \delta(\rv_1-\rv_3)\,,
\label{eq_Greens_op_inv}\eeq
with $\Gcal(\rv_1,\rv_2) = \Gcal(\rv_2,\rv_1)$.

In order to proceed, we recall that, for an $N\times N$ matrix $A_{ij}$ and fields $K_i$,  $\sigma_j$, the following fundamental result for multidimensional Gaussian integrals holds \cite{zinn-justin_quantum_2002}:
\beq \int \Dcal\sigma \exp\left(-\onehalf \sigma_i A_{ij} \sigma_j + K_i\sigma_i\right) = \frac{(2\pi)^{N/2}}{(\det A)^{1/2}}\exp\left(\onehalf K_i A_{ij}^{-1} K_j\right)
\label{eq_gaussint}\eeq 
(with summation over repeated indices), as well as the identity $\ln\det A=\tr\ln A$.
With the aid of these relations, the linear and quadratic parts of the action in \cref{eq_Z_constr} can now be integrated over $\sigma$, yielding
\begin{multline} \Zcalc(\constrOP_{\wt};[K]) =\int_{-\infty}^\infty \frac{\d J}{2\pi} \exp\left\{-\int_\rv \Vcal\left(\rv; \left[\Mc, \sigma\to \frac{\delta}{\delta K(\rv)} \right] \right) \right\} \exp\Bigg\{ -\Hcal[\Mc]  - \onehalf \tr \ln \Hc2 \\ + \onehalf \left(\dHdMc-K- \im  J \wt,\Gcal,\dHdMc-K - \im J \wt\right)  +\im J (\wt,\Mc) - \im J \constrOP_{\wt} \Bigg\} .
\label{eq_genfunc0}
\end{multline}
In the exponent in \cref{eq_genfunc0} we have neglected the term $(\sfrac{N}{2})\ln (2\pi)$ stemming from the prefactor on the right hand side of \cref{eq_gaussint}. This term turns infinite in the continuum limit and leads to an unimportant additive shift of the free energy.
If $(\wt,\Gcal,\wt)\neq 0$, one obtains, after performing the Gaussian integration over $J$, the constrained generating functional
\begin{multline} \Zcalc(\constrOP_{\wt}; [K]) = \frac{1}{\sqrt{2\pi}} \exp\left\{-\int_\rv \Vcal\left(\rv; \left[\Mc, \sigma\to \frac{\delta}{\delta K(\rv)} \right] \right) \right\} \exp\Bigg\{ -\Hcal[\Mc] - \onehalf \tr \ln \Hc2 \\ 
+ \onehalf \left(\dHdMc-K,\Gcal,\dHdMc-K\right) - \onehalf \ln (\wt,\Gcal,\wt) 
-\onehalf \frac{\left(\dHdMc-K,\Gcal,\wt\right)^2}{(\wt,\Gcal,\wt)} \\
+ \frac{\left(\dHdMc-K,\Gcal,\wt\right)\left[(\wt,\Mc)-\constrOP_{\wt}\right]  }{(\wt,\Gcal,\wt)} - \onehalf \frac{[(\wt,\Mc)-\constrOP_{\wt}]^2}{(\wt,\Gcal,\wt)} \Bigg\}.
\label{eq_genfunc}
\end{multline}
Due to the constraint expressed by \cref{eq_Mc_constr}, the last two terms in $\Zcalc[K]$ vanish so that
\begin{multline} \Zcalc(\constrOP_{\wt}; [K]) = \frac{1}{\sqrt{2\pi}} \exp\left\{-\int_\rv \Vcal\left(\rv; \left[\Mc, \sigma\to \frac{\delta}{\delta K(\rv)} \right]\right) \right\} \exp\Bigg\{ -\Hcal[\Mc] - \onehalf \tr \ln \Hc2 \\
+ \onehalf \left(\dHdMc-K,\Gcal,\dHdMc-K\right)  - \onehalf \ln(\wt,\Gcal,\wt)
-\onehalf \frac{\left(\dHdMc-K,\Gcal,\wt\right)^2}{(\wt,\Gcal,\wt)}  \Bigg\}.
\label{eq_genfunc_constr}
\end{multline}
The last two terms in \cref{eq_genfunc_constr} emerge as a direct consequence of the constraint.
Introducing a Green function $\Gcalc$ which accounts for the constraint as
\beq \Gcalc(\rv_1,\rv_2) \equiv \Gcal(\rv_1,\rv_2) - \frac{(\Gcal,\wt)_{\rv_1} (\Gcal,\wt)_{\rv_2}}{(\wt,\Gcal,\wt)} ,
\label{eq_G_redef}\eeq 
the constrained generating functional in \cref{eq_genfunc_constr} finally reduces to 
\begin{multline} \Zcalc(\constrOP_{\wt}; [K]) =\frac{1}{\sqrt{2\pi}}  \exp\left\{-\int_\rv \Vcal\left(\rv; \left[\Mc, \sigma\to \frac{\delta}{\delta K(\rv)} \right] \right) \right\} \exp\Bigg\{ -\Hcal[\Mc] - \onehalf \tr \ln \Hc2 \\
+ \onehalf \left(\dHdMc-K, \Gcalc,\dHdMc-K\right) 
- \onehalf \ln(\wt,\Gcal,\wt)  \Bigg\}.
\label{eq_genfunc_redef}
\end{multline}
It is useful to note that
\beq (\Gcalc,\wt)_\rv = \int_{\rv'} \Gcalc(\rv,\rv')\wt(\rv') =0
\label{eq_Gcalc_int}\eeq
for all $\rv$, which follows immediately from \cref{eq_G_redef}.
Returning to \cref{eq_genfunc0}, we find that, if $(\wt,\Gcal,\wt)=0$, the integral over $J$ is readily obtained as
\begin{multline} \Zcalc_0(\constrOP_\wt; [K]) \equiv \exp\left\{-\int_\rv \Vcal\left(\rv; \left[\Mc, \sigma\to \frac{\delta}{\delta K(\rv)} \right] \right) \right\} \exp\Bigg\{ -\Hcal[\Mc] - \onehalf \tr \ln \Hc2 
+ \onehalf \left(\dHdMc-K,\Gcal,\dHdMc-K\right) \Bigg\}
\label{eq_genfunc_redef0}
\end{multline}
instead of \cref{eq_genfunc_redef}.
The case $(\wt,\Gcal,\wt)=0$ occurs for models where the complete set of fluctuation modes [see \cref{eq_Greens_inhom,eq_inhom_fGf} below] respect the constraint from the outset. 
For the specific systems investigated in the present study (see Sec.\ \ref{sec_spec}) one actually has $(\wt,\Gcal,\wt)\neq 0$ and therefore the constrained partition function is the one in \cref{eq_genfunc_redef}. Aside from occasional comments, we shall therefore no longer consider the case $(\wt,\Gcal,\wt)=0$.
Finally, repeating the above derivation for the grand canonical partition function in \cref{eq_Z_unconstr}, one obtains the well-known generating functional \cite{domb_field_1976, zinn-justin_quantum_2002}
\begin{multline} \Zcal(h; [K]) = \exp\left\{-\int_\rv \Vcal\left(\rv; \left[\Mc, \sigma\to \frac{\delta}{\delta K(\rv)} \right]\right) \right\} \exp\Bigg\{ -\Hcal(h;[\Mc]) - \onehalf \tr \ln \Hc2 \\
+ \onehalf \left(\dHgcdMc-K,\Gcal,\dHgcdMc-K\right) \Bigg\}.
\label{eq_genfunc_gc}
\end{multline}
\end{widetext}
In the case $\wt=1$, corresponding to a constraint on the total OP, we observe that $\Zcalc(\constrOP;[K])$ in \cref{eq_genfunc_redef} has, apart from the last term in the curly brackets, the same expression as $\Zcal(h;[K])$ in \cref{eq_genfunc_gc} provided one replaces $\Gcal$ by $\Gcalc$ and $\Hcal(h;[\Mc])$ by $\Hcal[\Mc]$. 
Accordingly, also the perturbation theory for the constrained case based on \cref{eq_genfunc_redef} or \cref{eq_genfunc_redef0} leads to expressions formally analogous to those in the unconstrained case based on \cref{eq_genfunc_gc}.
Note that, even if the constraint does not explicitly appear in the expression of $\Zcalc_0$, it still acts via Eqs.~\eqref{eq_Mc_constr} and \eqref{eq_fluct_constr}, which have to be fulfilled in the construction of $\Mc$ and $\sigma$ (see Sec.~\ref{sec_Gaussian} below).

\subsection{Gaussian approximation}
\label{sec_Gaussian}
Here, we investigate the constrained generating functional in \cref{eq_genfunc_redef} within the Gaussian approximation, i.e., neglecting the non-quadratic interactions collected summarily in the potential $\Vcal$ [\cref{eq_potential}].
Within this approximation, the condition in \cref{eq_sigma_avg} results in 
\beq 0 = \bra \sigma(\rv)\ket = -\frac{\delta \ln \Zcalc(\constrOP_{\wt}; [K])}{\delta K(\rv)}\Big|_{K=0} = \left(\Gcalc,\dHdMc\right)_\rv.
\label{eq_sigma_avg_Z}\eeq
Due to the property of $\Gcalc$ expressed in \cref{eq_Gcalc_int}, this condition can be satisfied by requiring \cite{rudnick_constraint_1981} 
\beq  \frac{\delta \Hcal}{\delta \Mc(\rv)} = \lagparam \wt(\rv),
\label{eq_meanfield}\eeq 
where the spatially constant $\lagparam$ can be interpreted as a Lagrange multiplier which must be chosen in order to satisfy the constraint in \cref{eq_Mc_constr}, which leads to $\lagparam \constrOP_\wt = \left( \dHdMcsf,\Mc\right)$ \footnote{Note that the choice $\delta \Hcal / \delta \Mc =0$, which also satisfies \cref{eq_sigma_avg_Z}, does not lead to a dependence of $\Mc$ on $\lagparam$, making it impossible to satisfy \cref{eq_Mc_constr} in general.}.
Owing to the dependence of $\Hc2$ on $\Mc$ [see \cref{eq_Greens_op_def}], the constraint also affects the fluctuations described by the theory, which will be discussed further in Sec.\ \ref{sec_spec}.
In the case $\wt=1$, which corresponds to total OP conservation, \cref{eq_meanfield} represents the equation of state within mean field approximation and $\lagparam$ plays the role of a bulk field or of the chemical potential.
If $(\wt,\Gcal,\wt)=0$, \cref{eq_sigma_avg_Z} must be evaluated with $\Zcalc$ [see \cref{eq_genfunc_redef0}] replaced by $\Zcalc_0$, which yields $(\Gcal,\delta\Hcal/\delta\Mc)=0$. This condition can be fulfilled by \cref{eq_meanfield} with $\lagparam=0$, as in the grand canonical case.

Once the mean OP $\psi$ is fixed according to \cref{eq_meanfield}, the constrained generating functional in \cref{eq_genfunc_redef} reduces to
\begin{widetext}
\beq \Zcalc(\constrOP_{\wt};[K]) = \frac{1}{\sqrt{2\pi}} \exp\left\{ -\Hcal[\Mc] - \onehalf \tr \ln\Hc2 -\onehalf \ln (\wt,\Gcal,\wt)  + \onehalf(K,\Gcalc,K) \right\}.
\label{eq_genfunc_OPfixed}\eeq 
As remarked above, terms involving $\wt$ are absent in the analogous expression for the partition function in the unconstrained case or if $(\wt,\Gcal,\wt)=0$.
From \cref{eq_genfunc_OPfixed}, the \emph{constrained free energy} $\Fcalc$ within the Gaussian approximation (i.e., at one-loop order) follows as 
\beq \Fcalc(\constrOP_{\wt})\equiv -\ln \Zcalc(\constrOP_{\wt}; K=0) = \Hcal[\Mc] + \onehalf \tr \ln\Hcal^{(2)} + \onehalf \ln \left[ 2\pi (\wt,\Gcal,\wt) \right].
\label{eq_freeE_Gauss}\eeq
\end{widetext}
For comparison, we also report here the corresponding expression for the \emph{unconstrained free energy} $\Fcal$, which, according to \cref{eq_genfunc_gc}, is given by \cite{amit_field_2005, zinn-justin_quantum_2002, domb_field_1976}
\beq \Fcal(h)\equiv -\ln \Zcal(h;K=0) = \Hcal(h;[\Mc]) + \onehalf \tr \ln\Hcal^{(2)}.
\label{eq_freeE_Gauss_gc}\eeq
In the expression for the free energy [\cref{eq_freeE_Gauss}] we keep numerical constants such as $(1/2)\ln (2\pi)$, because they are required for a consistent relation between the canonical and the grand canonical ensembles according to \cref{eq_Z_gc_c_rel} (see also Ref.\ \cite{dohm_critical_2011}).
Explicit expressions of $\Fcalc$ and $\Fcal$ will be presented below in Sec.~\ref{sec_spec}, where also the required regularization is discussed.
The constraint induced two-point \emph{correlation function} $\Ccalc$ of the OP fluctuations $\sigma$ follows from \cref{eq_genfunc_OPfixed} as
\begin{multline} \Ccalc(\rv_1,\rv_2) \equiv \bra \sigma(\rv_1)\sigma(\rv_2)\ket = \frac{\delta^2 \ln \Zcalc(\constrOP_{\wt};[K])}{\delta K(\rv_1)\delta K(\rv_2)}\Bigg|_{K=0} \\ = \Gcalc(\rv_1,\rv_2) = \Gcal(\rv_1,\rv_2) - \frac{(\Gcal,\wt)_{\rv_1} (\Gcal,\wt)_{\rv_2}}{(\wt,\Gcal,\wt)},
\label{eq_correl_sigma}\end{multline}
where, as before, the last term is only present if $(\wt,\Gcal,\wt)\neq 0$.
From \cref{eq_correl_sigma} it follows directly that 
\beq \int_{\rv} \wt(\rv) \Ccalc(\rv, \rv') = \int_{\rv'} \wt(\rv') \Ccalc(\rv, \rv') = 0
\label{eq_correl_int0}\eeq 
for all $\rv$, consistently with \cref{eq_Gcalc_int}.
In the unconstrained case, the two-point correlation function $\Ccal$ coincides with the Green function, i.e., $\Ccal=\Gcal$ \cite{zinn-justin_quantum_2002, amit_field_2005}.
In contrast, within the Gaussian approximation, the constraint affects the free energy [\cref{eq_freeE_Gauss}] and the correlation function [\cref{eq_correl_sigma}] in two ways: explicitly, via the generation of correction terms involving $\wt$ and, implicitly, via the dependence of $\Mc$ on $\lagparam \wt$ and $\constrOP_{\wt}$ as required by \cref{eq_meanfield}. 
The latter dependence is a consequence of the fact that the operator $\Hc2$ and, therefore, also the Green function $\Gcal$, which is its functional inverse [\cref{eq_Greens_op_inv}], are affected by the constraint only via their dependence on $\psi$. 
However, the analytic form of $\Hc2$ and $\Gcal$, as well as the spectrum and the form of the eigenfunctions of $\Hc2$, are identical in the constrained and the unconstrained cases (see \cref{sec_spec} below).
The fact that the constraint restricts the allowed modes of a fluctuation [see \cref{eq_fluct_constr}] is accounted for by additive corrections to the free energy [\cref{eq_freeE_Gauss}] and the correlation function [\cref{eq_correl_sigma}].
The meaning of these terms will be further elucidated in \cref{sec_spec}, where we apply the present framework to specific systems.

\subsection{Perturbation theory}
In order to be able to illustrate the perturbative calculation of corrections beyond the Gaussian approximation, an expression for the interaction potential $\Vcal$ in \cref{eq_genfunc_redef} has to be specified.
We assume in the following that the corresponding interaction term in $\Lcal$ [\cref{eq_Hamiltonian}] is of the form $g\phi(\rv)^4/4!$ [see also \cref{eq_landau_fe} below], where $g>0$ is a coupling constant.
It is well-known that a model based on such a density $\Lcal$ captures properly the universal features associated with critical phenomena in the Ising universality class \cite{zinn-justin_quantum_2002,amit_field_2005}.
Apart from this interaction, no additional non-quadratic terms in $\phi$ are assumed to appear in $\Lcal$. 
(This is in line with the vanishing of the coupling constants of the other higher order terms under renormalization group flow.)
For this choice of $\Lcal$, the potential $\Vcal$ defined in \cref{eq_potential} becomes
\begin{multline} \Vcal(\rv; [\Mc,\sigma]) = \frac{1}{3!} g\, \Mc(\rv) \sigma^3(\rv)+ \frac{1}{4!} g\, \sigma^4(\rv) \\ = \int_\sv \delta(\rv-\sv) \left[ \frac{1}{3!} g\, \Mc(\sv) \sigma^3(\sv)+ \frac{1}{4!} g\, \sigma^4(\sv) \right],
\label{eq_potential_phi4}\end{multline}
where the last expression serves to reveal the functional form and pointlike interaction character of $\Vcal$.
Note the appearance of a three-point vertex proportional to the mean field OP $\Mc$.

\subsubsection{Mean order parameter}

As a first application, we calculate the perturbative correction to $\Ocal(g)$ of the mean field expression for $\Mc$.
Using \cref{eq_potential_phi4}, the generating functional in \cref{eq_genfunc_redef} up to $\Ocal(g)$ becomes
\begin{widetext}
\begin{multline} \Zcalc(\constrOP_{\wt}; [K]) \simeq \left[1-\frac{g}{3!} \int_\yv \Mc(\yv) \left(\frac{\delta}{\delta K(\yv)}\right)^3 - \frac{g}{4!} \int_\yv \left(\frac{\delta}{\delta K(\yv)}\right)^4 \right] \exp \Bigg\{ \onehalf (K,\Gcalc, K) - \left(K,\Gcalc, \dHdMc\right) \\ + \text{terms independent of $K$}\Bigg\},
\label{eq_genfunc_Mc_pert}\end{multline}
from which the condition in \cref{eq_sigma_avg}, which defines $\Mc$, results as
\beq\begin{split} 0 &=\bra \sigma(\rv)\ket 
= -\frac{\delta \ln \Zcalc(\constrOP_{\wt};[K])}{\delta K(\rv)}\Big|_{K=0} \\
 &=  \int_\yv \Gcalc(\rv,\yv) \Bigg[ \frac{\delta\Hcal}{\delta \Mc(\yv)} + \onehalf g \Mc(\yv) \Gcalc(\yv,\yv)+ \onehalf g \Mc(\yv) \left(\Gcalc, \dHdMc\right)_\yv^2 - \onehalf g \Gcalc(\yv,\yv) \left(\Gcalc, \dHdMc\right)_\yv  - \frac{1}{6} g\left(\Gcalc, \dHdMc\right)_\yv^3 \Bigg]  .
\end{split}\label{eq_Mc_cond_1loop}\eeq
\end{widetext}
In obtaining the r.h.s.\ of the last equation, we used $\ln(1+  X)\simeq X$.
Analogously to \cref{eq_sigma_avg_Z}, \cref{eq_Mc_cond_1loop} must be solved for $\dHdMcsf$ up to $\Ocal(g)$. 
Importantly, at this stage, no assumption concerning the order in $g$ of $\Mc$ should be made, i.e., $\Mc$ should be formally assumed to be of $\Ocal(g^0)$. It is only \emph{after} imposing the corresponding equation of state [see \cref{eq_meanfield_fluct} below] that $\Mc$ turns into a quantity of $\Ocal(g^{-1/2})$ [compare also \cref{eq_phi_bulk_scalfunc_eps0} and the associated discussion]. 
The solution of \cref{eq_Mc_cond_1loop} can thus be iteratively constructed as a series in $g$ by considering $\dHdMcsf$ and $\Mc$ in \cref{eq_Mc_cond_1loop} to be formally of $\Ocal(g^0)$.
This yields a perturbatively corrected version of \cref{eq_meanfield}:
\beq  \frac{\delta{\Hcal}}{\delta \Mc(\rv)} + \onehalf g \Mc(\rv) \Gcalc(\rv,\rv) = \lagparam \wt(\rv).
\label{eq_meanfield_fluct}\eeq 
Making use of \cref{eq_Gcalc_int}, we find that this expression of $\dHdMcsf$ indeed solves \cref{eq_Mc_cond_1loop} up to and including $\Ocal(g)$ and therefore it implicitly provides the desired leading-order perturbative correction to the mean OP $\Mc$.
Note that \cref{eq_meanfield_fluct} in fact coincides (upon interpreting $\lagparam$ as an external field) with the corresponding expression in the grand canonical ensemble \cite{wilson_surface_1980, diehl_field-theoretical_1981}.
As in \cref{eq_meanfield}, the parameter $\lagparam$ in \cref{eq_meanfield_fluct} has to be chosen such that the constraint on $\Mc$ in \cref{eq_Mc_constr} is fulfilled. 
We recall that $\Gcalc$ itself depends on $\Mc(\rv)$ through its definition in \cref{eq_Greens_op_inv} as the inverse of $\Hc2$.
In practice, \cref{eq_meanfield_fluct} must therefore be solved iteratively (see \cref{sec_spec_bcs} for further discussion).

In order to obtain the perturbative corrections at $\Ocal(g)$ to the free energy or to a correlation function, \cref{eq_meanfield_fluct} has to be imposed as an implicit definition of $\Mc$. As a consequence, $\Mc$ becomes a quantity of $\Ocal(g^{-1/2})$.
Inserting \cref{eq_meanfield_fluct} for $\dHdMcsf$ into \cref{eq_genfunc_redef} and using \cref{eq_Gcalc_int,eq_potential_phi4}, yields the generating functional valid up to $\Ocal(g)$:
\begin{widetext}
\begin{multline} \Zcalc(\constrOP_{\wt}; [K]) \simeq \left\{1-\int_\yv \Vcal\left(\yv; \left[\Mc, \sigma\to \frac{\delta}{\delta K(\yv)} \right]\right) + \frac{1}{72} g^2 \left[ \int_{\yv} \Mc(\yv) \frac{\delta^3}{\delta K(\yv)^3} \right]^2 \right\} \exp \Bigg[ \onehalf (K,\Gcalc, K) \\
+ \onehalf g(K,\Gcalc,\Mc \Gcalc) + \frac{1}{8}g^2 \left(\Mc \Gcalc, \Gcalc, \Mc\Gcalc\right) -\Hcal[\Mc] - \onehalf \tr \ln \Hc2 - \onehalf \ln \left(2\pi(\wt,\Gcal,\wt) \right)\Bigg],
\label{eq_Z_constr_meanOP}\end{multline}
\end{widetext}
where we have used the compact notation introduced in \cref{eq_uGv_notation}, e.g., $(\Mc\Gcalc,\Gcalc,K)=\int_\rv \int_{\rv'} \Mc(\rv)\Gcalc(\rv,\rv)\Gcalc(\rv,\rv') K(\rv')$. 
The term in curly brackets in \cref{eq_Z_constr_meanOP} arises from an expansion of the first exponential term in \cref{eq_genfunc_redef}, keeping only those terms which contribute up to $\Ocal(g)$, taking into account that $\Mc\sim \Ocal(g^{-1/2})$.
It is interesting to specialize \cref{eq_Z_constr_meanOP} to the case $\wt=1$ and a translationally invariant system, e.g., a uniform system with periodic \bcs in all directions. In this case, one has a spatially constant $\Mc(\rv)=\avOP$ as well as $\Gcalc(\rv,\rv)=\Gcalc(\bv0)$, i.e., also the Green function evaluated at coinciding arguments does not depend on the spatial location \footnote{The Green function $\Gcal(\rv,\rv)$ is formally infinite and therefore requires a suitable regularization \cite{zinn-justin_quantum_2002}. However, this aspect does not affect the conclusions in the present paragraph.}. 
Accordingly, using \cref{eq_Gcalc_int} with $\wt=1$, one obtains $(\Mc\Gcalc,\Gcalc,\ldots)=\avOP \Gcalc(\bv0) (1,\Gcalc,\ldots) = 0$, implying that the second and the third term in the second exponential of \cref{eq_Z_constr_meanOP} vanish in this case. 

\subsubsection{Free energy}
The constrained free energy to $\Ocal(g)$ (recall that $\Mc\sim \Ocal(g^{-1/2})$) follows from \cref{eq_Z_constr_meanOP} as  
\begin{widetext}
\beq\begin{split} \Fcalc(\constrOP_\wt) &= -\ln \Zcalc(\constrOP_\wt;K=0)\\
&= -\ln \Bigg\{ \left[1-\frac{1}{8}g\int_\yv [\Gcalc(\yv,\yv)]^2 
- \frac{1}{8}g^2 \left(\Mc \Gcalc, \Gcalc, \Mc \Gcalc\right) 
+ \frac{1}{12}g^2 \int_\xv \int_\yv \Mc(\xv) \Mc(\yv) [\Gcalc(\xv,\yv)]^3 \right] \\
&\qquad \exp\left[-\Hcal[\Mc] - \onehalf \tr \ln \Hc2 - \onehalf \ln \left[2\pi(\wt,\Gcal,\wt)\right] + \frac{1}{8}g^2 \left(\Mc \Gcalc, \Gcalc, \Mc\Gcalc\right) \right] \Bigg\} \\
&= \Hcal[\Mc] + \onehalf \tr \ln \Hc2 + \onehalf \ln \left[2\pi(\wt,\Gcal,\wt)\right] + \frac{1}{8}g\int_\yv [\Gcalc(\yv,\yv)]^2 -\frac{1}{12}g^2 \int_\xv \int_\yv \Mc(\xv) \Mc(\yv) [\Gcalc(\xv,\yv)]^3,
\end{split}\label{eq_free_en2loop}\eeq 
\end{widetext}
where, as before, we have approximated $\ln(1+X)\simeq  X$ in order to evaluate the contribution to $\Ocal(g)$ from the logarithm in the second equation \footnote{
The term $(\sfrac{1}{8}) g^2 \left(\Mc \Gcalc, \Gcalc, \Mc\Gcalc\right)$ in \cref{eq_Z_constr_meanOP} is canceled by the perturbative corrections generated by the last term in curly brackets in \cref{eq_Z_constr_meanOP}.}.
We remark that the expression in \cref{eq_free_en2loop} reduces to the corresponding two-loop result for periodic \bcs obtained in a different context in Refs.\  \cite{guo_hyperuniversality_1987,dohm_diagrammatic_1989}. 
The fourth term in the last equation of \cref{eq_free_en2loop} involving the constraint-induced Green function $\Gcalc$ can be rewritten as
\begin{multline} \int_\yv [\Gcalc(\yv,\yv)]^2 = \int_\yv \Bigg\{[\Gcal(\yv,\yv)]^2 - 2\Gcal(\yv,\yv) \frac{(\Gcal,\wt)_{\yv}^2 }{(\wt,\Gcal,\wt)} \\ + \frac{(\Gcal,\wt)_{\yv}^4}{(\wt,\Gcal,\wt)^2}\Bigg\},\label{eq_free_en2loop_constrcorr_ex}\end{multline}
where have we used \cref{eq_G_redef} as well as the symmetry of $\Gcal$ with respect to an exchange of its arguments. 
An analogous expression applies also to the last term in \cref{eq_free_en2loop}.
Equation \eqref{eq_free_en2loop_constrcorr_ex} explicitly shows the higher-order contributions to the free energy stemming from the constraint. 
The two-loop constrained free energy including the required renormalization will be discussed further elsewhere.

\subsection{Summary}
\label{sec_fieldtheory_summary}
In this section, a statistical field theory for an OP field subject to the integral constraint given in \cref{eq_constr} has been developed based on the approach introduced in Ref.\ \cite{rudnick_constraint_1981}.
The special case $\wt=1$ of the weight function, which enters into the definition of the constraint, leads to a theoretical description within the canonical ensemble.
In order to estimate the typical magnitude of the constraint induced corrections to the free energy [\cref{eq_freeE_Gauss}] and to the correlation function [\cref{eq_correl_sigma}], we consider, having periodic \bcs in mind, a film geometry of volume $V=AL$, where $A$ is the transverse area and $L$ the film thickness.
Since the Green function $\Gcal$ in fact represents the correlation function, one finds the estimate
\begin{subequations}\begin{align} 
(\Gcal,1) &\sim \chi
\intertext{and therefore}
(1,\Gcal,1) &\sim \chi V\,,\label{eq_constr_correl_int_est_1g1}
\end{align}\label{eq_constr_correl_int_est}\end{subequations} 
where $\chi$ denotes the global OP susceptibility.
Accordingly, one obtains an estimate for the correction term on the r.h.s.\ in \cref{eq_G_redef}: 
\beq 
\frac{(\Gcal,1)^2 }{(1,\Gcal,1)} \sim \frac{\chi}{V}.
\label{eq_constr_correl_correct}\eeq
These relations are confirmed below in Sec.~\ref{sec_spec} by means of analytical calculations for the Landau-Ginzburg model [see, e.g., \cref{eq_pbc_Gw} below].
Based on \cref{eq_constr_correl_correct} we conclude that the constraint correction to the Green function vanishes for a system of infinite volume: 
\beq \Gcalc(\rv,\rv')\to \Gcal(\rv,\rv')\quad \text{for } V\to\infty .
\label{eq_Greens_constr_infvol}\eeq 
Extending this analysis to the free energy, we note that (after introducing a suitable regularization, see \cref{sec_Fres_calc}) the first two terms on the r.h.s.\ of \cref{eq_freeE_Gauss} scale $\propto V$ at leading order. According to \cref{eq_constr_correl_int_est_1g1}, the constraint correction $\propto \ln(1,\Gcal,1)$, instead, scales $\propto \ln V$ and, therefore, the constraint correction becomes irrelevant in the thermodynamic limit ($AL\to \infty$) and ensemble equivalence is recovered. In this case, the canonical and the grand canonical free energies are related via a Legendre transform.

Note that the infinite-volume limit encompasses the case in which fewer than the $d$ dimensions of the system become infinite---in particular also the case $A\to\infty$ at fixed $L$ (thin-film limit).
Consider, for instance, for fixed $L$ and $A\to \infty$, the situation at the critical point: assuming that the correlation length $\xi$ scales with the largest size in the system and that the system exhibits critical behavior of a $(d-1)$-dimensional system, we have $\chi\propto \xi^{2-\eta} \propto A^{(2-\eta)/(d-1)}$ with the usual critical exponent $\eta$. Hence, provided $d\geq 3-\eta$, the result in \cref{eq_Greens_constr_infvol} is expected to hold also near criticality.

We emphasize that this analysis does not imply that in the thin-film limit the \resFE or the CCF are generally equivalent among the various ensembles.
Indeed, as has been shown in Ref.\ \cite{gross_critical_2016}, this is not the case for systems with inhomogeneities caused by external bulk or surface fields. The reason for the ensemble inequivalence of the CCF in such systems is that the CCF refers to a bulk system, the coupling of which to the film itself depends on the ensemble \cite{gross_critical_2016}.
In fact, if one considers the thin-film limit, which is natural for MFT, it is reasonable to define the constraint with respect to the total ``mass'' $\intOP$ \emph{per transverse area} $A$, such that \cref{eq_constr} reduces to
$ \int_L \d z\, \phi(z) = \constrOP/A=\const.,
$ 
with formally $\wt=1/A$.
This definition is motivated by the idea that the thermodynamic limit should in general be performed by keeping the \emph{mean} OP 
\beq \avOP=\frac{\constrOP}{V}
\label{eq_meanOP}\eeq 
(e.g., the particle density) constant.

In a finite volume, the OP constraint generally modifies the fluctuations, reflecting the fact that those fluctuations which change the total number of particles in the system [or, in general, the value of the integral in \cref{eq_constr}] are not permitted within the canonical ensemble.
This also means that the canonical free energy [\cref{eq_freeE_Gauss}] is no longer the Legendre transform of the grand canonical one [\cref{eq_freeE_Gauss_gc}], but it exhibits additive corrections [represented by the last term in \cref{eq_freeE_Gauss}] \footnote{Indeed, applying a Legendre transform to the grand canonical free energy [\cref{eq_freeE_Gauss_gc}] would yield a canonical free energy without the constraint-induced correction terms, which is incorrect in finite volumes. In order to obtain the correct canonical finite-size free energy, the constraint must be imposed at the level of the partition function [\cref{eq_Z_constr}].}.
While the presence of these finite-volume corrections is in principle known \cite{white_density-functional_2000, roman_ensemble_2008, binney_theory_1992}, they have so far not been systematically discussed within a statistical field theory and their significance for critical Casimir forces in the canonical ensemble has not been elucidated.

We close this section by summarizing the essential consequences of the OP constraint.
\benum
\item The constraint causes the presence of a bulk field-like parameter $\lagparam$ in the equation which determines the mean OP $\Mc$ [see Eqs.~\eqref{eq_meanfield} and \eqref{eq_meanfield_fluct}] \cite{rudnick_constraint_1981}. This parameter essentially corresponds to the Lagrange multiplier associated with the constrained minimization of the action within the mean field approximation \cite{gross_critical_2016}.
\item The Green function $\Gcal$, defined as the inverse of the quadratic part of the action [see \cref{eq_Greens_op_def}], is affected by the constraint only \emph{implicitly}, via its dependence on the mean OP $\Mc$.
\item Within the Gaussian approximation, the constrained free energy $\ring\Fcal$ differs from the unconstrained one $\Fcal$ by an additive correction of the form $(1/2) \ln \left[2\pi (\wt,\Gcal,\wt)\right]$  [\cref{eq_freeE_Gauss}]. 
\item By introducing a constrained Green function $\Gcalc$ [see \cref{eq_G_redef}], the generating functional in \cref{eq_genfunc_redef} assumes the same form as in the unconstrained case with a nonzero mean OP. 
As a consequence, perturbation theory can be introduced analogously, implying that perturbative results formally carry over from the unconstrained to the constrained theory by simply replacing the usual Green function $\Gcal$ with $\Gcalc$.
\item Within the Gaussian approximation, the two-point correlation function $\ring\Ccal$ of the OP fluctuations [\cref{eq_correl_sigma}] is modified compared to the unconstrained case such that its weighted integral vanishes [\cref{eq_correl_int0}]. The constrained Gaussian correlation function is, in fact, identical to the constrained Green function, i.e., $\Gcalc=\ring\Ccal$. Contributions from higher loop orders provide further constraint corrections to the free energy and to correlation functions. 
Specific examples are given in \cref{eq_meanfield_fluct} for the one-loop equation determining the constraint-induced OP and in \cref{eq_free_en2loop} for the two-loop free energy.
\item In the limit of infinite volume $V\to\infty$, the constraint-induced fluctuation corrections to the free energy and the correlation function vanish and $\Gcalc$ reduces to $\Gcal$ [see \cref{eq_Greens_constr_infvol}].
As remarked above, for the idealized case of a thin film with a transverse area $A\to\infty$ at fixed thickness $L$, mean-field contributions to the model can still be affected by the constraint \cite{gross_critical_2016}. 
\eenum 

\begin{figure}[t]\centering
	\includegraphics[width=0.4\linewidth]{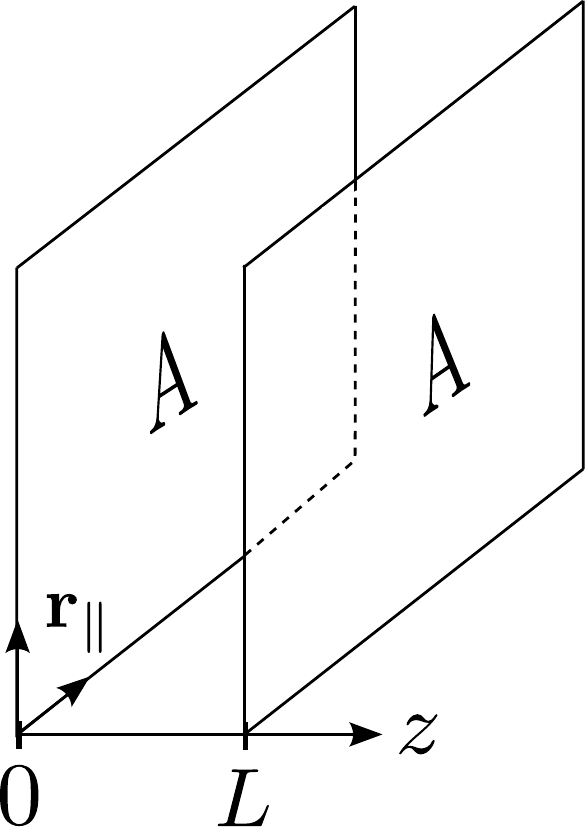}
	\caption{We consider a film of finite volume $V=AL$ in $d$ spatial dimensions, where $A$ is the $(d-1)$-dimensional transverse area and $L$ is the thickness of the film. The coordinate along the transverse direction is denoted by $z$, while the lateral coordinates along the confining surfaces are collectively denoted by $\rv_\parallel$. Depending on the specific system under consideration, periodic, Dirichlet, or Neumann \bcs are applied at $z=0$ and $L$ (see \cref{sec_spec_bcs}). In all cases, periodic \bcs are assumed along all lateral directions.}
	\label{fig_setup}
\end{figure}

\section{Application to the Landau-Ginzburg model}
\label{sec_spec}

\subsection{Model and general results}

In the remaining part of this study, we consider a $d$-dimensional film of volume $V=AL$ which is translationally invariant and has periodic \bcs along the first $(d-1)$ lateral directions, but which can be inhomogeneous in the remaining direction ($z$) of extent $L$, as sketched in \cref{fig_setup}. The boundaries of the film are taken to be located at $z=0$ and $z= L$, while we indicate the coordinates along the lateral directions by the subscript $\parallel$, i.e., we decompose the generic position vector as $\rv = (\rv_{\parallel},z)$. 
We shall interchangeably use the notation $\Gcal(\rv_\parallel,\rv'_\parallel, z,z')$ for the Green function $\Gcal(\rv,\rv')$.
The subsequent discussion shows how the field-theoretical formalism, which is well-known in the grand canonical ensemble \cite{diehl_field-theoretical_1986,domb_field_1976,krech_free_1992}, carries over to the canonical case.

In the following we focus on the one-loop (Gaussian) approximation of the field theory developed in \cref{sec_fieldtheory}. This approximation already displays the essential effects induced by the constraint.
Specifically, we consider the scalar Landau-Ginzburg form of the effective free energy density [\cref{eq_Hamiltonian}], i.e., 
\begin{widetext}
\beq \begin{split}
\Lcal(\rv,\tau, g;  [\phi]) &= \onehalf \left[\nabla\phi(\rv)\right]^2 + \onehalf \tau\phi^2(\rv) +\frac{1}{4!} g \phi^4(\rv) + \left[- h_1\phi(\rv)+ \onehalf c \phi^2(\rv)\right]\left[\delta(z)+\delta(z-L)\right]  \\
&\equiv \onehalf \left(\nabla\phi\right)^2 + \Lcal_b(\rv,\tau,g; [\phi]) + \Lcal_s(\rv, h_1, c; [\phi]) \left[\delta(z)+\delta(z-L)\right],
\label{eq_landau_fe}\end{split}\eeq 
\end{widetext}
where the second equation defines the effective bulk and surface free energy densities $\Lcal_b$ and $\Lcal_s$, respectively.
The parameter $\tau$ is proportional to the reduced temperature 
\beq t\equiv \frac{T-T_c}{T_c},
\label{eq_t_red}
\eeq 
where $T_c$ is the bulk critical temperature; $g>0$ is a coupling constant, $h_1$ is a surface field, and $c$ is the so-called surface enhancement \cite{diehl_field-theoretical_1986}.
The interaction potential $\Vcal$ [see \cref{eq_potential}] which pertains to the action in \cref{eq_landau_fe} has already been reported in \cref{eq_potential_phi4}.
In the grand canonical ensemble, we additionally consider a bulk field $h$ and define
\beq \begin{split} \Lcal(\rv, \tau,g,h; [\phi]) &\equiv \Lcal(\rv, \tau,g; [\phi]) - h\phi(\rv),\quad \text{and}\\ \Lcal_b(\rv, \tau,g,h; [\phi]) &\equiv \Lcal_b(\rv, \tau,g; [\phi]) - h\phi(\rv).
\end{split} \label{eq_landau_fe_gc}\eeq 
In order to simplify the notation, we occasionally suppress the dependence of $\Lcal$, $\Lcal_b$, and $\Lcal_s$ on the parameters $\tau$, $g$, $h_1$, and $c$, and write $\Lcal[\phi(\rv)]\equiv \Lcal(\rv; [\phi])$ (analogously for $\Lcal_b$ and $\Lcal_s$).
Henceforth, in the notation of \cref{eq_constr} we set
\beq \wt= 1\,,
\label{eq_weightfnc_unity}\eeq
i.e., as it is the case for the canonical ensemble, a constraint is imposed on the spatial integral of the OP [see \cref{eq_Mc_constr,eq_fluct_constr}]: 
\beq \int_\rv \phi(\rv)= \int_\rv \Mc(\rv) = \constrOP_1\equiv \constrOP, \label{eq_constr_spec}\eeq 
where $\constrOP$ is the imposed total mass in the system.
Since we assume translational invariance in the lateral directions, the mean profile $\Mc(\rv) = \Mc(z)$ is a function of $z$ only. 

At the leading order, which corresponds to the mean field approximation, in the \emph{canonical} ensemble $\Mc(z)$ is determined by \cref{eq_meanfield}, which yields, for $\Lcal$ given in \cref{eq_landau_fe},  
\begin{widetext}
\beq 
\lagparam = \frac{ \delta \Hcal}{\delta \Mc(\rv)} = -\nabla^2 \Mc(\rv) + \Lcal_b'[\Mc(\rv)] + \left(\pd_z \Mc(\rv) + \Lcal_s'[\Mc(\rv)]\right)\delta\left(z-L\right) + \left(-\pd_z \Mc(\rv) + \Lcal_s'[\Mc(\rv)]\right)\delta\left(z\right).
\label{eq_meanfield_spec}\eeq 
This expression implies the Euler-Lagrange equation 
\beq
 \lagparam = -\pd_z^2 \Mc(z) + \Lcal_b'[\Mc(z)] = -\pd_z^2\Mc(z) + \tau \Mc(z) + \frac{1}{6}g \Mc^3(z) \label{eq_MFT_EOS}
\eeq 
\end{widetext}
and the boundary conditions
\beq\begin{split}
 \pd_z \Mc(z)\Big|_{z=0}  &= \Lcal_s'[\Mc(z)]\Big|_{z=0} = -h_1 + c\psi(z)\Big|_{z=0},\\ 
  -\pd_z \Mc(z)\Big|_{z=L} &= \Lcal_s'[\Mc(z)]\Big|_{z=L} = -h_1 + c\psi(z)\Big|_{z=L}. 
\end{split}\label{eq_MFT_EOS_BC}\eeq
The parameter $\lagparam$ is the Lagrange multiplier required to satisfy the OP constraint in \cref{eq_constr_spec}.
Dirichlet \bcs [$\psi(0)=\psi(L)=0$] are realized for $|h_1|<\infty$ and $c\to\infty$, while (within MFT) Neumann \bcs hold [$\pd_z\psi(0)=\pd_z\psi(L)=0$] for $h_1=0$ and $c=0$.
Upon accounting for the one-loop corrections, \cref{eq_MFT_EOS} is modified as in \cref{eq_meanfield_fluct} and it turns into \footnote{Concerning methods to solve \cref{eq_EOS_oneloop_can} we refer to Refs.\ \cite{diehl_field-theoretical_1981,diehl_critical_1993}, which discuss the analogous grand canonical case.}
\beq \lagparam = -\pd_z^2\Mc(z) + \tau \Mc(z) + \frac{1}{6}g \Mc^3(z) + \onehalf g \Mc(z) \Gcalc(z,z).
\label{eq_EOS_oneloop_can}\eeq
Here, for simplicity, we use the notation $\Gcalc(z,z) \equiv \Gcalc(\rv_\parallel,\rv_\parallel,z,z)$ which, due to translation invariance, does actually not depend on $\rv_\parallel$. 
We anticipate that consistency with the $\epsilon$-expansion of the one-loop free energy [which includes terms up to $\Ocal(\epsilon^0)$] requires to use the mean-field Euler-Lagrange equation in \cref{eq_MFT_EOS} instead of \cref{eq_EOS_oneloop_can} in order to obtain $\psi(z)$.
The reason is that $\Mc$ itself is a quantity of $\Ocal(g^{-1/2})$, implying that the last term on the r.h.s.\ in \cref{eq_EOS_oneloop_can} becomes formally of $\Ocal(\epsilon^{1/2})$ (see, e.g., Ref.\ \cite{amit_field_2005} and the discussion in \cref{sec_renorm_F_gc}). 
Accordingly, in the \emph{grand canonical} ensemble one has, analogously to \cref{eq_MFT_EOS}, the mean-field equation of state
\beq h  = -\pd_z^2\Mc(z) + \tau \Mc(z) + \frac{1}{6}g \Mc^3(z) .
\label{eq_MFT_EOS_gc}\eeq 
Equation \eqref{eq_MFT_EOS_BC} continues to hold for the \bcs in the grand canonical case.

Unless specified otherwise, the following expressions apply to both the canonical and the grand canonical ensemble, because neither the constraint-induced field $\lagparam$ nor the bulk field $h$ appear explicitly in them. Instead, the information about the constraint or the external field is implicitly contained in the mean-field contribution $\Mc$ (see also the discussion in \cref{sec_fieldtheory_summary}).
The expression of $\Hc2$ [\cref{eq_Greens_op_def}] follows from a second functional differentiation of the r.h.s.\ of \cref{eq_meanfield_spec} with respect to $\Mc(\rv')$: 
\begin{widetext}
\beq\begin{split} 
\Hc2(\rv,\rv';[\Mc]) &=  \Big\{ -\nabla^2 + \Lcal_b''[\Mc(z)]  + \delta\left(z-L\right)\left(\pd_z  + \Lcal_s''[\Mc(z)]\right) + \delta\left(z\right) \left(-\pd_z + \Lcal_s''[\Mc(z)]\right) \Big\} \delta(\rv-\rv')\\
&= \Big\{-\nabla_{\rvp}^2 - \pd^2_z + \tau + \onehalf g \Mc^2(z)  + \delta(z-L)\left( \pd_z + c \right) + \delta(z)\left(-\pd_z + c\right) \Big\} \delta(\rvp-\rvp')\delta(z-z').
\label{eq_Greens_op_inhom}
\end{split}\eeq 
Accordingly, the definition in \cref{eq_Greens_op_inv} yields the following differential equation 
for the (unconstrained) Green function $\Gcal$:
\beq \left[-\nabla_{\rvp}^2 - \pd^2_z + \tau + \onehalf g \Mc^2(z) \right] \Gcal(\rvp,\rvp',z,z') = \delta(\rvp-\rvp')\delta(z-z'),
\label{eq_Greens_func_inhom_DE}\eeq 
\end{widetext}
together with the boundary conditions
\beq \pd_z \Gcal(\rvp,\rvp',z,z')\big|_{z=z_s} = \pm \Lcal_s''[\Mc(z)] \Gcal(\rvp,\rvp',z_s,z'),
\label{eq_Greens_func_inhom_bc}\eeq
where $z_s\in \{0,L\}$ denotes the position of one of the surfaces, $z'$ is off the surface and the minus (plus) sign applies to the case $z_s=L$ ($z_s=0$). 
In the case of Dirichlet \bcs, \cref{eq_Greens_func_inhom_bc} reduces to $\Gcal(\rvp,\rvp',z_s,z')=\Gcal(\rvp,\rvp',z,z_s')=0$, while for Neumann \bcs, one has $\pd_z \Gcal(\rvp,\rvp',z,z')\big|_{z=z_s}$ = 0. In the case of periodic \bcs in the transverse direction, instead of \cref{eq_Greens_func_inhom_bc}, one has $\Gcal(\rvp,\rvp',z+L,z'+L) = \Gcal(\rvp,\rvp',z,z')$ for all $z$ and $z'$.

In order to proceed, we introduce a complete set of orthonormal eigenfunctions $\sigma_{\kv_\parallel,n}$ of the operator contained in the curly brackets in $\Hc2$ [\cref{eq_Greens_op_inhom}]:
\beq \sigma_{\kv_\parallel,n}(\rv) = \frac{1}{\sqrt{A}} \exp\left(\im \kv_\parallel \cdot \rvp\right) \eigenf_n(z),
\label{eq_sigma_inhom_eigenf}\eeq 
where and $A=\prod_{\alpha=1}^{d-1} L_\alpha$ and $\kv_\parallel$ is determined by the periodic \bcs in all lateral directions $\alpha=1,\ldots, d-1$ of extent $L_\alpha$ as
\beq k_{\parallel \alpha} =  \frac{2\pi n_\alpha}{L_\alpha},\quad \text{with} \quad n_\alpha=0,\pm 1,\pm 2,\ldots.
\label{eq_wavenum}\eeq
The eigenfunctions $\eigenf_n$ and the corresponding index $n$ pertain to the transverse direction and the associated \bcs.
Denoting the eigenvalue of the operator $-\pd_z^2 + (g/2) \Mc^2(z)$ as $E_n$, the bulk term in \cref{eq_Greens_op_inhom} yields the eigenvalue equation for $\sigma_{\kv_\parallel,n}$:
\begin{multline} \left[-\nabla_{\rvp}^2 - \pd^2_z + \tau + \onehalf g \Mc^2(z) \right] \sigma_{\kv_\parallel,n}(\rvp,z) \\ = (\kv_\parallel^2+\tau+ E_n)\sigma_{\kv_\parallel,n}(\rvp,z),
 \label{eq_sigma_inhom_eigenval_full}\end{multline}
which, using \cref{eq_sigma_inhom_eigenf}, results in an eigenvalue equation for $\eigenf_n$:
\beq \left[\kv^2_\parallel - \pd^2_z + \tau + \onehalf g \Mc^2(z) \right] \eigenf_{n}(z) = (\kv_\parallel^2+\tau+ E_n)\eigenf_{n}(z).
\label{eq_sigma_eigeneq}\eeq 
The boundary terms in \cref{eq_Greens_op_inhom} imply the \bcs 
\beq \pd_z \eigenf_n(z_s) = \pm \Lcal_s''[\Mc(z_s)] \eigenf_n(z_s) = \pm c\, \eigenf_n(z_s),
\label{eq_sigma_inhom_bc}\eeq
where, as before, $z_s\in \{0,L\}$ and the minus (plus) sign applies to the case $z_s=L$ ($z_s=0$).
Periodic \bcs in the transverse directions imply $\eigenf_n(z+L)=\eigenf_n(z)$ for all $z$, replacing \cref{eq_sigma_inhom_bc}.
Also the functions $\eigenf_n$ fulfill completeness and orthonormality relations, i.e.,
\begin{subequations}\begin{align}
 \sum_n \eigenf_n^*(z) \eigenf_n(z') &= \delta(z-z') ,\label{eq_sigma_complete}\\
 \int_0^L \d z\, \eigenf_n^*(z) \eigenf_m(z) &= \delta_{n,m} .\label{eq_sigma_orthonorm}
\end{align}\label{eq_sigma_complete_ortho}\end{subequations}
The formal solution of \cref{eq_Greens_func_inhom_DE,eq_Greens_func_inhom_bc} can now be given in terms of the spectral representation of the Green function:
\beq \Gcal(\rvp,\rvp',z,z') = \frac{1}{A}\sum_{\kv_\parallel,n}\frac{e^{\im \kv_\parallel\cdot(\rvp-\rvp')} \eigenf_n(z)\eigenf_n^*(z')}{\kv_\parallel^2+\tau+E_n}.
\label{eq_Greens_inhom}\eeq
Due to the assumed translational invariance along the lateral directions, $\Gcal$ in fact depends only on the difference $\rv_\parallel-\rv'_\parallel$. It is therefore convenient to introduce its Fourier transform $\hat\Gcal$ along the lateral coordinates,
\begin{multline} \hat \Gcal(\pv,z,z') = A\int_A \d^{d-1} r_\parallel \, \exp\left(-\im \pv\cdot \rv_\parallel\right) \Gcal(\rv_\parallel,\bv0,z,z') 
\\ = A \sum_n \frac{\eigenf_n(z)\eigenf_n^*(z')}{\pv^2+\tau+E_n},
\label{eq_Greens_pz}\end{multline}
referred to as the $pz$-representation of $\Gcal$.
Here, consistently with the periodicity of $\Gcal(\rv_\parallel,\bv0,z,z')$ along the lateral directions, the components of $\pv$ take the discrete values $p_\alpha=2\pi n_\alpha /L_\alpha$, with $n_\alpha=0,\pm 1,\pm 2,\ldots$ for $\alpha=1,\ldots,d-1$.
In obtaining the last expression in \cref{eq_Greens_pz}, we have furthermore used \cref{eq_expIK_int}. 
The transverse area $A$ appears as a prefactor because here we consider $\hat\Gcal$ to be a function of only a single wave vector $\pv$, whereas, in real space, $\Gcal$ is defined as a function of two positions, $\rvp$ and $\rvp'$ (see Appendix \ref{app_Fourier}).
The fluctuating field $\sigma$ [see \cref{eq_phi_Mc}] can also be expanded in terms of the eigenfunctions in \cref{eq_sigma_inhom_eigenf}:
\beq \sigma(\rvp,z) = \frac{1}{\sqrt{A}} \sum_{\kv_\parallel,n}\, c_n(\kv_\parallel) \exp(\im \kv_\parallel\cdot\rv_\parallel) \eigenf_n(z),
\label{eq_sigma_inhom_exp}\eeq
with the coefficients $c_n$ given by $c_n(\kv_\parallel)= (1/\sqrt{A}) \int_0^L \d z \int_A\d^{d-1} r_\parallel \exp(-\im \kv_\parallel\cdot\rv_\parallel) \eigenf_n^*(z) \sigma(\rvp,z) $.
Since \cref{eq_fluct_constr} constrains the function $\sigma$ as a whole, nothing can be stated at this point about each individual $c_n$, except that 
\begin{multline} 0 = \int_A \d^{d-1}r_\parallel \int_0^L \d z\, \sigma(\rv_\parallel,z) \\ = \sqrt{A} \sum_n c_n(\b0) \int_0^L \d z\, \eigenf_n(z).
\label{eq_fluct_constr_inhom}\end{multline}
In particular, we emphasize that it is not justified to include in the expansion in \cref{eq_sigma_inhom_exp} only those eigenfunctions $\eigenf_n$ which satisfy the constraint in \cref{eq_fluct_constr} [with $\wt=1$, see \cref{eq_constr_spec}]. 

In order to be able to calculate the constrained Green function $\Gcalc$ [\cref{eq_G_redef}] and the constrained free energy $\Fcalc$ [\cref{eq_freeE_Gauss}], expressions for the quantities $(\Gcal,1)$ and $(1,\Gcal,1)$ have to be worked out.
Making use of the spectral representation of $\Gcal$ [\cref{eq_Greens_inhom}] as well as of Eqs.\ \eqref{eq_expIK_int} and \eqref{eq_Greens_pz}, we eventually find:
\begin{multline} (\Gcal,1)_{z} = \int_A \d^{d-1} r_\parallel' \int_0^L \d z' \, \Gcal(\rvp,\rvp',z,z') \\ = \frac{1}{A}\int_0^L \d z' \hat\Gcal(\pv=0,z,z').
\label{eq_inhom_fG}\end{multline}
As a consequence of translational invariance, this expression does not depend on the lateral coordinate $\rvp$.
The quantity $(1,\Gcal,1)$ follows as
\begin{multline} (1,\Gcal,1) \\ = \int_A \d^{d-1} r_\parallel \int_A \d^{d-1} r_\parallel' \int_0^L \d z \int_0^L \d z'\, \Gcal(\rvp,\rvp',z,z') \\ = \int_0^L \d z \int_0^L \d z' \hat\Gcal(\pv=0,z,z')\,,
\label{eq_inhom_fGf}\end{multline}
which, in general, does not vanish (see \cref{sec_spec_bcs} below), so that the generating functional defined in \cref{eq_genfunc_redef} applies to the constrained case.
The $pz$-representation of the constrained correlation function in \cref{eq_correl_sigma} follows, analogously to \cref{eq_Greens_pz}, as 
\beq \hat\Ccalc(\pv,z,z') = \hat\Gcal(\pv,z,z') - A^2 \delta_{\pv,0} \frac{(\Gcal,1)_z (\Gcal,1)_{z'}}{(1,\Gcal,1)}.
\label{eq_correl_sigma_pz}\eeq
The constrained (canonical) free energy $\Fcalc$ within the Gaussian approximation [see \cref{eq_freeE_Gauss}] takes the form 
\beq \Fcalc(\constrOP) = \Hcal[\Mc] + \onehalf \sum_{\kv_\parallel}\sum_n \ln(\kv_\parallel^2+\tau+E_n) +\onehalf \ln \left[2\pi(1,\Gcal,1)\right].
\label{eq_freeE_inhom}\eeq
The r.h.s.\ of this expression depends on $\constrOP$ via the mean field $\Mc$ [\cref{eq_MFT_EOS,eq_constr_spec}], the eigenvalues $E_n$ [\cref{eq_sigma_eigeneq}], and the Green function [\cref{eq_Greens_func_inhom_DE}].
The second term in \cref{eq_freeE_inhom} requires a suitable regularization in order to render a physically meaningful, finite result. This issue as well as the relevance of the constraint correction will be discussed in \cref{sec_Fres_calc} below.
We recall that in \cref{eq_freeE_inhom} we have suppressed the lattice constant $a$, the presence of which is implied within the corresponding discrete field theory via the definition of the functional integral in \cref{eq_func_int}.
Accordingly, the arguments of the first and second logarithm in \cref{eq_freeE_inhom} would have to be multiplied by $a^2$ and $a^{2+d}$, respectively, which renders them dimensionless (see, e.g., Refs.\ \cite{dohm_diversity_2008, dohm_critical_2011} and Appendix \ref{app_lattice}). 
However, a physical observable with universal features, such as the CCF, is independent of the lattice constant.

\subsection{Specialization to various boundary conditions}
\label{sec_spec_bcs}
We now specialize the general expressions derived above to \emph{single-phase} systems having periodic, Dirichlet, or Neumann boundary conditions at both boundaries $z=0$ and $z=L$. 
Within the Gaussian approximation, the latter two \bcs are realized by setting $h_1=0, c=\infty$ and $h_1=0, c=0$, respectively, in $\Lcal_s$ [\cref{eq_landau_fe}].
In the case of periodic \bcs, instead, one has $\Lcal_s=0$ and requires $\phi(\rv_\parallel,0)=\phi(\rv_\parallel,L)$.
In all cases, periodic \bcs along the lateral directions are applied (see \cref{fig_setup}).
The calculation of the regularized free energy [see \cref{eq_freeE_inhom}] is deferred to Sec.\ \ref{sec_Fres_calc}.

\subsubsection{Periodic boundary conditions}
\label{sec_periodic_bc}
For periodic boundary conditions along the $z$ direction, the system is homogeneous in all directions, with the mean OP [see \cref{eq_Mc_constr,eq_meanOP}]
\beq \avOP\equiv \frac{\constrOP}{V} = \Mc(\rv),
\label{eq_constr_pbc} \eeq
which does not vary spatially.
Within the mean-field approximation, $\avOP$ is determined by \cref{eq_MFT_EOS}, which, for periodic \bcs, turns into:
\beq \tau \avOP + \frac{1}{6}g \avOP^3 = \lagparam.
\label{eq_EOS_pbc}\eeq 
The parameter $\lagparam$ must be chosen such that the constraint in \cref{eq_constr_pbc} is obeyed by the solution of \cref{eq_EOS_pbc} for $\avOP$.
The orthonormal eigenfunctions $\sigma_\kv\equiv \sigma_{\kv_\parallel,n}$ [see \cref{eq_sigma_inhom_eigenf}] are given by
\begin{multline} \sigma_\kv(\rv) = \frac{1}{\sqrt{V}} \exp\left(\im \kv \cdot \rv\right),\\  \text{with}\quad k_\alpha = \frac{2\pi n_\alpha}{L_\alpha},\quad\text{and}\quad n_\alpha=0,\pm 1,\pm 2,\ldots,
\label{eq_eigenf_per}\end{multline}
for $\alpha=1,2,\ldots,d$ and $V=\prod_{\alpha=1}^d L_\alpha$.
Note that here we have simplified the notation used in \cref{eq_sigma_inhom_eigenf}. 
The functions $\sigma_\kv$ in \cref{eq_eigenf_per} fulfill the eigenvalue equation in \cref{eq_sigma_inhom_eigenval_full} with  $E_n=k_z^2+(g/2)\avOP^2$ and $n\equiv n_z$ (denoting by $z$ the last of the $d$ Cartesian coordinates). 
The temperature parameter $\tau$ enters these expressions in combination with the mean OP density $\avOP$ in the form of an effective, shifted temperature
\beq \mtau\equiv \tau+ \onehalf g\avOP^2.
\label{eq_mtau}\eeq 
Due to \cref{eq_expIK_int}, the eigenfunctions in \cref{eq_eigenf_per} for periodic \bcs include a single zero mode $\sigma_{\kv=\bv0}$, which is spatially constant. 
Accordingly, the Green function $\Gcal\pbc$ has the spectral representation [see \cref{eq_Greens_inhom}]
\beq \Gcal\pbc(\rv,\rv') = \frac{1}{V} \sum_\kv \frac{e^{\im\kv\cdot(\rv-\rv')}}{\kv^2+\mtau}.
\label{eq_Greens_per}\eeq 
By using this equation together with \cref{eq_expIK_int}, one readily finds 
\beq (\Gcal\pbc,1)_\rv = \int_{\rv'} \Gcal\pbc(\rv,\rv') =  \frac{1}{\mtau} ,
\label{eq_pbc_Gw}\eeq
and
\beq (1,\Gcal\pbc,1) = \int_\rv \int_{\rv'} \Gcal\pbc(\rv,\rv') =  \frac{V}{\mtau} .
\label{eq_pbc_wGw}\eeq
Since $\chi=1/\mtau$ is the susceptibility within MFT, these results confirm the estimates in \cref{eq_constr_correl_int_est}.

In order to gain further insight into the effect of the constraint on the free energy, we insert \cref{eq_pbc_wGw} into \cref{eq_freeE_inhom} and obtain
\beq\begin{split} 
\Fcalc\pbc(\constrOP) &= \Hcal[\avOP] + \onehalf\sum_{\kv} \ln(\kv^2+\mtau) +\onehalf \ln \left(\frac{2\pi V}{\mtau}\right)\\
&= \Hcal[\avOP] + \onehalf\sum_{\substack{\kv\\ \kv\neq \b0}} \ln(\kv^2+\mtau) +\onehalf \ln \left(2\pi V\right).
\end{split}\label{eq_free_en_per}\eeq
As expected for this particular case, the effect of the constraint consists of, apart from generating an additional term $\propto \ln V$, the cancellation of the zero mode contribution from the free energy. (Regarding the dimensions of the last two terms in \cref{eq_free_en_per} see the discussions after \cref{eq_freeE_inhom,eq_func_int}.)
Constrained free energies of the type given in \cref{eq_free_en_per} have in fact  been studied previously in the context of finite-size criticality within the grand canonical ensemble \cite{rudnick_finite-size_1985, brezin_finite_1985, guo_hyperuniversality_1987, dohm_diagrammatic_1989, caracciolo_finite-size_2001, dohm_diversity_2008, dohm_critical_2011}. 
Here, we have obtained \cref{eq_free_en_per} by explicitly taking into account the OP constraint. In particular, the contribution $\propto \ln V$ in \cref{eq_free_en_per} is relevant for the calculation of the canonical CCF.
Finite-size properties of the free energy will be investigated further in Sec.\ \ref{sec_Fres_calc} below.

According to \cref{eq_correl_sigma}, the correlation function $\Ccalc\pbc$ of a constrained  system  with periodic \bcs is, within the Gaussian approximation, given by
\begin{multline} \Ccalc\pbc(\rv-\rv')=\Gcalc\pbc(\rv-\rv') = \bra\sigma(\rv)\sigma(\rv')\ket \\ = \Gcal\pbc(\rv-\rv') - \frac{1}{\mtau V} = \frac{1}{V} \sum_{\substack{\kv\\ \kv\neq \b0}} \frac{e^{\im\kv\cdot(\rv-\rv')}}{\kv^2+\mtau}.
\label{eq_correl_per}\end{multline}
As expected from \cref{eq_correl_int0}, one has $\int_\rv \Ccalc\pbc(\rv)=0$. Since $\Ccal\pbc(\rv)$ typically vanishes exponentially upon increasing $|\rv|$, the fact that $\Ccalc\pbc$ is shifted by the amount $-1/(\mtau V)$ relative to $\Ccal\pbc=\Gcal\pbc$ means that the constraint induces \emph{anti-correlations} of fluctuations at large distances. However, at least within the Gaussian approximation, the constraint does not cause $\Ccalc\pbc(\rv)$ to approach its limit for large $|\rv|$ differently than in the unconstrained system.
We finally note that, in the continuum limit, i.e., with $\sum_\kv \to \frac{V}{(2\pi)^d} \int_\kv$, \cref{eq_Greens_per} becomes $\Gcal\pbc\simeq (2\pi)^{-d}\int_\kv e^{\im\kv\cdot(\rv-\rv')}/(\kv^2+\mtau)$, showing that $\Gcal\pbc$ is a quantity of $\Ocal(V^0)$. 
Hence, for $V\to\infty$, which includes the case of a film with $A\to\infty$ at finite $L$, it follows from \cref{eq_correl_per} that $\Gcalc\pbc=\Gcal\pbc$, as anticipated in \cref{eq_Greens_constr_infvol}. 

\subsubsection{Dirichlet boundary conditions}

\begin{figure}[b]\centering
	\includegraphics[width=0.9\linewidth]{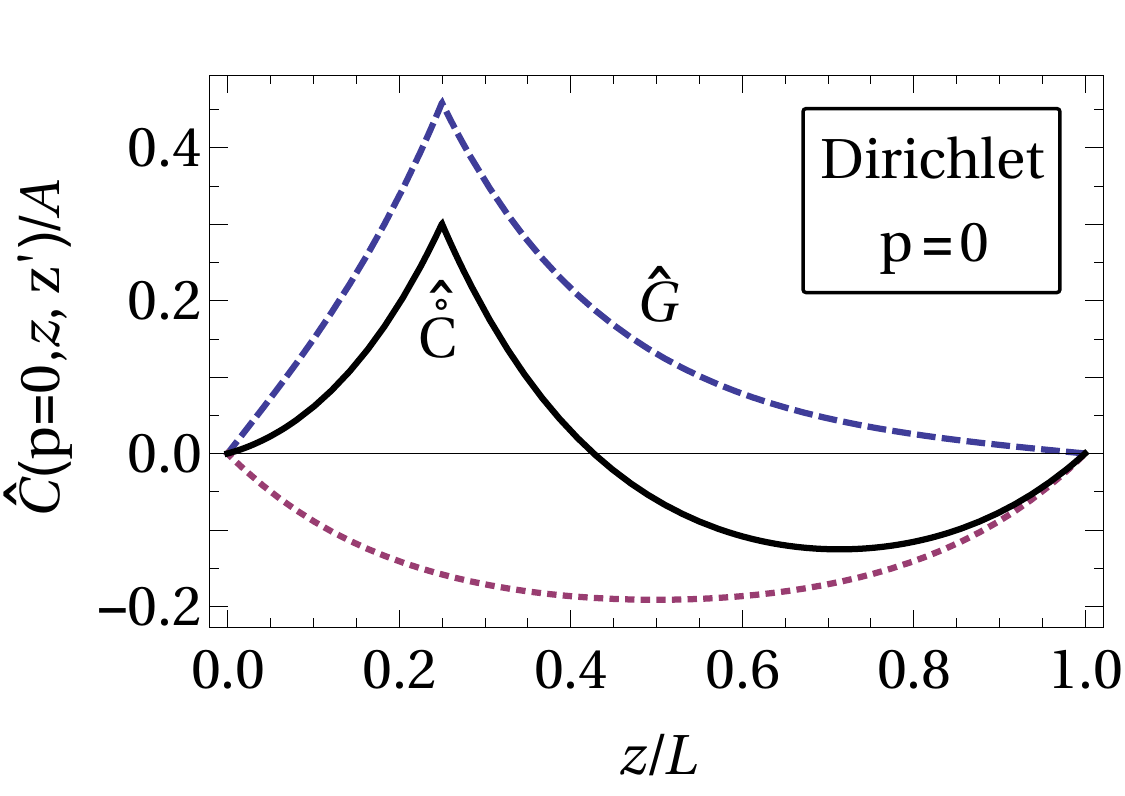}
	\caption{Correlation function $\hat\Ccalc\Dbc(\pv,z,z')$ [\cref{eq_correl_Dirichl}, solid black line], Green function $\hat\Gcal\Dbc(\pv,z,z')$ [\cref{eq_Greens_Dirichl}, dashed blue line, corresponding to the correlation function in the unconstrained case], and the correction term due to the constraint given by the second term on the r.h.s.\ of \cref{eq_correl_Dirichl} (dotted red line) for Dirichlet \bcs and $\pv=0$. For illustrative purposes, we have chosen here $\tau L^2=25$ and $z'=L/4$, but the qualitative features of the various curves (such as the cusp at $z=z'$) do not depend on this specific choice. Note that both terms on the r.h.s.\ of \cref{eq_correl_Dirichl} are proportional to $A$.}
	\label{fig_Greens_dirichl}
\end{figure}

\label{sec_Dir_gen}
For a system with Dirichlet boundary conditions at $z=0$ and $L$, the mean OP $\Mc$ within MFT is determined by [\cref{eq_MFT_EOS}]
\begin{equation}\begin{split} -\Mc''(z) + \tau \Mc(z) + \frac{1}{6}g \Mc^3(z) &= \lagparam ,\\ \text{with}\qquad \Mc(0) = \Mc(L) &= 0.
\label{eq_MFT_EOS_Dir} \end{split}\end{equation}
According to \cref{eq_sigma_inhom_bc,eq_sigma_inhom_exp}, the fluctuating component $\sigma$ of the OP [see \cref{eq_phi_Mc}] also fulfills Dirichlet \bcs, i.e.,
$\sigma(\rv_\parallel,0)=\sigma(\rv_\parallel,L)=0$.
For non-vanishing $\lagparam$, an explicit analytical solution of \cref{eq_MFT_EOS_Dir} is not available \footnote{For $\mu=0$, instead, see, e.g., the Appendix in Ref.\ \cite{gambassi_critical_2006}}. 
Although in principle \cref{eq_MFT_EOS_Dir} can be solved numerically, this poses additional challenges due to the presence of a spatially varying profile $\psi(z)$.
For the purpose of highlighting the effects of the constraint, in the following we focus on the simpler case $\lagparam=0$ (and $\tau\geq 0$), corresponding to a vanishing total mass $\constrOP=0$, for which \cref{eq_MFT_EOS_Dir} is solved by 
\beq \Mc=0=\avOP.
\label{eq_Dirichlet_avOP}\eeq 
Consequently, the set of orthonormal eigenfunctions $\eigenf_n$ [see \cref{eq_sigma_inhom_eigenf}] is given by
\beq \eigenf_n\Dbc(z) = \sqrt{\frac{2}{L}} \sin\left(\frac{\pi}{L} n z\right),\qquad n=1,2,\ldots,
\label{eq_Dirichlet_eigenf}\eeq
with eigenvalues [see \cref{eq_sigma_eigeneq}]
\beq E_n\Dbc = \left(\frac{\pi}{L}n\right)^2,
\label{eq_Dirichlet_eigenval}\eeq 
as in the grand canonical ensemble \cite{krech_free_1992}.
Since  
\beq \int_0^L \d z\, \eigenf_n\Dbc(z) = 
\begin{cases}
 \displaystyle \frac{2\sqrt{2 L}}{\pi\,n},\quad &\text{odd $n$}\\
 0,\quad &\text{even $n$} ,                             
\end{cases}
\eeq 
all eigenfunctions $\eigenf_n\Dbc$ with odd $n$ contribute to the fluctuation constraint in \cref{eq_fluct_constr_inhom}, in contrast to the case of periodic \bcs, in which only the mode [see \cref{eq_eigenf_per}] with $\kv=0$ contributes.
The Green function can be straightforwardly obtained by solving the differential equation in \cref{eq_Greens_func_inhom_DE}, subject to the \bcs given in \cref{eq_Greens_func_inhom_bc} (with $\psi=0$ and $\Lcal_s''[\psi]=c=\infty$), in the $pz$-representation [see \cref{eq_Greens_pz}]. This yields \cite{diehl_field-theoretical_1986, gambassi_critical_2006, kleinert_path_2009}
\begin{widetext}
\beq\begin{split} \hat\Gcal\Dbc(\pv,z,z') &= A\frac{\cosh(\kappa(L-|z-z'|)) - \cosh(\kappa(L-z-z'))}{2\kappa \sinh(\kappa L)},\qquad \text{with}\quad \kappa\equiv \sqrt{\pv^2+\tau} .
\end{split}
\label{eq_Greens_Dirichl}\eeq 
$\hat\Gcal\Dbc$ is symmetric with respect to $z\leftrightarrow z'$ and it has a finite limit for $\kappa\to 0$:
\beq \hat\Gcal\Dbc(\pv=\bv0, z,z')\Big|_{\tau=0} = A\, \min(z,z')\left(1-\frac{\max(z,z')}{L}\right).
\eeq 
The evaluation of the two-point correlation function $\Ccalc$ according to \cref{eq_correl_sigma} requires the calculation of the term $(\Gcal\Dbc,1)$ defined in  \cref{eq_inhom_fG}, which is easily inferred from \cref{eq_Greens_Dirichl}:
\begin{equation} (\Gcal\Dbc,1)_{z}= \frac{1}{A}\int_0^L \d z' \hat\Gcal\Dbc(\pv=0,z,z')  = \frac{\sinh(L\sqtau)-\sinh((L-z)\sqtau)-\sinh(z\sqtau)}{\tau \sinh(L\sqtau)}.
\label{eq_Dirichl_Gw}\end{equation}
This quantity is \emph{finite} for all $\tau\geq 0$ and, for $\tau\to 0$, it turns into $(\Gcal\Dbc,1)_z|_{\tau\to 0} = (L-z)z/2$. 
Further integrations over $\rv_\parallel$ and $z$ of \cref{eq_Dirichl_Gw} yield [see \cref{eq_inhom_fGf}]
\beq (1,\Gcal\Dbc,1) =\int_0^L \d z \int_0^L  \d z'\, \hat\Gcal\Dbc(\pv=0,z,z')= A L^3  \left[\frac{1}{\tau L^2} - \frac{2}{(\tau L^2)^{3/2}}\tanh\left(L\sqrt{\tau}/2\right) \right]. 
\label{eq_Dirichl_wGw}\eeq 
\end{widetext}
For $\tau L^2\to 0$, $(1,\Gcal\Dbc,1)$ is finite with the expansion
\beq (1,\Gcal\Dbc,1) \simeq 
\begin{cases}
\displaystyle AL^3 \left(\frac{1}{12} - \frac{1}{120}\tau L^2 \right)  &\text{for } \tau L^2 \to 0,\\[2mm]
\displaystyle \frac{AL^3 }{\tau L^2} &\text{for }  \tau L^2\to \infty ,
\end{cases}\label{eq_Dirichl_wGw_limits}
\eeq 
where the latter behavior also applies to the case $\tau\to \infty$ at fixed $L$. 
In contrast, for $L\to \infty$ at fixed $\tau$, one has $(1,\Gcal\Dbc,1)\simeq A L/\tau $.

In the $pz$-representation, the constraint-induced correlation function $\hat\Ccalc\Dbc$ [\cref{eq_correl_sigma_pz}] for Dirichlet \bcs is then given by
\begin{multline} \hat\Ccalc\Dbc(\pv,z,z')\\ = \hat\Gcal\Dbc(\pv,z,z') - A^2 \delta_{\pv,0}\frac{(1,\Gcal\Dbc)_{z}(1,\Gcal\Dbc)_{z'}}{(1,\Gcal\Dbc,1)}.
\label{eq_correl_Dirichl}\end{multline}
Note that both terms on the r.h.s.\ of \cref{eq_correl_Dirichl} are proportional to $A$.
Figure \ref{fig_Greens_dirichl} illustrates the typical behavior of $\hat\Ccalc\Dbc(\pv=\bv0,z,z')$ as a function of $z$ for a fixed value of $z'$.
In contrast to the corresponding unconstrained correlation function $\hat \Ccal$, which takes only positive values and is identical to $\hat\Gcal\Dbc$, $\hat\Ccalc\Dbc(\pv=\bv0,z,z')$ is modified such that, in accordance with \cref{eq_correl_int0}, the integral over either of its arguments $z$ and $z'$ vanishes.

\subsubsection{Neumann \bcs}
\label{sec_Neumann_bcs}

In the case of Neumann \bcs the mean OP $\Mc$ is determined by \cref{eq_MFT_EOS}.
For $\tau\geq 0$, this equation is solved by a constant OP profile $\Mc(z) = \avOP$, which fulfills
\beq \tau \avOP  + \frac{1}{6}g \avOP^3 = \lagparam
\label{eq_MFT_EOS_Neu}\eeq 
and satisfies the \bcs $\Mc'(0) = \Mc'(L)= 0$.
Equations \eqref{eq_sigma_eigeneq} and \eqref{eq_sigma_inhom_bc} for the eigenfunctions $\eigenf_n(z)$ turn into
\beq\begin{split} \left(-\pd_z^2 + \tau + \onehalf g\avOP^2 \right)\eigenf_n(z) &= (\tau + E_n)\eigenf_n(z), \\
\eigenf_n'(0)=\eigenf_n'(L) &=0.
\end{split}
\label{eq_eigenEq_Neu}\eeq 
As in the case of periodic \bcs [see \cref{eq_mtau}], the temperature parameter $\tau$ enters these expressions in combination with the mean OP $\avOP$ in the form given by \cref{eq_mtau}.
Equation \eqref{eq_eigenEq_Neu} is solved by the eigenfunctions
\beq 
\eigenf_n(z) = \begin{cases} \displaystyle
              \frac{1}{\sqrt{L}},\quad &n=0,\\[4mm]
	      \displaystyle \sqrt{\frac{2}{L}}\cos\left(\frac{\pi}{L}nz\right),\quad &n=1,2,\ldots
             \end{cases}
\label{eq_Neumann_eigenf}\eeq
with eigenvalues
\beq E_n = \onehalf g\avOP^2+ \left(\frac{\pi}{L}n\right)^2.
\eeq             
Since
\beq \int_0^L \d z\, \eigenf_n(z) = 
\begin{cases}
 \sqrt{L},\quad &n=0\\
 0,\quad & n=1,2,\ldots,                              
\end{cases}
\eeq 
Neumann \bcs entail a well defined zero mode, $\sigma_{\kv_\parallel=0,n=0}$, similarly to the case of periodic \bcs.
\Cref{eq_inhom_fG,eq_inhom_fGf}, upon using \cref{eq_Greens_inhom}, render the expressions
\beq (\Gcal\Nbc,1)_\rv = \frac{1}{\mtau} 
\label{eq_Neu_Gw}\eeq
and
\beq (1,\Gcal\Nbc,1) = \frac{V}{\mtau} ,
\label{eq_Neu_wGw}\eeq
which coincide with the ones obtained for periodic \bcs and reported in \cref{eq_pbc_Gw,eq_pbc_wGw}.
The (unconstrained) Green function in the $pz$-representation [see \cref{eq_Greens_pz}] is given by \cite{diehl_field-theoretical_1986}
\begin{multline} \hat\Gcal\Nbc(\pv,z,z') \\ = A\frac{\cosh(\kappa(L-|z-z'|)) + \cosh(\kappa(L-z-z'))}{2\kappa \sinh(\kappa L)},
\label{eq_Greens_Neu}\end{multline}
with $\kappa\equiv \sqrt{\pv^2+\mtau}$.
According to \cref{eq_Neu_Gw,eq_Neu_wGw,eq_G_redef}, the presence of the constraint simply gives rise to an overall $\tau$-dependent shift of the unconstrained correlation function, as it is the case for the periodic \bcs discussed above.

\subsection{Canonical free energy}
\label{sec_Fres_calc}
Here, we discuss, within the one-loop (Gaussian) approximation, the canonical free energy $\Fcalc$ [\cref{eq_freeE_inhom}] for finite systems with aspect ratio 
\beq \rho\equiv \frac{L}{A^{1/(d-1)}}
\label{eq_aspectratio}\eeq 
and exhibiting periodic, Dirichlet, or Neumann \bcs at both surfaces, located at $z=0$ and $z=L$.
In all three cases periodic \bcs are imposed in the remaining lateral directions and phase separation is excluded.
Analytical results for the finite-size free energy of constrained systems with periodic \bcs have been presented for $\rho=1$, e.g., in Refs.\ \cite{brezin_investigation_1982, rudnick_finite-size_1985, eisenriegler_finite_1985, eisenriegler_helmholtz_1987, esser_field_1995} and, for $\avOP=0$ and arbitrary $\rho$, in Refs.\ \cite{dohm_diversity_2008, dohm_critical_2011}. The finite-size free energy for Dirichlet \bcs and cubical volumes (i.e., $\rho=1$) has been studied, e.g., in Refs.\ \cite{dohm_diagrammatic_1989, dohm_pronounced_2014}.
With the exception of Ref.\ \cite{eisenriegler_helmholtz_1987}, these studies aimed, however, for the grand canonical free energy [which, according to \cref{eq_Z_gc_c_rel}, can be constructed from the canonical free energy discussed here]. 
Instead, here we focus on the canonical ensemble and, extending previous studies, we allow also for a nonzero mean OP $\avOP$ in the case of periodic and Neumann \bcs.
The details of the corresponding perturbative calculation are deferred to \cref{app_calc_fcan}.
The results reported here are subsequently improved in Sec.~\ref{sec_scaling} by means of renormalization-group theory. 

Analogously to what is expected for the grand canonical free energy $\Fcal$ \cite{privman_finite-size_1990, brankov_theory_2000}, the \emph{canonical} free energy $\Fcalc$ of a $d$-dimensional system of volume $V=AL$ decomposes into a bulk ($\,\fcalc_b$), a surface ($\,\fcalc_s$) and a residual finite size contribution $\Fcalc\res$ [see also \cref{eq_Fgc_decomp} below]:
\beq \Fcalc(\tau,\avOP,A,L) = A L \ring\fcal_b(\tau,\avOP) + A \ring\fcal_{s}(\tau,\avOP) +  A \Fcalc\res\left(\tau ,\avOP ,\rho ,L\right) .
\label{eq_Fcan_decomp}\eeq 
We anticipate that in our case the \resFE (per area $A$) $\Fcalc\res$ depends on the area only via the aspect ratio $\rho$. 
Explicitly, from \cref{eq_freeE_inhom}, the total, regularized free energies for periodic, Dirichlet, and Neumann \bcs turn out to be [see \cref{eq_finiteAR_F_pbc,eq_finiteAR_F_Dir,eq_finiteAR_F_Neu}]
\begin{widetext}
\begin{subequations}\begin{align} 
\Fcalc\pbc &= A L \left[\Lcal_b(\avOP)-\frac{A_d}{d} \mtau^{d/2} \right] +  \onehalf AL^{-d+1} \modesum\pbc_{d,\reg}(\mtau L^2,\rho)  + \delta F\pbc(\mtau,A,L),\\
\Fcalc\Dbc &=  - A L \frac{A_d}{d} \tau^{d/2} + \frac{A}{2} \frac{A_{d-1}}{d-1} \tau^{(d-1)/2}  + \onehalf A L^{-d+1} \modesum\Dbc_{d,\reg}(\tau L^2,\rho) + \delta F\Dbc(\tau,A,L), \\
\Fcalc\Nbc &= A L\left[\Lcal_b(\avOP) - \frac{A_d}{d} \mtau^{d/2}\right]  - \frac{A}{2} \frac{A_{d-1}}{d-1} \mtau^{(d-1)/2} + \onehalf A L^{-d+1} \modesum\Nbc_{d,\reg}(\mtau L^2,\rho)  + \delta F\Nbc(\mtau,A,L), \label{eq_F_can_pert_Neu}
\end{align}\label{eq_F_can_pert}\end{subequations}
where 
\beq A_d \equiv -(4\pi)^{-d/2}\Gamma\left(1-d/2\right)
\label{eq_Ad}
\eeq
and $\mtau$ is defined in \cref{eq_mtau} (recall that, for Dirichlet \bcs, $\mtau=\tau$ because in that case we focus on the choice $\avOP=0$ [see \cref{eq_Dirichlet_avOP}]). The quantities $\modesum\pbc_{d,\reg}$, $\modesum\Dbc_{d,\reg}$, and $\modesum\Nbc_{d,\reg}$ represent the regularized dimensionless expression of the corresponding mode sum, i.e., the second term in \cref{eq_freeE_inhom}. They are given by [see \cref{eq_modesum_per,eq_modesum_D_reg_scalf,eq_modesum_N_reg}] 
\begin{subequations}
\begin{align} 
\modesum\pbc_{d,\reg}(\mtscal,\rho) &= \int_0^\infty \d y\, y^{-1} \exp\left(-\frac{\mtscal y}{4\pi^2}\right) \left\{\left(\frac{\pi}{y}\right)^{d/2} - \left[\rho\, \vartheta(\rho^2 y)\right]^{d-1} \vartheta(y) \right\},\label{eq_S_pbc}\\
\modesum\Dbc_{d,\reg}(\tscal,\rho) &= 2^{-d} \modesum\pbc_{d,\reg}(4\tscal, 2\rho) - \onehalf \rho^{d-1} \modesum\pbc_{d-1,\reg}(\tscal/\rho^2, 1),\label{eq_S_Dbc} \\
\modesum\Nbc_{d,\reg}(\mtscal,\rho) &= 2^{-d} \modesum\pbc_{d,\reg}(4\mtscal, 2\rho) + \onehalf \rho^{d-1} \modesum\pbc_{d-1,\reg}(\mtscal/\rho^2, 1), \label{eq_S_Nbc}
\end{align}\label{eq_Sreg_coll}\end{subequations}
\end{widetext}
for periodic, Dirichlet, and Neumann \bcs, respectively. In \cref{eq_S_pbc}, $\vartheta$ is a Jacobi theta function [see \cref{eq_thetafunc}].
In \cref{eq_Sreg_coll}, we introduced the notions $\tscal=\tau L^2$ and $\mtscal=\mtau L^2$.
For all these boundary conditions, far from criticality one has 
\beq \modesum_{d,\reg}(\mtscal\to \infty,\rho)\to 0.
\label{eq_modesum_convTInfty} \eeq 
For periodic and Neumann \bcs, $\modesum_{d,\reg}(\mtscal,\rho)$ diverges logarithmically upon approaching bulk criticality, i.e., for $\mtscal\to 0$ [see \cref{eq_modesum_per_div}]:
\beq \modesum^{\text{(p,N)}}_{d,\reg}(\mtscal\to 0,\rho)\simeq \rho^{d-1}\ln \mtscal,
\label{eq_modesum_divTc} \eeq 
while $\modesum\Dbc_{d,\reg}$ is finite in that limit.
We shall return to this aspect in \cref{sec_F_gc}.
The quantities $\delta F^{(p,D,N)}$ in \cref{eq_F_can_pert} represent the correction stemming from the OP constraint in \cref{eq_constr_spec}. Within the one-loop approximation, one has [see Eqs.~\eqref{eq_pbc_wGw}, \eqref{eq_Dirichl_wGw}, and \eqref{eq_Neu_wGw}] 
\begin{widetext}
\begin{subequations}
  \begin{empheq}[left={\displaystyle \delta F(\mtau,A,L)=\onehalf \ln \left[ 2\pi(1,\Gcal,1) \right]= \empheqlbrace}]{align}
  & \onehalf \ln  \frac{2\pi V}{\mtau}, & \text{periodic,\,\,\,} \label{eq_constrcorr_per}\\
  & \onehalf \ln \left(2\pi V L^2  \left[\frac{1}{\tau L^2} - \frac{2}{(\tau L^2)^{3/2}}\tanh\left(L\sqrt{\tau}/2\right) \right]\right), & \text{Dirichlet,   } \label{eq_constrcorr_Dir}\\
  & \onehalf \ln \frac{2\pi V}{\mtau}, & \text{Neumann,} \label{eq_constrcorr_Neu}
   \end{empheq}\label{eq_constrcorr}
\end{subequations}
for the indicated \bcs.
It is interesting to note that the same form of the constraint correction $\delta F\pbc$ as in \cref{eq_constrcorr_per} is obtained also for an uncorrelated Gaussian field [which is described by the Hamiltonian $\int_V \d^d r\, \tau \phi^2(\rv)/2$ instead of the one in \cref{eq_landau_fe}] in a finite volume (see \cref{app_lattice}).
Since the perturbation theory in the canonical ensemble is based on the modified Green function $\Gcalc$ [see \cref{eq_G_redef}], higher-order constraint corrections are, however, sensitive to the presence of a finite correlation length in the system  [see, e.g., \cref{eq_free_en2loop}].

In view of the formulation of the scaling theory in \cref{sec_scaling} below, it is convenient to cast $\delta F$ given by \cref{eq_constrcorr} into the form
\beq \delta F(\mtau,A,L) = \begin{cases}
 \delta F\st{s}\pNbc(\mtau L^2,\rho)+\delta F\st{ns}(L), \quad &\text{periodic, Neumann,}\\
 \delta F\st{s}\Dbc(\tau L^2,\rho)+\delta F\st{ns}(L), \quad &\text{Dirichlet,}
 \end{cases}
\label{eq_constrcorr_def_sns}\eeq 
where
\begin{subequations}
 \begin{align}
 \delta F\st{s}\pNbc(\mtscal,\rho) &=  -\frac{1}{2} \ln \frac{\rho^{d-1} \mtscal}{2\pi} , \label{eq_constrcorr_s_perNeu}\\
 \delta F\st{s}\Dbc(\tscal,\rho) &=  \onehalf \ln \left(\left(\frac{1}{\tscal} - \frac{2}{\tscal^{3/2}} \tanh\left(\sqrt{\tscal}/2\right) \right) 2\pi\rho^{-d+1} \right), \label{eq_constrcorr_s_Dir}
 \end{align}\label{eq_constrcorr_scal}
\end{subequations}
\end{widetext}
is a ``scaling'' contribution, which is specific to each boundary condition, while 
\beq \delta F\st{ns}(L) = \onehalf \ln L^{d+2}
\label{eq_constrcorr_nsterm}\eeq 
is a ``non-scaling'' contribution, which is common to all \bcs considered here.
Upon reinstating the lattice spacing $a$, which we formally disregarded in taking the continuum limit [see \cref{eq_func_int}], the arguments of the logarithms in \cref{eq_constrcorr,eq_constrcorr_nsterm} are divided by a factor $a^{d+2}$, which renders them dimensionless [compare \cref{eq_lattice_freeE}].

According to \cref{eq_Fcan_decomp}, the \resFE \emph{per volume}, $\fcalc\res$, is given by
\beq \fcalc\st{res} = \frac{\Fcalc - AL \fcalc_b - A \fcalc_s}{A L} = \frac{\Fcalc\res}{L} ,
\label{eq_Fres_def}\eeq 
which, upon using \cref{eq_F_can_pert,eq_constrcorr_def_sns} and noting that
$AL=L^d/\rho^{d-1}$ [see \cref{eq_aspectratio}], can be expressed as
\beq \ring\fcal\res(\tau,\avOP,\rho,L) =  L^{-d} \left[ \ring\Theta(\mtau L^2,\rho) + \rho^{d-1} \delta F\st{ns}( L)\right].
\label{eq_Fres_c}\eeq
The scaling function $\ring\Theta$ introduced in this expression contains the contribution $\delta F\st{s}$ from the constraint correction given in \cref{eq_constrcorr_scal}:
\beq \ring\Theta(\mtscal,\rho) = \onehalf\modesum_{d,\reg}(\mtscal,\rho) + \rho^{d-1} \delta F\st{s}(\mtscal,\rho).
\label{eq_Fres_c_scalfunc}\eeq 
Note that the divergence of $\modesum_{d,\reg}^{\text{(p,N)}}$ for $\mtscal\to 0$ [see \cref{eq_modesum_divTc}] is canceled by that of $\delta F\st{s}^{\text{(p,N)}}$ [\cref{eq_constrcorr_s_perNeu}], such that $\ring\Theta$ remains finite for $\mtscal=0$.
In the case of Dirichlet \bcs, neither $\modesum\Dbc_{d,\reg}$ [\cref{eq_S_Dbc}] nor $\delta F\st{s}\Dbc$ [\cref{eq_constrcorr_s_Dir}] diverge in the limit $\tscal\to 0$.
Accordingly, \emph{at} bulk criticality, $\fcalc\res$ [\cref{eq_Fres_c}] generally scales $\propto L^{-d}$ while $\Fcalc\res$ [\cref{eq_Fcan_decomp,eq_Fres_def}] scales $\propto L^{-(d-1)}$, as in the grand canonical ensemble.
Since $\modesum_{d,\reg}(\mtscal\to \infty)\to 0$, the canonical scaling function $\ring\Theta$ turns out to diverge logarithmically for $\mtscal\gg 1$, i.e.,
\beq \ring\Theta(\mtscal\gg 1,\rho) \simeq -\onehalf\rho^{d-1} \ln \frac{\mtscal \rho^{d-1}}{2\pi},
\label{eq_Fres_largeTau}\eeq 
due to the term $\delta F\st{s}$ in \cref{eq_constrcorr_scal}. This behavior applies to all three \bcs considered here. In the thin-film limit, one has $\ring\Theta(\mtscal\gg 1, \rho\to 0)\to 0$.
The \resFE will be discussed further in \cref{sec_discussion}.

\subsection{Grand canonical free energy}
\label{sec_F_gc}

For comparison, here we report the corresponding expressions for the one-loop free energy $\Fcal(\tau,h,A,L)$ in the grand canonical ensemble. Field theory yields the well-known perturbative expression for $\Fcal$ in \cref{eq_freeE_Gauss_gc} \cite{brezin_investigation_1982, rudnick_finite-size_1985, eisenriegler_finite_1985, eisenriegler_helmholtz_1987, krech_free_1992, esser_field_1995, dohm_diversity_2008, dohm_critical_2011, dohm_diagrammatic_1989, dohm_pronounced_2014}.
The corresponding regularized forms of $\Fcal$ coincide with the ones in \cref{eq_F_can_pert} for the various \bcs, except for the fact that $\Lcal_b(\avOP)$ is replaced by $\Lcal_b(\avOP,h)\equiv \Lcal_b(\avOP)-h\avOP$ [see \cref{eq_landau_fe_gc}] and that the constraint induced term $\delta F$ [\cref{eq_constrcorr}] is absent:
\begin{widetext}
\begin{subequations}\begin{align} 
\Fcal\pbc &= A L \left(\Lcal_b(\avOP,h)-\frac{A_d}{d} \mtau^{d/2} \right) +  \onehalf AL^{-d+1} \modesum\pbc_{d,\reg}(\mtau L^2,\rho), \label{eq_F_gc_pert_per} \\
\Fcal\Dbc &=  - A L \frac{A_d}{d} \tau^{d/2} + \frac{A}{2} \frac{A_{d-1}}{d-1} \tau^{(d-1)/2}  + \onehalf A L^{-d+1} \modesum\Dbc_{d,\reg}(\tau L^2,\rho),  \\
\Fcal\Nbc &= A L\left(\Lcal_b(\avOP,h) - \frac{A_d}{d} \mtau^{d/2}\right)  - \frac{A}{2} \frac{A_{d-1}}{d-1} \mtau^{(d-1)/2} + \onehalf A L^{-d+1} \modesum\Nbc_{d,\reg}(\mtau L^2,\rho). \label{eq_F_gc_pert_Neu}
\end{align}\label{eq_F_gc_pert}\end{subequations}
\end{widetext}
The expressions for $\modesum_{d,\reg}$ are reported in \cref{eq_Sreg_coll}.
In the case of periodic and Neumann \bcs, the mean OP $\avOP$ is a function of the bulk field $h$ via the equation of state, which, within the presently considered approximation, is given by \cref{eq_MFT_EOS_gc}:
\beq h = \pd_\avOP \Lcal_b(\avOP) =  \tau \avOP + \frac{1}{6}g \avOP^3 .
\label{eq_EOS_gc_MFT}\eeq 
Equation \eqref{eq_EOS_gc_MFT} takes already into account that for periodic and Neumann \bcs the system is spatially homogeneous so that $\Mc(\rv)=\avOP$. For Dirichlet \bcs, as explained below \cref{eq_MFT_EOS_Dir}, we focus on the simple case $h=0$, i.e., $\avOP=0$.
Note that \cref{eq_EOS_gc_MFT} in fact coincides with the equation of state for the corresponding \emph{bulk} system. Finite-size corrections enter through higher loop orders, analogous to \cref{eq_EOS_oneloop_can}.
The renormalized forms of these expressions will be discussed in \cref{sec_scaling} below.

As it is well known from general finite-size scaling arguments \cite{barber_finite-size_1983, privman_finite-size_1990, brankov_theory_2000}, the grand canonical free energy $\Fcal$ of a confined $d$-dimensional system of volume $V=AL$ decomposes into a bulk ($\fcal_b$), a surface ($\fcal_s$) and a residual finite size $\Fcal\res$ contribution [compare \cref{eq_Fcan_decomp}]:
\beq \Fcal(\tau,h,A,L) = A L \fcal_b(\tau,h) + A \fcal_{s}(\tau,h) +  A \Fcal\res\left(\tau ,h ,\rho ,L\right) .
\label{eq_Fgc_decomp}\eeq 
Crucially, in order to be able to cast \cref{eq_F_gc_pert} into the form prescribed by \cref{eq_Fgc_decomp}, $\Fcal$ must be first expressed as a function of the bulk field $h$, which is the relevant thermodynamic control parameter in the grand canonical ensemble. 
In their present form, the expressions in \cref{eq_F_gc_pert} are still explicit functions of $\avOP$.
To proceed, any $\avOP$ occurring in \cref{eq_F_gc_pert} must therefore be replaced by the $\avOP(h)$ determined from the equation of state.
Before turning to the specific approximation for the latter as given by \cref{eq_EOS_gc_MFT}, for the time being we adopt a generic equation of state of the form $\avOP=\avOP(\tau,h,\rho,L)$.
In this case, bulk and surface free energies can be identified based on their scaling behavior with $L$ according to \cref{eq_Fgc_decomp}.
In particular, the \emph{bulk} limit is obtained by taking $L\to\infty$ and by assuming $A$ to be either constant or to scale with a certain positive power of $L$ (the precise formulation does not matter here). 
Accordingly, from \cref{eq_F_gc_pert} the bulk free energy follows as
\begin{multline} f_b(\tau,h) = \lim_{L\to\infty}\frac{\Fcal}{AL} =  \fcalc_b(\tau,\avOP_b) - h \avOP_b  \\ = \Lcal_b(\avOP_b,h) - \frac{A_d}{d} (\mtau(\tau,\avOP_b))^{d/2},
\label{eq_Fgc_bulk}\end{multline}
with the bulk OP given by 
\beq \avOP_b=\avOP_b(\tau,h)=\lim_{L\to\infty}\avOP(\tau,h,\rho,L).
\eeq 
The \emph{surface} free energy (per area $A$) is defined as the $L$-independent part of the total free energy. Therefore it can be obtained as the dominant contribution to $\Fcal$ in the limit $L\to\infty$ after subtracting the bulk contribution $\fcal_b$:
\beq \fcal_s(\tau,h) = \frac{1}{A}\lim_{L\to\infty} \left[\Fcal(\tau,h,\rho,L) - AL\fcal_b(\tau,h)\right] .
\eeq
Note that the limit $L\to\infty$ implies again the use of the bulk OP for the evaluation of $\fcal_s$.
Neumann \bcs are the only case considered in this study for which $\fcal_s$ does not vanish, and \cref{eq_F_gc_pert_Neu} yields
\beq \fcal_s\Nbc(\tau,h) = - \frac{1}{2} \frac{A_{d-1}}{d-1} \mtau^{(d-1)/2}\Big|_{\avOP=\avOP_b}.
\label{eq_Fgc_surf_Neu}\eeq 
The \emph{residual} finite-size free energy per volume $\fcal\res=\Fcal\res/V$ in the grand canonical ensemble follows according to \cref{eq_Fgc_decomp} as
\begin{multline} \fcal\res(\tau,h,\rho,L) \\ = \fcal(\tau,h,\rho,L)\big|_{\avOP(h)} - \fcal_b(\tau,h)\big|_{\avOP_b(h)} - \frac{1}{L} \fcal_s(\tau,h)\big|_{\avOP_b(h)},
\label{eq_Fres_gc_def}\end{multline}
with $\fcal\equiv \Fcal/V$.
Since one generally expects $\avOP\neq \avOP_b$ due to finite-size effects (see, e.g., Refs.\ \cite{chen_nonmonotonic_2000, gross_critical_2016}), it follows from \cref{eq_Fres_gc_def} that the \resFE in this case does not necessarily coincide with the last term in each of the Eqs.\ \eqref{eq_F_gc_pert_per}$-$\eqref{eq_F_gc_pert_Neu}, in spite of the fact that they apparently display the appropriate scaling as a function of $L$, but for fixed $\avOP$ only.
However, within the presently considered approximation of $\Ocal(g^0)$ for the equation of state in \cref{eq_EOS_gc_MFT}, finite-size effects are absent and therefore 
\beq \avOP(h)=\avOP_b(h)+ \Ocal(g).
\label{eq_gc_bulkOP}\eeq 
Accordingly, \cref{eq_Fres_gc_def,eq_F_gc_pert} immediately yield
\beq \fcal\res (\tau,h,\rho,L) =L^{-d}\Theta(\mtau L^2,\rho)\big|_{\avOP(h)} 
\label{eq_Fres_gc_meanfield}\eeq 
with the scaling function 
\beq \Theta(\mtscal,\rho) = \onehalf\modesum_{d,\reg}(\mtscal,\rho) + \Ocal(g).
\label{eq_Fres_gc_scalfunc}\eeq 
The subscript on the r.h.s.\ of \cref{eq_Fres_gc_meanfield} indicates that $\mtau$ [\cref{eq_mtau}] is to be evaluated by using $\avOP=\avOP(h)$. 

According to \cref{eq_Fres_gc_scalfunc,eq_modesum_divTc}, the grand canonical \resFE for periodic and Neumann \bcs diverges logarithmically for $\mtscal\to 0$ in the case $\rho>0$, while this divergence is absent in the thin-film limit ($\rho=0$).
This behavior is a well-known artifact of perturbation theory and stems from the contribution of the zero mode to the free energy \cite{rudnick_finite-size_1985, brezin_finite_1985, schloms_minimal_1989,diehl_fluctuation-induced_2006, dohm_diversity_2008, gruneberg_thermodynamic_2008} (see also \cref{eq_free_en_per} and the related discussion).
In order to overcome this problem, the zero mode must be treated non-perturbatively, which results in a finite \resFE for $\mtscal=0$. 
Since here we are interested in a comparison between the canonical and grand-canonical ensemble, we do not consider such improvements of the theory further. 
Instead, we note that, for the grand canonical ensemble in the case $\rho>0$, the perturbative expressions of the \resFE and the CCF for periodic and Neumann \bcs are reliable only for $\mtscal\gtrsim 1$. Since Dirichlet \bcs do not involve zero-mode fluctuations, the perturbative results for $\Theta\Dbc$ are well behaved for all $\tscal\geq 0$.

\section{Renormalization and scaling}
\label{sec_scaling}

\subsection{Residual finite-size free energy}
\subsubsection{Canonical \resFE}
\label{sec_renorm_F_c}

In order to be applicable in the critical regime, the perturbative results of Sec.~\ref{sec_spec} must be renormalized \cite{zinn-justin_quantum_2002, amit_field_2005}.
On general grounds, it is expected that the short-distance singularities of field theory are not affected by the finiteness of the volume of the system \cite{brezin_investigation_1982, guo_hyperuniversality_1987,guo_erratum:_1989, amit_field_2005, zinn-justin_quantum_2002}. 
As it has been shown in Ref.\ \cite{eisenriegler_helmholtz_1987}, renormalization based on minimal subtraction of dimensional poles in conjunction with an expansion in $\epsilon=4-d$ is applicable also in the canonical ensemble and it requires the same additive and multiplicative counterterms which are known from the grand canonical case \cite{diehl_field-theoretical_1986, krech_free_1992}.
In particular, the findings of Ref.\ \cite{eisenriegler_helmholtz_1987} apply also to the present study, because here we focus on planar surfaces only (compare Ref.\ \cite{eisenriegler_finite_1985}) and do not consider surface correlation functions.
Furthermore, because off the surfaces neither Dirichlet nor Neumann \bcs introduce new dimensional poles as $d\nearrow 4$, the same counterterms as for periodic \bcs can be used in these cases as well.
The renormalized (grand) canonical free energy can be constructed by following the same steps as in Refs.\ \cite{diehl_field-theoretical_1986,eisenriegler_helmholtz_1987, krech_free_1992}; for further details we refer to these studies \footnote{Conventionally, the definition of the renormalized \emph{total} free energy involves the subtraction of the bare free energy and of its first two temperature derivatives taken at a certain reference temperature (see, e.g., Eq.~(3.1) in Ref.\ \cite{krech_free_1992}). With such a prescription, the non-scaling contribution $\delta F\st{ns}$ [see \cref{eq_F_can_pert,eq_constrcorr_nsterm}] to the canonical free energy is eliminated. However, such a definition also introduces a shift of the renormalized \resFE, which is undesired for our purposes. Consistently with the literature (see, e.g., Refs.\ \cite{krech_free_1992,krech_casimir_1994,gruneberg_thermodynamic_2008}), we therefore proceed by studying the un-subtracted but renormalized \resFE.}. 
Along these lines one obtains the expected scaling laws for the free energy near the infrared renormalization-group (RG) fixed point, at which, within the $\epsilon$-expansion, the renormalized coupling constant $u$ is given by 
\beq u^* = \frac{1}{3}\epsilon + \Ocal(\epsilon^2).
\label{eq_u_fixpt}\eeq 
We adopt the same conventions as in Refs.\ \cite{diehl_field-theoretical_1986,krech_free_1992} and define $g=\mu^\epsilon Z_u r u$, where $\mu$ is the RG momentum scale, $Z_u$ is the standard $Z$-factor for the coupling constant, and
\beq r\equiv (4\pi)^{d/2}.
\label{eq_u_numfact}\eeq

In the following, we focus directly on the renormalization of the \resFE, which turns out to not require any additive renormalization in order to cancel its dimensional poles (see also Ref.\ \cite{gruneberg_thermodynamic_2008}). 
From \cref{eq_Fres_c}, one obtains the following finite-size scaling form of the \resFE per volume and per $k_B T_c$:
\begin{multline} \fcalc\resR(\tren,\avOP_R, \rho,L) = L^{-d} \\ \times \Bigg[ \,\ring\Theta\left(\mtscal\left(\tscal = \left(\frac{L}{\amplXip}\right)^{1/\nu} t , \phiscal= \left(\frac{L}{\amplXiPhi}\right)^{\beta/\nu} \avOP_R\right), \rho \right) \\ + \rho^{d-1}\delta F\st{ns}(L) \Bigg] ,
\label{eq_Fres_c_scal_L} \end{multline}
where $\tren=(T-T_c)/T_c$ is the reduced temperature [\cref{eq_t_red}]. We recall that $\delta F\st{ns}(L) = (1/2)\ln L^{d+2}$ [see \cref{eq_constrcorr_nsterm}] is a contribution stemming from the constraint which cannot be expressed solely in terms of scaling variables.
The subscript $R$ indicates a dimensionless, renormalized quantity.
Specifically, $\tren$ and $\avOP_R$ can be related to the correlation length $\xi(\tren,\avOP_R)$ via the non-universal critical amplitudes $\amplXip$ and $\amplXiPhi$ \cite{pelissetto_critical_2002}:
\begin{subequations}\begin{align}
 \xi(\tren\to 0^+, \avOP_R=0) &= \amplXip \tren^{-\nu}, \label{eq_def_xip}\\
 \xi(\tren=0, \avOP_R\to 0) &= \amplXiPhi \avOP_R^{-\nu/\beta}. \label{eq_def_xiPhi}
\end{align}\label{eq_ampl_def1}\end{subequations}
The amplitude $\amplXiPhi$ can be related to the amplitude $\amplXimu$ of the correlation length at $T_c$ as a function of the bulk field [see \cref{eq_amplXiPhi_def} below].
The scaling variable corresponding to $\mtau$, which has been introduced in \cref{eq_mtau}, is defined by
\beq \mtscal(\tscal,\phiscal) \equiv  \tscal+\onehalf r u^* \phiscal^2.
\label{eq_tscal_mod}\eeq
The expressions of the scaling functions $\ring\Theta$ are reported in \cref{eq_Fres_c_scalfunc,eq_constrcorr_scal,eq_Sreg_coll} for the respective \bcs. Consistently with the considered one-loop approximation for the free energy, the scaling functions $\ring\Theta$ are to be evaluated to $\Ocal(\epsilon^0)$, i.e., for $d=4$.
Since the constraint-induced terms $\delta F\st{s,ns}$ turn out to be $\propto \rho^{d-1}$, they are negligible in the thin-film limit $\rho\to 0$.
They are also negligible, together with $\ring\fcal\res$, in the thermodynamic limit obtained for $L\to \infty$.
We note that $\phiscal^2\sim O(u^{-1})$, such that the last term in \cref{eq_tscal_mod} is actually of $\Ocal(\epsilon^0)$.
This expression and the equation of state [see \cref{eq_EOS_MFT_scalf} below] are the only instances in which within the approximation $\Ocal(\epsilon^0)$ the renormalized coupling constant $u$ appears in the final expressions of the \resFE and of the CCF.

\subsubsection{Grand canonical \resFE}
\label{sec_renorm_F_gc}

Here, we summarize the scaling forms obtained for the grand canonical \emph{residual} free energy based on the renormalization of the perturbative results in \cref{sec_F_gc}.
In particular, at the fixed point, the RG yields the scaling property of the renormalized grand canonical residual free energy per volume and per $k_B T_c$ (see, e.g., Refs.\ \cite{diehl_field-theoretical_1986, krech_free_1992, gruneberg_thermodynamic_2008})
\begin{multline}
\fcal\resR(\tren,h_R,\rho,L) \\ = L^{-d}\, \widetilde\Theta\left(\tscal=\left(\frac{L}{\amplXip}\right)^{1/\nu} t, \hscal= \left(\frac{L}{\amplXimu}\right)^{\beta\delta/\nu}  h_R, \rho \right).
\label{eq_Fres_gc_scal_L}\end{multline}
The scaling function $\widetilde\Theta(\tscal,\hscal,\rho)$ is related to $\Theta(\mtscal(\tscal,\phiscal),\rho)$ in \cref{eq_Fres_gc_scalfunc} via 
\beq \widetilde\Theta(\tscal,\hscal,\rho) = \Theta(\mtscal\left(\tscal,\phiscal(\tscal,\hscal,\rho)\right),\rho),
\label{eq_Fres_gc_pert_equal}\eeq 
and $\phiscal(\tscal,\hscal,\rho)$ is the scaling form of the equation of state (see \cref{eq_phi_bulk_scalfunc} below).
The renormalized bulk field $h_R$ can be introduced on the basis of the correlation length \cite{pelissetto_critical_2002}:
\beq 
 \xi(t=0, h_R\to 0) = \amplXimu h_R^{-\nu/\Delta}, \label{eq_defgc_ximu}
\eeq 
which also serves as a definition of the amplitude $\amplXimu$. It is useful to recall the relation $\Delta=\delta\beta$ between standard bulk critical exponents.
We emphasize that \cref{eq_def_xiPhi} can be obtained from \cref{eq_defgc_ximu} and from the relation  $\avOP_R(t=0,h_R\to 0) = \amplPhimu h_R^{1/\delta}$ \cite{pelissetto_critical_2002}, which defines the amplitude $\amplPhimu$ and thereby yields the expression 
\beq \amplXiPhi = \amplXimu \left(\amplPhimu\right)^{\nu/\beta}
\label{eq_amplXiPhi_def}\eeq 
for the amplitude $\amplXiPhi$.

The equation of state exhibits the scaling form
\begin{multline} 
h_R(\tren,\avOP_R,\rho,L) \\= L^{-\beta\delta/\nu} \hscal\Bigg(\tscal=\left(\frac{L}{\amplXip}\right)^{1/\nu}t,   \phiscal= \left(\frac{L}{\amplXiPhi}\right)^{\beta/\nu} \avOP_R,\rho\Bigg),
\label{eq_EOS_fss_h}\end{multline}
where the scaling function $\hscal$ results from \cref{eq_EOS_gc_MFT} within the approximation $\Ocal(\epsilon^0)$ as
\beq\begin{split} \hscal(\tscal,\phiscal,\rho) = \tscal\phiscal + \frac{1}{6}r u^* \phiscal^3 ,
\end{split}\label{eq_EOS_MFT_scalf}\eeq 
which is, in fact, independent of $\rho$.
Within that approximation, this equation of state applies to all \bcs and it coincides with the one in the bulk [see \cref{eq_gc_bulkOP}].
An alternative form of the equation of state can be obtained from the total grand canonical free energy $\fcal_R$ via the basic thermodynamic relation $\avOP_R=\pd \fcal_R/\pd h_R$. This leads to the scaling form (see, e.g., Ref.\ \cite{gross_critical_2016}):
\begin{multline} \avOP_R(\tren,h_R,\rho,L) \\= L^{-\beta/\nu} \hat\phiscal\Bigg(\tscal=\left(\frac{L}{\amplXip}\right)^{1/\nu}t,  \phiscal= \left(\frac{L}{\amplXimu}\right)^{\beta \delta/\nu} h,\rho\Bigg).
\label{eq_EOS_fss}\end{multline}
It can be shown that the scaling function 
$\phiscal(\tscal,\hscal,\rho) \equiv \hat\phiscal(\tscal,\hscal,\rho) (\amplXimu)^{\beta\delta/\nu}$
is universal \cite{privman_universal_1984}.
In the bulk limit, i.e., for $\tscal\gg 1$ or $\hscal\gg 1$, the scaling function $\phiscal$ reduces to (see, e.g., Ref.\ \cite{amit_field_2005}) 
\beq \phiscal(\tscal ,\hscal ,\rho)= \hscal^{1/\delta} \phiscal_b( \tscal \hscal^{-1/(\delta\beta)},\rho).
\label{eq_phi_bulk_scalfunc}\eeq 
Within the considered approximation $\Ocal(\epsilon^0)$, \cref{eq_phi_bulk_scalfunc} holds for all $\tscal$ and $\hscal$, where $\phiscal_b$ follows from \cref{eq_EOS_MFT_scalf} as \footnote{It is useful to note that \cref{eq_phi_bulk_scalfunc} can expressed alternatively as $\phiscal(\tscal ,\hscal ,\rho)\simeq \hat{\hscal}^{1/\delta} \hat{\phiscal_b}( \tscal \hat{\hscal}^{-1/\delta\beta},\rho)$ with a scaling function $\hat{\phiscal_b}(y,\rho)=(ru^*)^{-1/2}\left[ 2y\left(\sqrt{9+8y^3}-3\right)^{1/3} - \left(\sqrt{9+8y^3}-3\right)^{1/3} \right]$ and $\hat{\hscal}\equiv (r u^*)^{1/2} \hscal $. This form shows explicitly that $\phiscal$ and $\hscal$ are quantities of $O(u^{-1/2})$.}
\begin{multline} \phiscal_b(y,\rho) = \Big\{ 2 y \left[\sqrt{9 (ru^*) +8y^3} - 3 (ru^*)^{1/2}\right]^{-1/3} \\ - \left[\sqrt{9(ru^*)+8y^3} - 3 (ru^*)^{1/2}\right]^{1/3} \Big\} / (ru^*)^{1/2},
\label{eq_phi_bulk_scalfunc_eps0}\end{multline} 
which is, in fact, independent of $\rho$.
Finite-size effects for a certain boundary condition enter the equation of state at $\Ocal(\epsilon)$. 
Within field theory, the finite-size scaling function $\phiscal(\tscal,\hscal,\rho)$ has been investigated further, e.g., in Ref.\ \cite{chen_nonmonotonic_2000}.

\subsection{Critical Casimir force}
\label{sec_renorm_CCF}
The \emph{critical Casimir force} $\Kcal$ (per area and per $k_B T_c$) is defined in terms of the \resFE $L\fcal\res$ per area $A$ and per $k_B T_c$ \cite{krech_casimir_1994, brankov_theory_2000,gambassi_casimir_2009}:
\beq \Kcal \equiv -\frac{\d\, (L\fcal\res)}{\d L}\Big|_{A=\const} .
\label{eq_CCF_def_Fres}\eeq 
We emphasize that this derivative is to be calculated by keeping the area as well as the appropriate thermodynamic control parameters of the respective ensemble constant: these are, in the grand canonical ensemble, the reduced temperature $t$ and the bulk field $h_R$, whereas in the canonical ensemble, these are $t$ and the total mass $\constrOP$ [\cref{eq_constr_spec}].
Furthermore, in order to obtain the CCF for a system with vanishing aspect ratio, in \cref{eq_CCF_def_Fres} the limit $\rho\to 0$ must be taken only at the end of the calculation.
Alternatively to \cref{eq_CCF_def_Fres}, the CCF can be defined as the pressure difference between the film and the surrounding fluid. While these definitions are equivalent in the grand canonical ensemble, this is not necessarily the case in the canonical ensemble \cite{gross_critical_2016}. We briefly discuss these aspects in \cref{app_CFF_pressure}, but continue to use the definition in \cref{eq_CCF_def_Fres} for the remainder of the present study.
The consequences of defining the CCF under the condition of a fixed total volume $V=AL$ instead of a fixed area are discussed in \cref{app_CCF_vol}.

As alluded to above, in order to evaluate \cref{eq_CCF_def_Fres} in the \emph{canonical} ensemble, we have to take into account the global OP constraint [\cref{eq_constr_spec}], $\avOP A L = \constrOP=\const.$, which immediately implies a dependence of the mean OP $\avOP$ on $L$ according to 
\beq \frac{\d\avOP}{\d L}\Big|_{A=\text{const}} = -\frac{\avOP}{L},
\label{eq_dPhi_dL}\eeq
assuming a fixed transverse area $A$. (We note that, as a consequence of this assumption, $\rho$ varies upon changing $L$.)
From \cref{eq_Fres_c_scal_L,eq_CCF_def_Fres} we then obtain the canonical \CCF (per area and per $k_B T_c$)
\begin{widetext}
\beq \ring\Kcal(\tren,\avOP_R,\rho,L) =  L^{-d} \ring\Xi\left(\tscal= \left(\frac{L}{\amplXip}\right)^{1/\nu} \tren, \phiscal= \left(\frac{L}{\amplXimu}\right)^{\beta/\nu} \avOP_R , \rho\right) ,
\label{eq_Casi_force_c}\eeq 
with the universal scaling function 
\beq 
\ring\Xi(\tscal, \phiscal, \rho) =  (d-1) \ring{\hat\Theta}(\tscal, \phiscal, \rho) - \frac{1}{\nu} \tscal \pd_{\tscal}\ring{\hat\Theta}(\tscal, \phiscal, \rho) - \left(\frac{\beta}{\nu}-1\right) \phiscal \pd_{\phiscal}\ring{\hat\Theta}(\tscal, \phiscal, \rho) - \rho \pd_{\rho}\ring{\hat\Theta}(\tscal, \phiscal, \rho) + \delta\ring\Xi\st{ns}(\rho),
\label{eq_Casi_force_c_scalf}\eeq
where $\ring{\hat\Theta}(\tscal,\phiscal,\rho)\equiv \ring\Theta(\mtscal(\tscal,\phiscal),\rho)$.
The contribution 
\beq \delta\ring\Xi\st{ns}(\rho) \equiv - \onehalf (d+2)\rho^{d-1}
\label{eq_CCF_can_nscal}\eeq 
stems from the non-scaling term $\delta F\st{ns}$ in \cref{eq_constrcorr_nsterm}.
Note that, while $\delta F\st{ns}$ is an explicitly $L$-dependent contribution to the \resFE [see \cref{eq_Fres_c_scal_L}], $\delta\ring\Xi\st{ns}$ can be expressed fully in terms of the scaling variable $\rho$ and therefore can be considered as a universal contribution to the CCF.

In the \emph{grand canonical} ensemble, assuming a fixed bulk field $h_R$, one obtains from \cref{eq_Fres_gc_scal_L,eq_CCF_def_Fres} the \CCF (per area and per $k_B T_c$)
\beq \Kcal(\tren,h_R,\rho,L) = L^{-d} \tilde\Xi\left(\tscal=\left(\frac{L}{\amplXip}\right)^{1/\nu} \tren, \hscal=\left(\frac{L}{\amplXimu}\right)^{\beta\delta/\nu} h_R, \rho \right),
\label{eq_Casi_force_gc}\eeq 
with the universal scaling function
\beq 
\tilde\Xi(\tscal, \hscal, \rho) =  (d-1) \tilde\Theta(\tscal, \hscal, \rho) - \frac{1}{\nu} \tscal \pd_{\tscal}\tilde\Theta(\tscal, \hscal, \rho) - \frac{\beta\delta}{\nu} \hscal \pd_{\hscal}\tilde\Theta (\tscal, \hscal, \rho)- \rho \pd_{\rho}\tilde\Theta(\tscal, \hscal, \rho)
\label{eq_Casi_force_gc_scalf}\eeq 
in terms of $\tilde\Theta(\tscal,\hscal,\rho)$ defined in \cref{eq_Fres_gc_scal_L}.
Furthermore, the r.h.s.\ of \cref{eq_Casi_force_gc_scalf} can be expressed in terms of $\hat\Theta(\tscal,\phiscal,\rho)= \Theta(\mtscal(\tscal,\phiscal),\rho)$ defined in \cref{eq_Fres_gc_pert_equal} as:
\beq 
\tilde\Xi(\tscal, \hscal, \rho) =  (d-1) \hat\Theta - \frac{1}{\nu} \tscal \left[ \pd_{\tscal}\hat\Theta + \frac{\pd \phiscal}{\pd\tscal} \pd_\phiscal\hat\Theta\right] - \frac{\beta \delta}{\nu} \hscal \frac{\pd \phiscal}{\pd\hscal} \pd_{\phiscal}\hat\Theta - \rho \left(\frac{\pd \phiscal}{\pd\rho}\pd_\phiscal+ \pd_{\rho}\right)\hat\Theta,
\label{eq_Casi_force_gc_scalf_mgen}\eeq 
where $\phiscal$ is determined as a function of $\tscal$ and $\hscal$ via the corresponding equation of state [see \cref{eq_EOS_fss}].

We now focus specifically on the approximation $\Ocal(\epsilon^0)$ of the CCF.
In this case, the scaling form of the equation of state in \cref{eq_phi_bulk_scalfunc} applies and can be used to express  \cref{eq_Casi_force_gc_scalf_mgen} as a function of $\phiscal$ instead of $\hscal$:
\beq 
\Xi(\tscal, \phiscal, \rho) =  (d-1) \hat\Theta(\tscal, \phiscal, \rho) - \frac{1}{\nu} \tscal \pd_{\tscal}\hat\Theta(\tscal, \phiscal, \rho) - \frac{\beta}{\nu} \phiscal \pd_{\phiscal}\hat\Theta(\tscal, \phiscal, \rho) - \rho \left(\frac{\pd \phiscal}{\pd\rho}\pd_\phiscal+ \pd_{\rho}\right)\hat\Theta(\tscal, \phiscal, \rho).
\label{eq_Casi_force_gc_scalf_m}\eeq 
Beyond $\Ocal(\epsilon^0)$, \cref{eq_phi_bulk_scalfunc} applies in general only in the bulk limit, i.e., for $\tscal,\hscal\gg 1$.
We note that \cref{eq_Casi_force_gc_scalf_m} presupposes that both in \cref{eq_phi_bulk_scalfunc} and in \cref{eq_Casi_force_gc_scalf_mgen} the same approximation for the values of the critical exponents is used.
Within the mean-field or Gaussian approximation considered here, one has in particular $\beta=\nu$ and, consequently, in \cref{eq_Casi_force_c_scalf} the term proportional to $\pd_\phiscal\ring{\hat\Theta}$ vanishes. 
Furthermore, upon using \cref{eq_tscal_mod}, the fact that $\beta=1/2$, and noting that $\pd_\rho\phiscal=0$ for the \bcs considered here [see \cref{eq_phi_bulk_scalfunc_eps0}], we can express  \cref{eq_Casi_force_gc_scalf_m} in terms of the scaling function $\Theta(\mtscal(\tscal,\phiscal),\rho) = \hat\Theta(\tscal,\phiscal,\rho)$ [see \cref{eq_Fres_gc_scalfunc}] as
\beq \Xi(\tscal, \phiscal, \rho) =  (d-1) \Theta(\mtscal, \rho) - \frac{1}{\nu} \mtscal  \pd_{\mtscal} \Theta(\mtscal, \rho) - \rho \pd_{\rho} \Theta(\mtscal, \rho) + \Ocal(\epsilon).
\label{eq_Casi_force_gc_scalf_Oeps}\eeq 

In order to analogously simplify the canonical CCF [\cref{eq_Casi_force_c_scalf}], we define $\delta\ring\Kcal$ as the total contribution to the canonical CCF $\ring\Kcal$ [\cref{eq_Casi_force_c}] stemming from the constraint-induced term $\delta F$ [\cref{eq_constrcorr}]:
\beq \delta \ring\Kcal(\tren,\avOP_R,A,L) = -\frac{1}{A} \frac{\d\, \delta F}{\d L}\Big|_{A=\const} = L^{-d} \delta \ring\Xi\left(\tscal= \left(\frac{L}{\amplXip}\right)^{1/\nu} \tren, \phiscal= \left(\frac{L}{\amplXimu}\right)^{\beta/\nu} \avOP_R, \rho\right)
\label{eq_Casi_constrcorr}\eeq
with the associated universal scaling function [see \cref{eq_tscal_mod}]
\begin{subequations}
  \begin{empheq}[left={\delta\ring \Xi(\tscal,\phiscal,\rho)= \empheqlbrace}]{align}
  & -\onehalf \rho^{d-1} \frac{\tscal + \frac{3}{2}r u^* \phiscal^2}{\mtscal}, & \text{periodic and Neumann,\,\,\,} \label{eq_Casi_constrcorr_scal_perNeu}\\
  & -\onehalf \rho ^{d-1} \frac{\sqtscal \tanh(\sqtscal/2)}{\sqtscal\coth(\sqtscal/2)-2}, & \text{Dirichlet.\hspace{2.4cm}}\label{eq_Casi_constrcorr_scal_Dir}
\end{empheq}\label{eq_Casi_constrcorr_scal}
\end{subequations}
We note that, in fact, $\delta\ring\Xi=\delta \ring\Xi\st{ns} + \delta\ring\Xi\st{s}$, where $\delta \ring\Xi\st{ns}$ is given in \cref{eq_CCF_can_nscal} and the scaling function $\delta \ring\Xi\st{s}$ is defined, analogously to \cref{eq_Casi_constrcorr}, by $L^{-d}\delta \ring\Xi\st{s}=-(1/A) \d\delta F\st{s}/\d L$ in terms of $\delta F\st{s}$ in \cref{eq_constrcorr_scal}. 
Using \cref{eq_Casi_constrcorr,eq_Fres_c_scalfunc,eq_Fres_gc_scalfunc}, $\ring\Xi$ in \cref{eq_Casi_force_c_scalf} can now be expressed in terms of $\Theta(\mtscal,\rho)$ as
\beq \ring\Xi(\tscal, \phiscal, \rho) =  (d-1) \Theta(\mtscal, \rho) - \frac{1}{\nu} \mtscal \pd_{\mtscal} \Theta(\mtscal, \rho) +r u^* \phiscal^2 \pd_{\mtscal}\Theta(\mtscal,\rho) - \pd_{\rho} \Theta(\mtscal, \rho) + \delta\ring\Xi(\tscal,\phiscal,\rho) + \Ocal(\epsilon).
\label{eq_Casi_force_c_scalf_Oeps}\eeq 
Comparing \cref{eq_Casi_force_c_scalf_Oeps,eq_Casi_force_gc_scalf_Oeps} reveals that the scaling functions of the canonical and grand canonical CCF are related  as
\beq\begin{split} \ring\Xi(\tscal,\phiscal,\rho) = \Xi(\tscal,\phiscal,\rho) + r u^* \phiscal^2 \pd_{\mtscal}\Theta(\mtscal,\rho) + \delta \ring\Xi(\tscal,\phiscal,\rho) + \Ocal(\epsilon),
\end{split}\label{eq_Casi_force_gc_c_rel_Oeps}\eeq 
\end{widetext}
where $\Theta(\mtscal,\rho) = \onehalf\modesum_{d,\reg}(\mtscal,\rho)$ with $\modesum_{d,\reg}$ given in \cref{eq_Sreg_coll}.
We also recall that, according to \cref{eq_tscal_mod}, $r u^* \phiscal^2 \pd_{\mtscal}\Theta(\mtscal,\rho) = \phiscal \pd_\phiscal \Theta(\mtscal(\tscal,\phiscal),\rho)$.
In the next section, the implications of \cref{eq_Casi_force_gc_c_rel_Oeps} are discussed further.
We remark that, if the CCF is defined under the condition of a fixed volume $V$ [see \cref{app_CCF_vol}], the resulting scaling functions for periodic and Neumann \bcs coincide in the two ensembles [see \cref{eq_Casi_force_c_gc_V_ident}]. In fact, in those cases the constraint-induced correction $\delta\ring\Xi$ in \cref{eq_Casi_constrcorr_scal} vanishes identically, whereas for Dirichlet \bcs [see \cref{eq_Casi_constrcorr_V_scal}] it is nonzero and takes a form different from \cref{eq_Casi_constrcorr_scal_Dir}.

\section{Discussion}
\label{sec_discussion}
Here we discuss the \resFE reported in \cref{eq_Fres_c_scal_L,eq_Fres_gc_scal_L}, with the scaling functions defined in \cref{eq_Fres_c_scalfunc,eq_Fres_gc_scalfunc}, respectively, and the associated \CCF given in \cref{eq_Casi_force_c_scalf_Oeps,eq_Casi_force_gc_scalf_Oeps} within the one-loop approximation, i.e., to $\Ocal(\epsilon^0)$. 
A meaningful comparison of the two ensembles requires to evaluate the scaling functions for the same value of the scaled mean OP $\phiscal$. 
This can be achieved by relating $\phiscal$ to $\hscal$ via the finite-size equation of state reported in \cref{eq_phi_bulk_scalfunc,eq_phi_bulk_scalfunc_eps0}. 
In \cref{sec_F_gc} it has been shown that, within the approximation $\Ocal(\epsilon^0)$ considered here, the grand canonical \resFE can be expressed as in \cref{eq_Fres_gc_scalfunc} [see also \cref{eq_Fres_gc_pert_equal}].
This result provides the desired grand canonical scaling function $\Theta(\mtscal(\tscal,\phiscal),\rho)$ expressed in terms of $\phiscal$.
Furthermore, in the discussion of the \resFE, we shall omit the non-scaling contribution $\delta F\st{ns}$ in \cref{eq_constrcorr_nsterm} stemming from the constraint. However, this latter contribution is taken into account for the scaling function of the CCF, because, as shown in \cref{eq_CCF_can_nscal}, it takes on a scaling form.
We recall here that the perturbative results in the present study refer to cubical systems with aspect ratios $0\leq \rho \lesssim 1$.
The description of rod-like geometries with $\rho\gg 1$ would require, \emph{inter alia}, a different set of scaling variables \cite{dohm_critical_2011, hucht_aspect-ratio_2011}.
In addition, certain features of the (grand canonical) CCF for $\rho\simeq 1$ near bulk criticality ($\tscal=\hscal=0$) \cite{hucht_aspect-ratio_2011} are not captured by our analytical expressions. Instead, they require more refined approaches, such as those described in Ref.\ \cite{dohm_critical_2011}.
In the subsequent discussion of the \resFE and CCF for the various \bcs, we shall therefore focus on the case $0\leq \rho\lesssim 1$. 

\subsection{Periodic \bcs}
\subsubsection{Residual finite-size free energy}
\label{sec_pbc_resFE}

\begin{figure*}[t]\centering
  \subfigure[]{\includegraphics[width=0.41\linewidth]{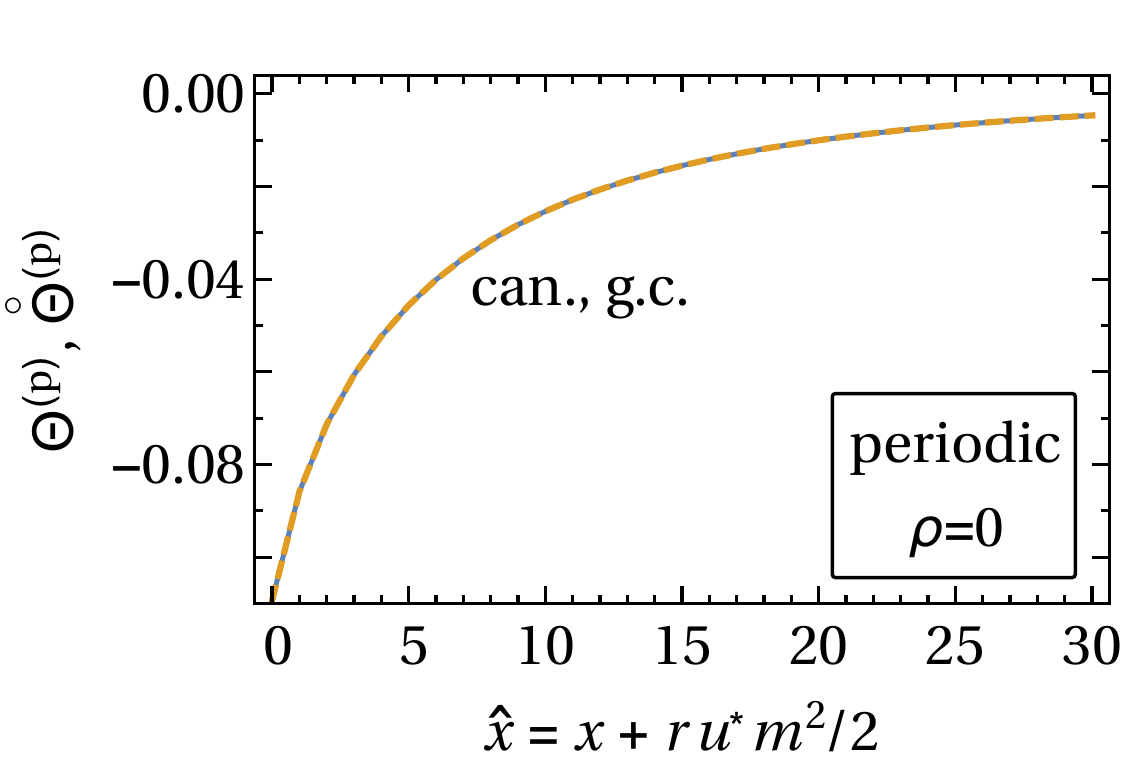} \label{fig_Fres_pbc_film}}\qquad
  \subfigure[]{\includegraphics[width=0.4\linewidth]{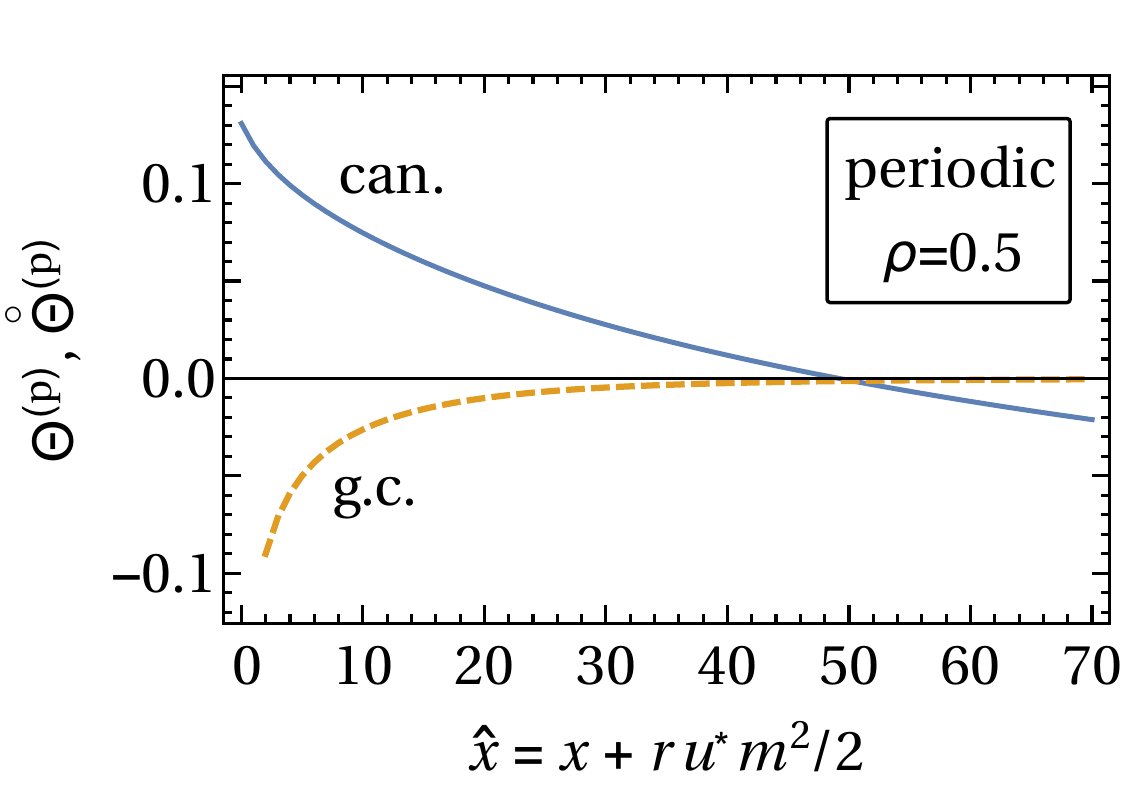}}
  \subfigure[]{\includegraphics[width=0.4\linewidth]{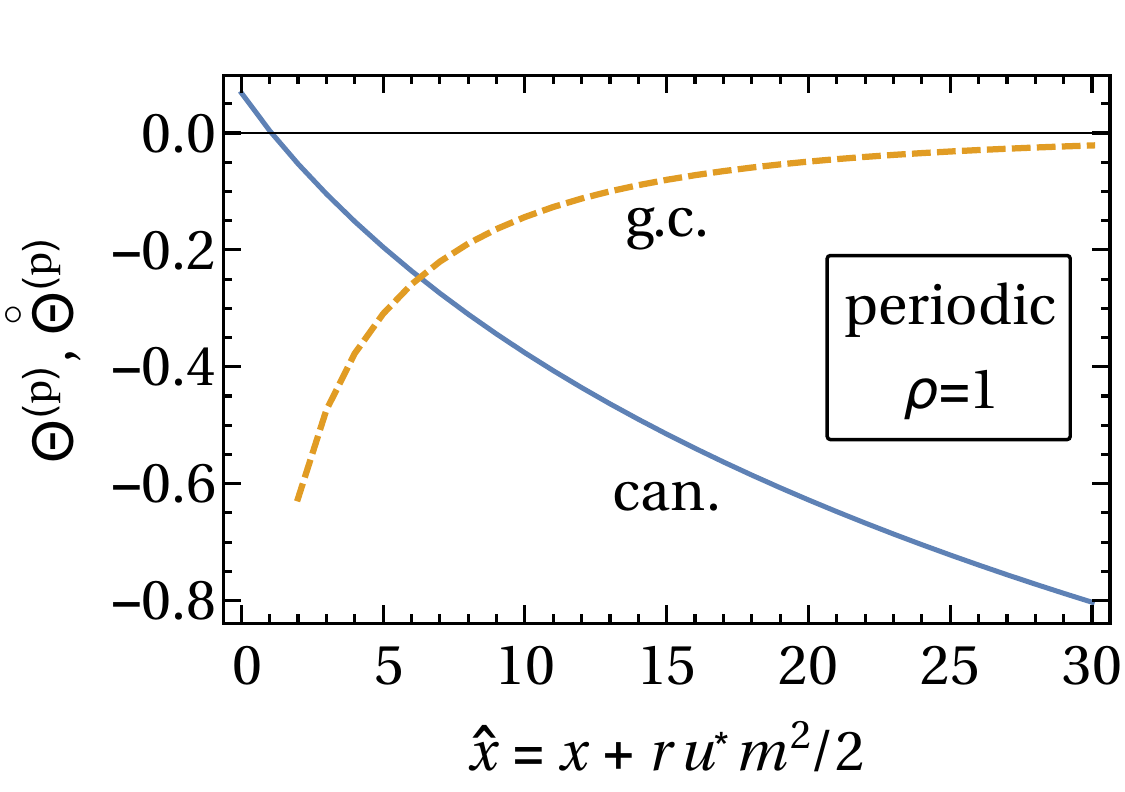}}\qquad
  \subfigure[]{\includegraphics[width=0.4\linewidth]{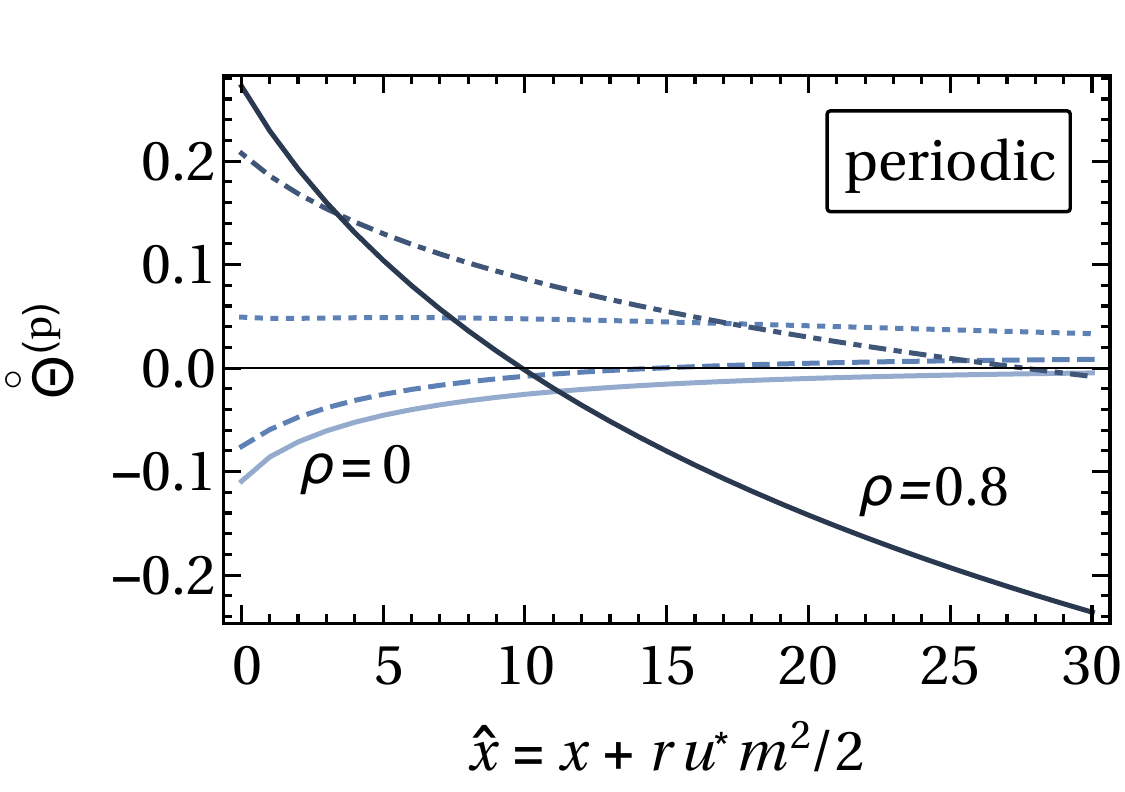}}
  \caption{Scaling functions $\Theta\pbc$ of the \resFE at $\Ocal(\epsilon^0)$ for periodic \bcs in the canonical [\cref{eq_ThetaC_pbc}, solid line] and the grand canonical [\cref{eq_ThetaGC_pbc}, dashed line] ensemble. In both ensembles, the scaling functions depend on the scaled temperature $\tscal$ and on the scaled mean OP $\phiscal$ via the quantity $\mtscal$ defined in \cref{eq_tscal_mod}. In the grand canonical ensemble, $\phiscal$ is related to the scaled bulk field $\hscal$ via \cref{eq_EOS_MFT_scalf}. For $\rho=0$ [panel (a)], the canonical and grand canonical scaling functions are identical [see \cref{eq_ThetaC_GC_equality}]. For $\rho>0$ [panels (b) and (c)], the perturbative expression for $\Theta$ [\cref{eq_ThetaGC_pbc}] applies only in the region $\mtscal\gtrsim 1$. The difference between $\ring\Theta\pbc$ and $\Theta\pbc$ stems solely from the constraint induced term in \cref{eq_ThetaC_pbc_constrcorr}, which leads to a logarithmic divergence $\ring\Theta\pbc\propto -\rho^{d-1}\ln \mtscal$ in the limit $\mtscal\to\infty$. Panel (d) illustrates how the dependence on $\mtscal$ of the canonical scaling function $\ring\Theta\pbc$ changes upon varying the aspect ratio $\rho$. The unlabeled dashed, dotted, and dash-dotted curves (with distinct blue shading) correspond to $\rho=0.2,0.4$, and 0.6, respectively.}
  \label{fig_Fres_pbc}
\end{figure*}

We recall that, in the \emph{grand canonical ensemble}, the scaling function of the renormalized \resFE is given by \cref{eq_Fres_gc_scalfunc,eq_Fres_gc_pert_equal}:
\beq \Theta\pbc(\mtscal,\rho) = \onehalf \modesum\pbc_{d,\reg}(\mtscal,\rho) ,
\label{eq_ThetaGC_pbc}\eeq 
whereas, in the \emph{canonical ensemble}, we have [see \cref{eq_Fres_c_scalfunc}]
\beq \ring\Theta\pbc(\mtscal,\rho) = \onehalf \modesum\pbc_{d,\reg}(\mtscal,\rho) + \delta\ring\Theta\pbc\st{s}(\mtscal,\rho).
\label{eq_ThetaC_pbc}\eeq 
The function $\modesum\pbc_{d,\reg}$ is reported in \cref{eq_S_pbc} and $\mtscal=\mtscal(\tscal,\phiscal)$ is defined in \cref{eq_tscal_mod}. 
In the canonical ensemble, the constraint contributes to the scaling function $\ring\Theta\pbc$ the expression [see \cref{eq_constrcorr_s_perNeu}]
\beq \delta\ring\Theta\pbc\st{s}(\mtscal,\rho) \equiv -\onehalf \rho^{d-1} \ln \frac{\rho^{d-1}\mtscal}{2\pi}.
\label{eq_ThetaC_pbc_constrcorr}\eeq 
Within the considered one-loop approximation, both $\Theta\pbc$ and $\ring\Theta\pbc$ have to be evaluated at $\epsilon=0$, i.e., $d=4$.
For a system with $\rho\neq 0$ and either periodic or Neumann \bcs, perturbative results for the grand canonical \resFE are applicable only for $\mtscal\gtrsim 1$ (see, in this respect, the discussion in \cref{sec_F_gc}). 
Accordingly, in these cases, the region $\mtscal\lesssim 1$ will be excluded from the corresponding plots.
For $\rho=0$, our perturbative results for periodic or Neumann \bcs are well behaved even for $\tscal,\hscal\lesssim 1$ and agree with the ones reported in Ref.\ \cite{krech_free_1992} (see also Refs.\ \cite{diehl_fluctuation-induced_2006, gruneberg_thermodynamic_2008} for further discussions).

Since the contribution $\delta\ring\Theta\st{s}$ due to the constraint [\cref{eq_ThetaC_pbc_constrcorr}] vanishes for $\rho\to 0$, the canonical and grand canonical scaling functions for periodic \bcs become identical in the thin-film limit, i.e., 
\beq \Theta\pbc(\mtscal,\rho=0) = \ring\Theta\pbc(\mtscal,\rho=0).
\label{eq_ThetaC_GC_equality}\eeq 
This is visualized in \cref{fig_Fres_pbc_film}, where $\Theta\pbc(\mtscal,\rho=0)$ is plotted as a function  of $\mtscal$.
In panels (b) and (c) of Fig.~\ref{fig_Fres_pbc}, we compare the dependence on $\mtscal$ of the scaling functions in the two ensembles for fixed nonzero aspect ratios $\rho$. 
The difference between $\ring\Theta\pbc$ (solid curve) and $\Theta\pbc$ (dashed curve) stems solely from the constraint-induced term $\delta\ring\Theta\st{s}\pbc$ in \cref{eq_ThetaC_pbc}, because the contribution from the regularized mode sum $\modesum\pbc_{d,\reg}$ is the same in both ensembles. 
Consequently, while $\Theta\pbc$ vanishes for $\mtscal\to\infty$, $|\ring\Theta\pbc|$ grows logarithmically upon increasing $\mtscal$ [see \cref{eq_Fres_largeTau}].
This behavior stems from the absence of the zero-mode fluctuations in the canonical ensemble [see also \cref{eq_free_en_per}], which, being spatially homogeneous, affect the residual free energy of a finite system for all values of $L$.
Figure \ref{fig_Fres_pbc}(d) illustrates that a change in the aspect ratio $\rho$ has a strong effect on the canonical \resFE, inducing, \emph{inter alia}, a change of sign of $\ring\Theta\pbc$ at small $\mtscal$. In contrast, in the grand canonical case (not shown), increasing $\rho$ leads, within the considered range of $\mtscal$, mainly to an increase in the overall strength of $\Theta\pbc$.

\begin{figure*}[t!]\centering
  \subfigure[]{\includegraphics[width=0.406\linewidth]{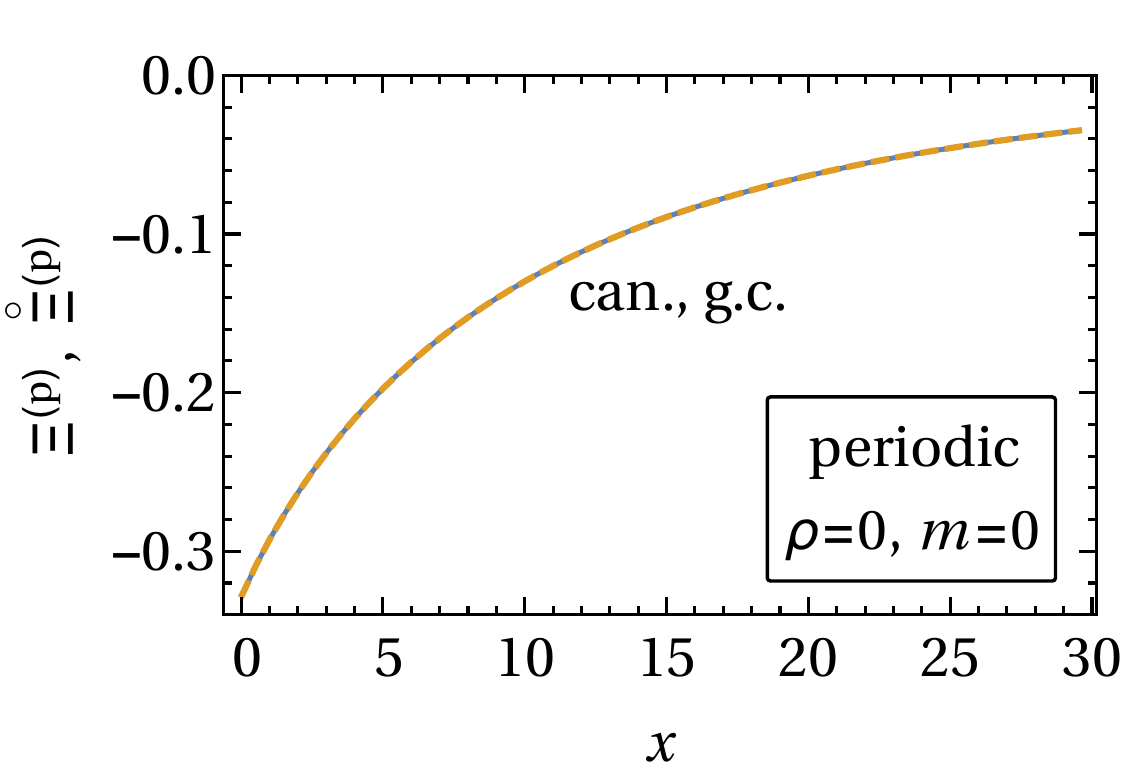}}\qquad
  \subfigure[]{\includegraphics[width=0.4\linewidth]{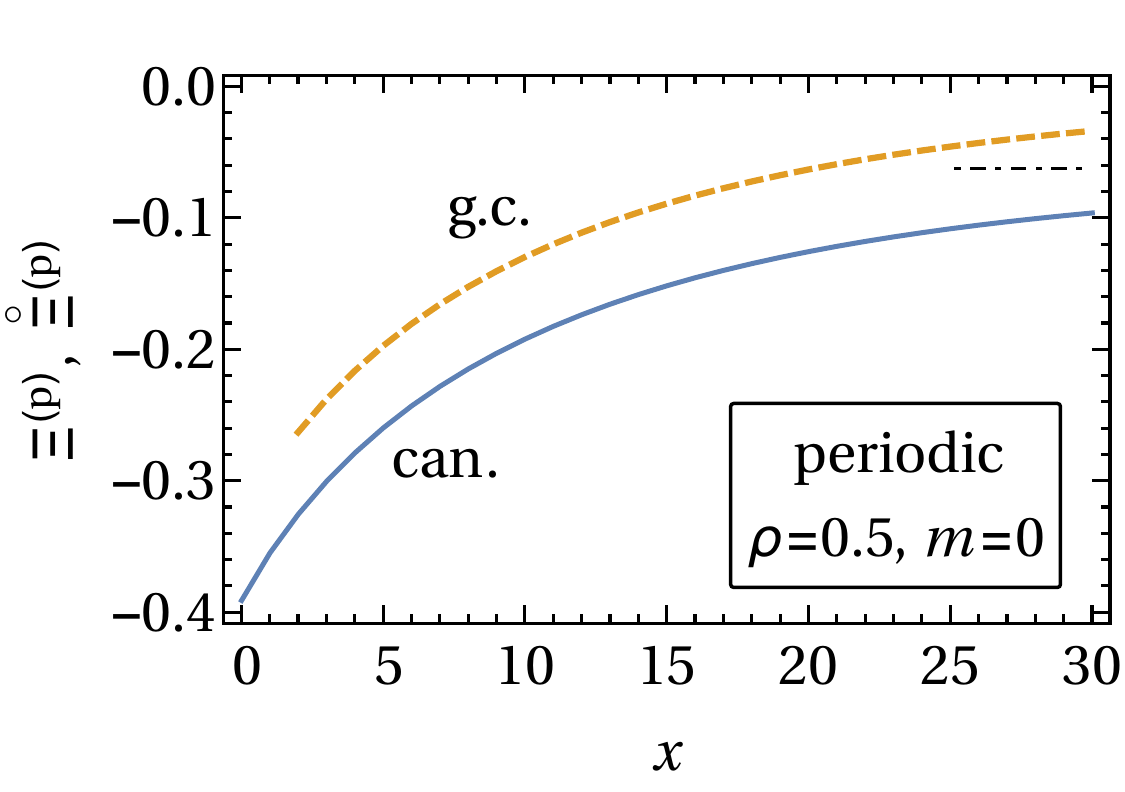}}
  \subfigure[]{\includegraphics[width=0.4\linewidth]{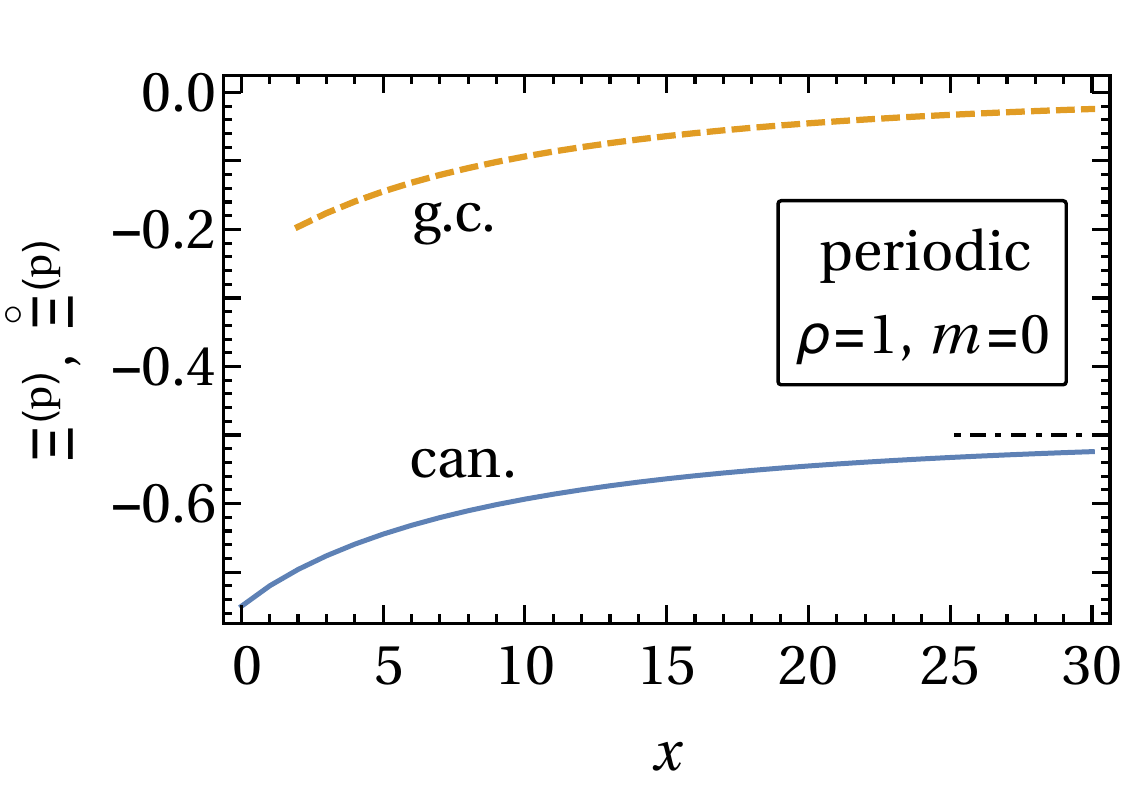}}\qquad
  \subfigure[]{\includegraphics[width=0.4\linewidth]{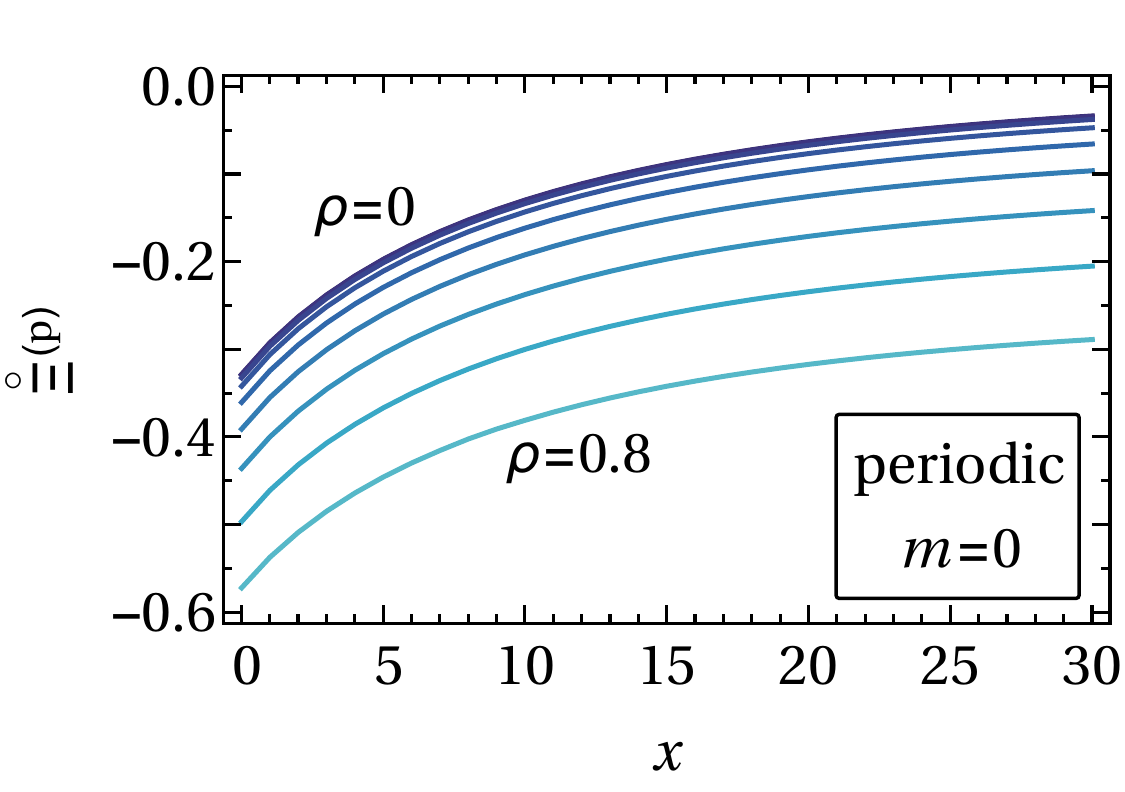}}
  \caption{(a)--(c) Scaling functions of the \CCF [\cref{eq_CCF_def_Fres}] at $\Ocal(\epsilon^0)$ for periodic \bcs in the canonical and the grand canonical ensembles [\cref{eq_Casi_force_c_scalf,eq_Casi_force_gc_scalf_m}, respectively] as functions of the scaled temperature $\tscal=(L/\amplXip)^{1/\nu}\tren $ for $\phiscal=0$ and three aspect ratios $\rho$. For $\phiscal=0$ one has $\mtscal=\tscal$ and $\ring\Xi\pbc=\Xi\pbc+\delta\ring\Xi\pbc$ with $\delta\ring\Xi\pbc$ given in \cref{eq_Casi_constrcorr_scal_perNeu}. For $\tscal\gg 1$ and $\phiscal\gg 1$, respectively, the canonical CCF attains the asymptotic values given in \cref{eq_Casi_constrcorr_pbc_lim} [short dash-dotted lines in (b) and (c)]. (d) Dependence of $\ring\Xi\pbc(\tscal,\phiscal=0,\rho)$ on $\tscal$ for various values of the aspect ratio $\rho$, increasing from 0 to 0.8 in steps of 0.1 from the top to the bottom curve (with distinct blue shading).}
  \label{fig_Xi_pbc_m0}
\end{figure*}

\subsubsection{Critical Casimir force}

\begin{figure*}[t!]\centering
  \subfigure[]{\includegraphics[width=0.41\linewidth]{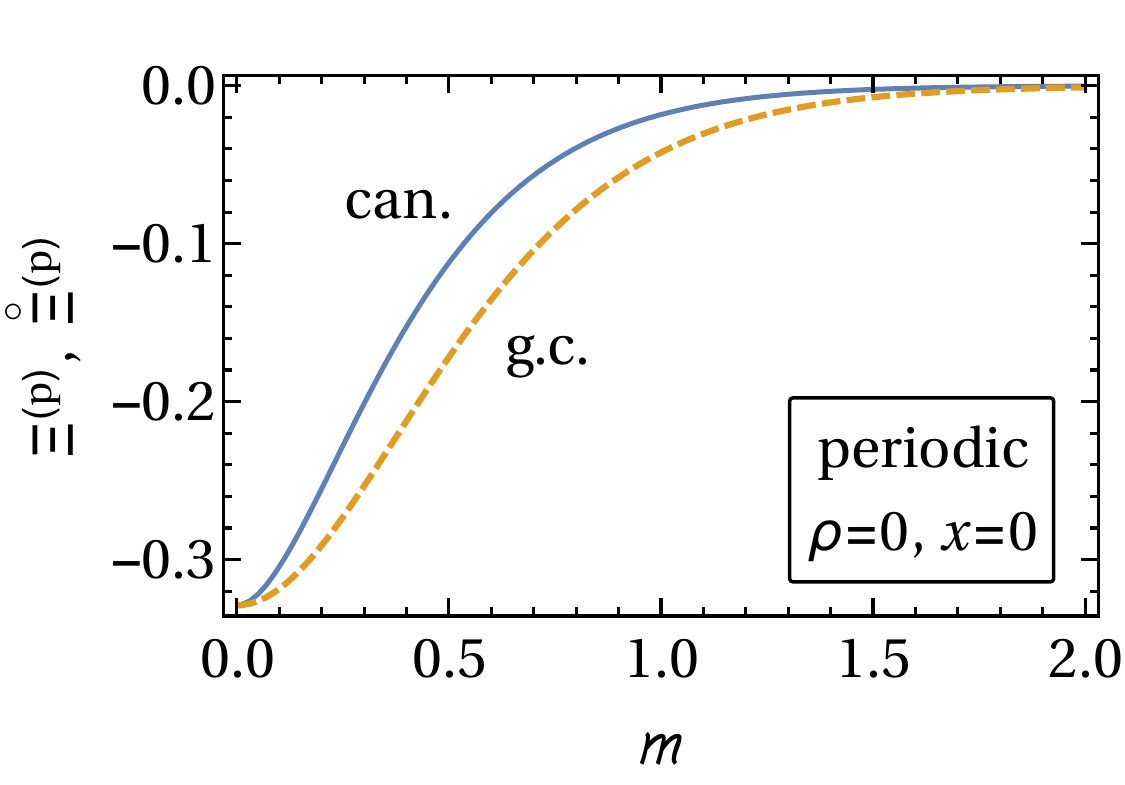} \label{fig_Xi_pbc_film_vs_m_at_Tc}}\qquad 
  \subfigure[]{\includegraphics[width=0.4\linewidth]{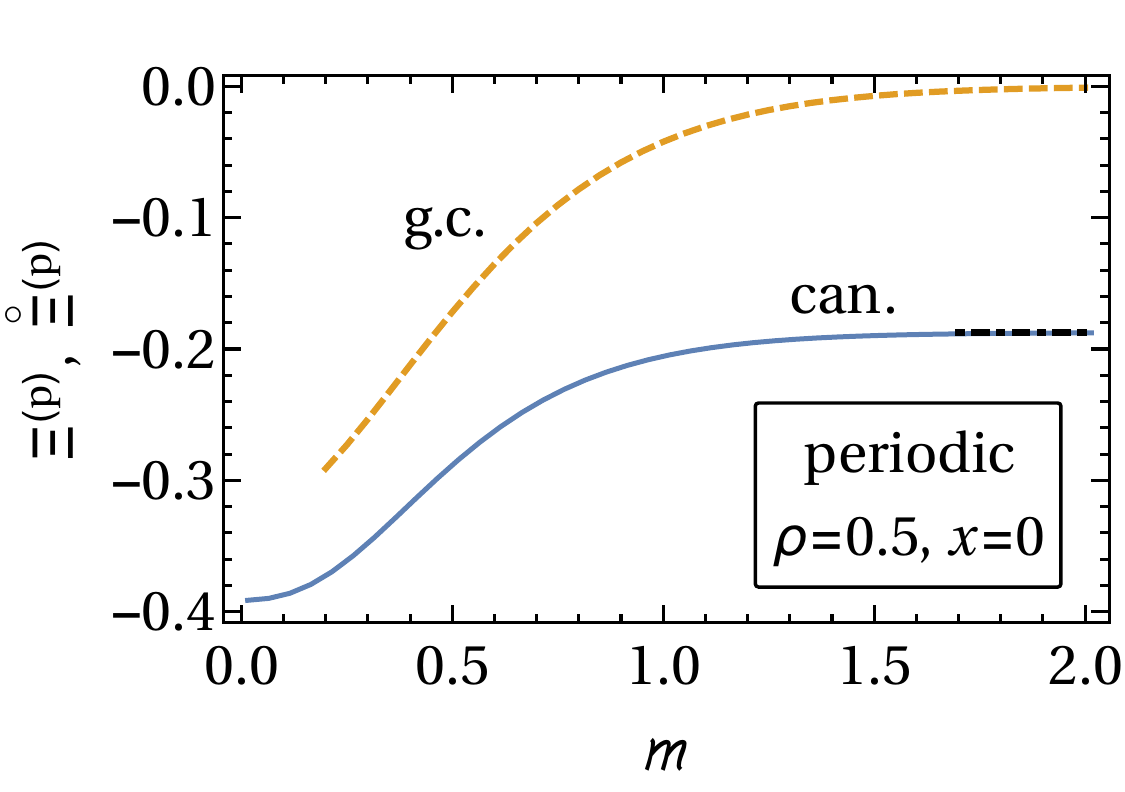}}
  \subfigure[]{\includegraphics[width=0.4\linewidth]{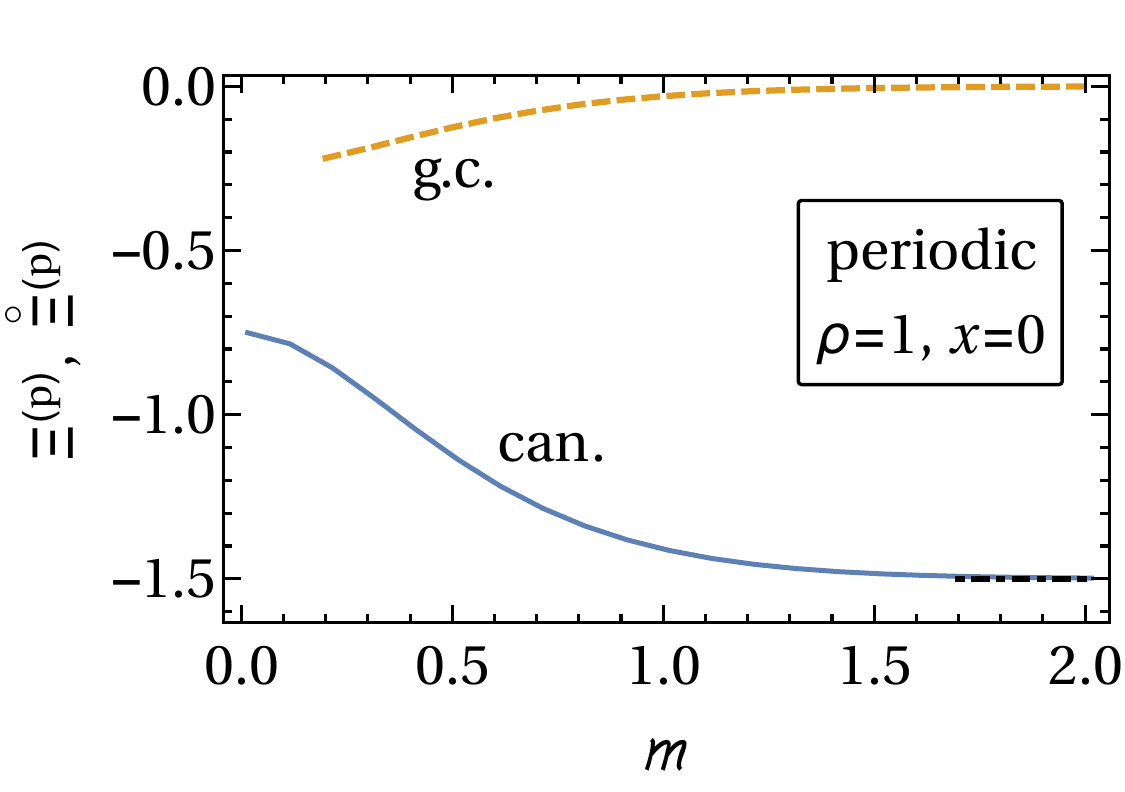}}\qquad
  \subfigure[]{\includegraphics[width=0.4\linewidth]{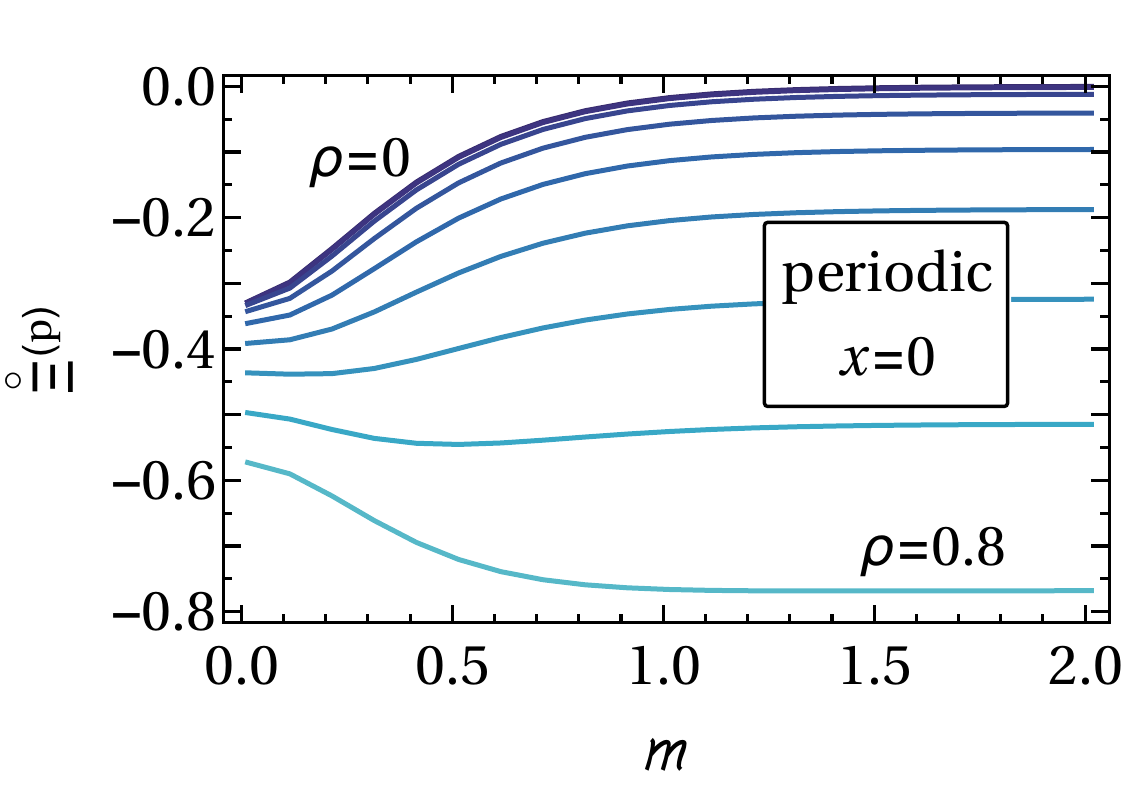}}
  \caption{(a)--(c) Scaling functions of the \CCF [\cref{eq_CCF_def_Fres}] at $\Ocal(\epsilon^0)$ for periodic \bcs in the canonical and the grand canonical ensembles [\cref{eq_Casi_force_c_scalf,eq_Casi_force_gc_scalf_m}, respectively] as functions of the scaled magnetization $\phiscal=(L/\amplXiPhi)^{\beta/\nu}\avOP_R$ for $\tscal=0$ and three aspect ratios $\rho$. For $\phiscal=0$ one has $\mtscal=\tscal$. Both for $\tscal\gg 1$ and for $\phiscal\gg 1$ the canonical CCF attains the asymptotic values given in \cref{eq_Casi_constrcorr_pbc_lim} [short dash-dotted lines in (b) and (c)]. (d) Dependence of $\ring\Xi\pbc(\tscal=0,\phiscal,\rho)$ on $\phiscal$ for various values of the aspect ratio $\rho$, increasing from 0 to 0.8 in steps of 0.1 from the top to the bottom curve (with distinct blue shading).}
  \label{fig_Xi_pbc_x0}
\end{figure*}

At $\Ocal(\epsilon^0)$, the difference between the canonical and grand canonical CCF is given by \cref{eq_Casi_force_gc_c_rel_Oeps}.
Since $\delta\ring\Xi\pbc(\mtscal,\rho=0)=0$ [see \cref{eq_Casi_constrcorr_scal_perNeu}], \cref{eq_Casi_force_gc_c_rel_Oeps} in the thin-film limit renders
\beq\begin{split} 
\Xi(\tscal, \phiscal, \rho=0) = \ring\Xi(\tscal,\phiscal, \rho=0) - \phiscal\pd_\phiscal \ring\Theta(\mtscal(\tscal,\phiscal), \rho=0),
\end{split}\label{eq_XiGC_C_sameTheta}\eeq 
where we have used \cref{eq_tscal_mod}.
We recall that the term $\phiscal\pd_\phiscal \ring\Theta$ in \cref{eq_XiGC_C_sameTheta} stems from \cref{eq_dPhi_dL}, which is a direct consequence of the OP constraint and of the assumption of a fixed transverse area $A$. 
For aspect ratios $\rho>0$, $\delta\ring\Xi\pbc$ [\cref{eq_Casi_constrcorr_scal_perNeu}] is in general nonzero and reduces to the limiting expressions 
\begin{subequations}\begin{align}
\delta\ring\Xi\pbc(\tscal,\phiscal=0,\rho) = \delta\ring\Xi\pbc(\tscal\to\infty,\phiscal,\rho) &= -\onehalf \rho^{d-1} \label{eq_Casi_constrcorr_pbc_m0}
\intertext{and}
\delta\ring\Xi\pbc(\tscal=0,\phiscal,\rho) = \delta\ring\Xi\pbc(\tscal,\phiscal\to\infty,\rho) &=- \frac{3}{2} \rho^{d-1}. \label{eq_Casi_constrcorr_pbc_x0}
\end{align}\label{eq_Casi_constrcorr_pbc_lim}\end{subequations}
In both limits, $\delta\ring\Xi\pbc$ is independent of $\phiscal$ and $\tscal$.
According to \cref{eq_Casi_force_gc_c_rel_Oeps,eq_Casi_constrcorr_scal_perNeu}, in general the constraint-induced contribution $\delta\ring\Xi\pbc$ enhances the attractive character of the \CCF compared to the unconstrained case. 
This is expected intuitively, because the constraint reduces the number of available fluctuation modes and thus the ``fluctuation pressure'' of the confined system compared to that of the bulk.
Interestingly, however, this effect is absent if the CCF is defined under the condition of a fixed total volume $V$ instead of a fixed transverse area (see \cref{app_CCF_vol}).
In this case, the CCF for periodic \bcs is identical in the two ensembles.
For $\phiscal=0$, the canonical and the grand canonical CCFs defined with fixed transverse area are related by a constant shift:
\begin{multline} \ring\Xi\pbc(\tscal,\phiscal=0,\rho) \\ =\Xi\pbc(\tscal,\phiscal=0,\rho)+\delta\ring\Xi\pbc(\tscal,\phiscal=0,\rho) + \Ocal(\epsilon).
\label{eq_Casi_pbc_m0_rel}\end{multline} 
Note that, beyond the approximation at $\Ocal(\epsilon^0)$, $\phiscal$ is in general expected to acquire a dependence on $\rho$ \cite{chen_nonmonotonic_2000}, such that \cref{eq_Casi_force_gc_scalf_m} has to be used.

\begin{figure*}[t]\centering
  \subfigure[]{\includegraphics[width=0.39\linewidth]{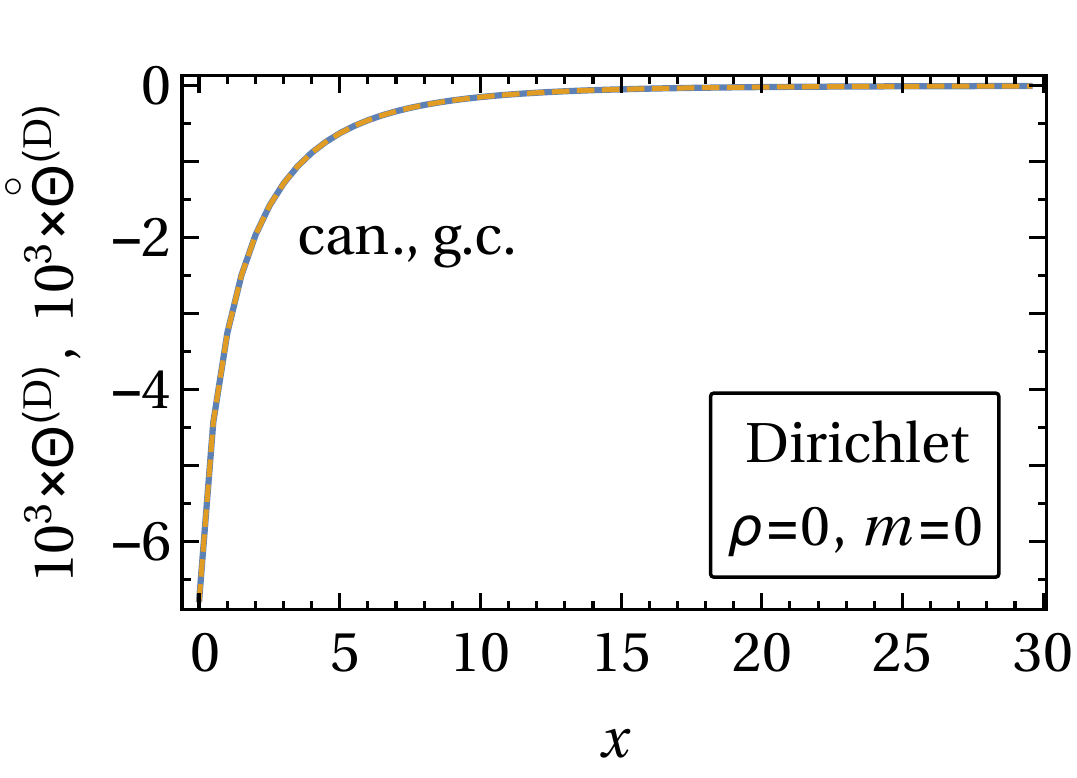}}\qquad
  \subfigure[]{\includegraphics[width=0.4\linewidth]{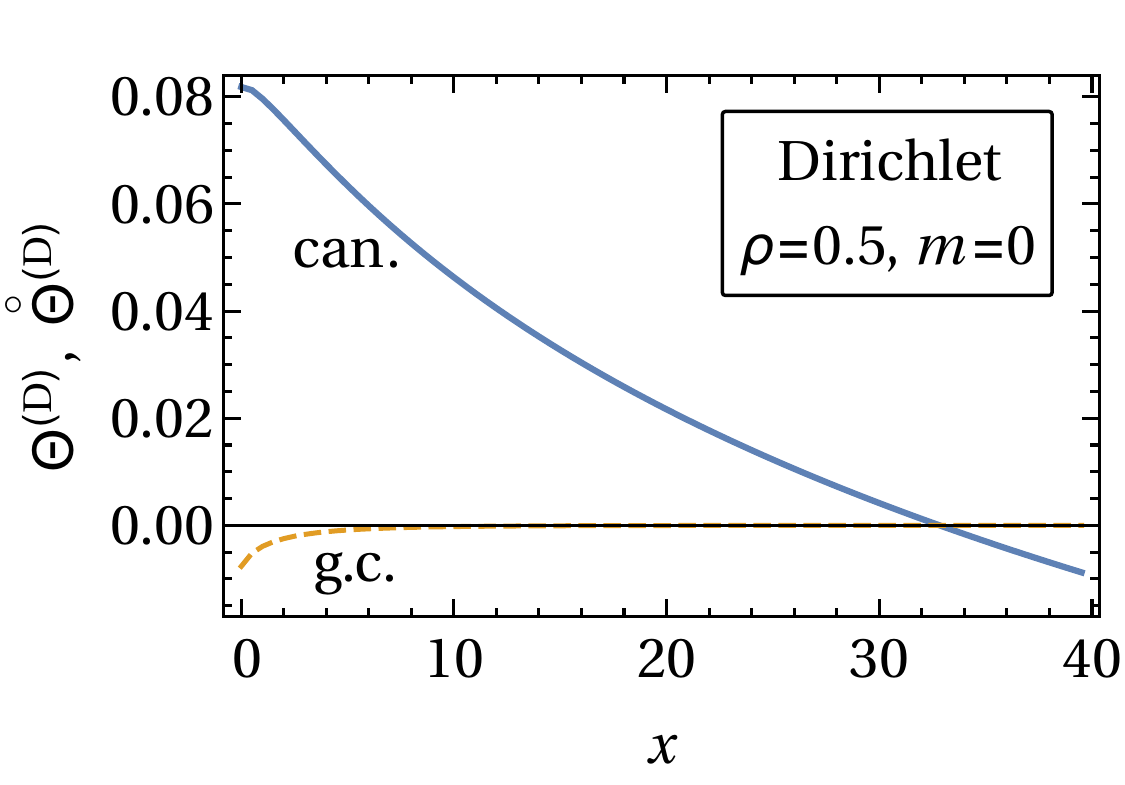}}
  \subfigure[]{\includegraphics[width=0.4\linewidth]{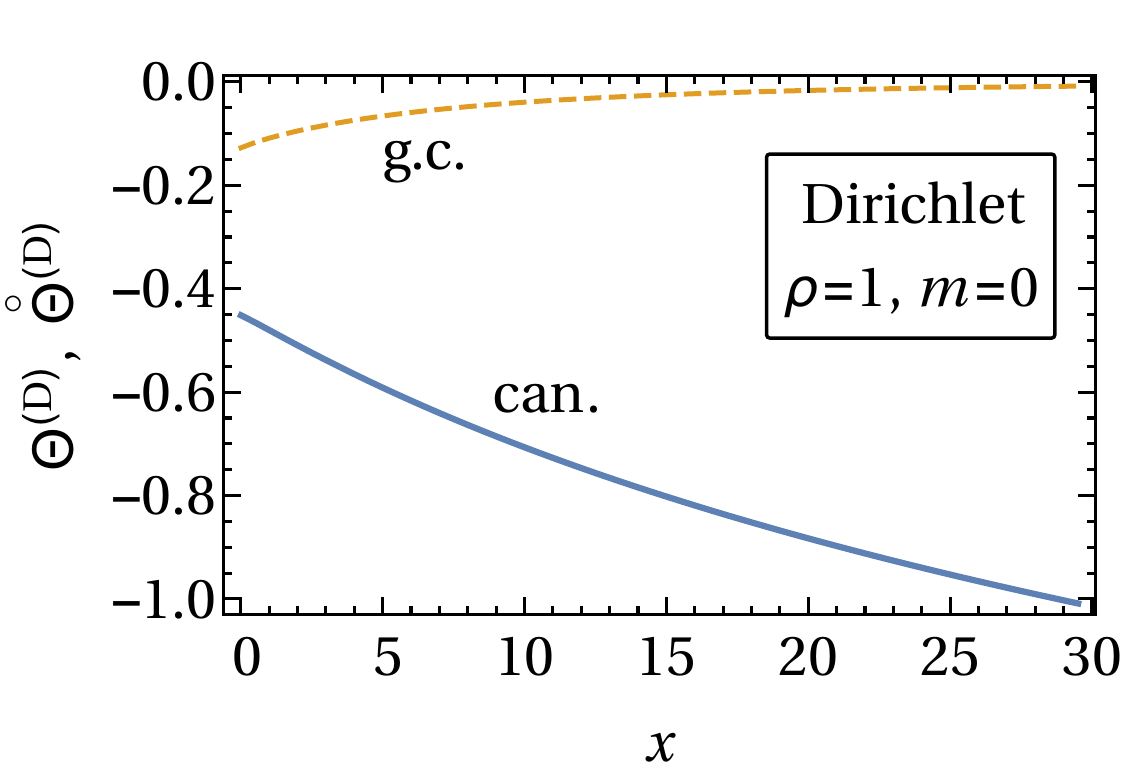}}\qquad
  \subfigure[]{\includegraphics[width=0.4\linewidth]{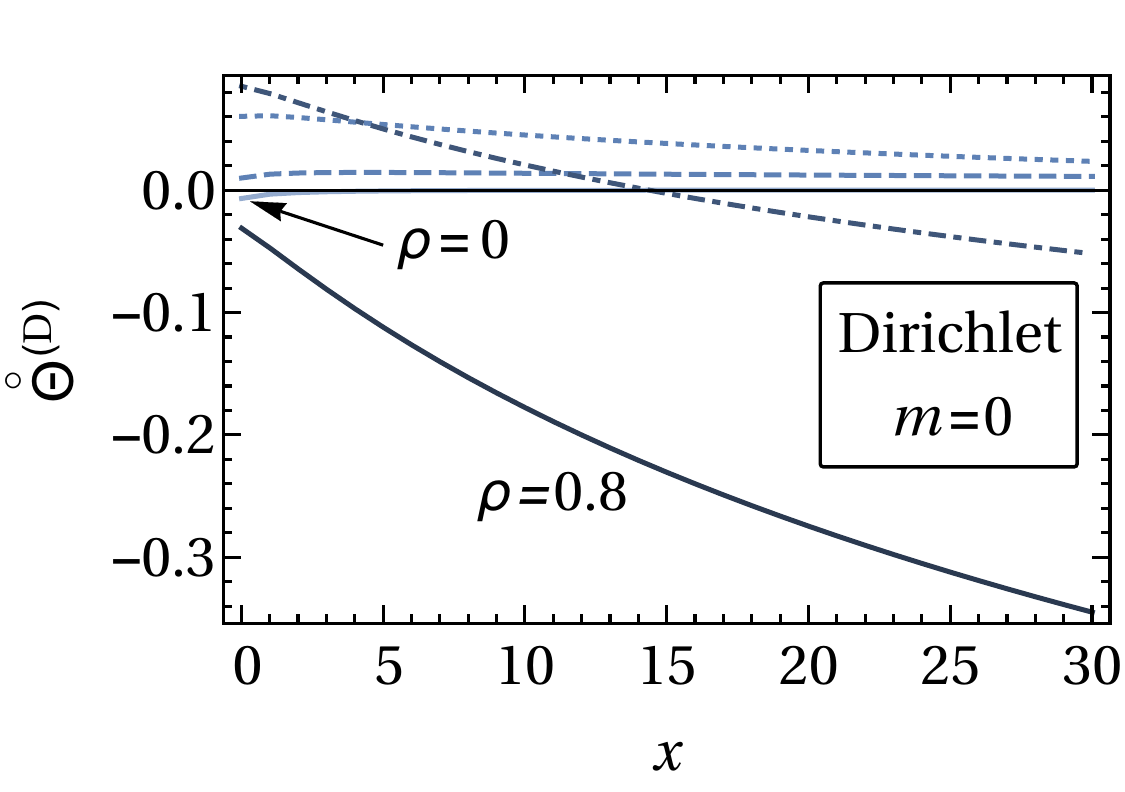}}
  \caption{(a)-(c) Scaling functions of the residual finite-size free energy at $\Ocal(\epsilon^0)$ for Dirichlet \bcs in the canonical [\cref{eq_S_Dbc,eq_Fres_c_scalfunc}] and the grand canonical ensemble [\cref{eq_S_Dbc,eq_Fres_gc_scalfunc}], as function of the scaling variable $\mtscal$ [\cref{eq_tscal_mod}] for three aspect ratios $\rho$. For the case $\phiscal=0$ considered here one actually has $\mtscal=\tscal$. For $\rho\neq 0$ and $\tscal\to\infty$, $\ring\Theta\Dbc$ diverges $\propto -\rho^{d-1}\ln \tscal$. Panel (d) illustrates how the dependence on $\tscal$ of the canonical scaling function $\ring\Theta\Dbc$ changes upon varying the aspect ratio $\rho$. The unlabeled dashed, dotted, and dash-dotted curves (with distinct blue shading) correspond to $\rho=0.2,0.4$, and 0.6, respectively.}
  \label{fig_Fres_Dir}
\end{figure*}

The scaling functions $\ring\Xi\pbc$ and $\Xi\pbc$ of the \CCF at $\Ocal(\epsilon^0)$ in the two ensembles are shown in \cref{fig_Xi_pbc_m0} for a vanishing mean OP, i.e., $\phiscal=0$. 
According to \cref{eq_XiGC_C_sameTheta}, $\ring\Xi\pbc$ and $\Xi\pbc$ become identical in the thin-film limit $\rho=0$, as shown in \cref{fig_Xi_pbc_m0}(a).
For $\rho>0$ [Figs.~\ref{fig_Xi_pbc_m0}(b)$-$(d)], $\ring\Xi\pbc$ approaches the constant in \cref{eq_Casi_constrcorr_pbc_m0} for large values $\tscal\gg 1$, while, correspondingly, $\Xi\pbc$ vanishes.
We recall that, for $\rho>0$, the results obtained perturbatively in the grand canonical ensemble are not expected to be reliable near the bulk critical point. Correspondingly, in spite of \cref{eq_Casi_pbc_m0_rel}, we plot the grand canonical CCF in this case only for $\mtscal\gtrsim 1$.
As \cref{fig_Xi_pbc_m0}(d) illustrates, upon increasing $\rho$ the absolute strength of $\ring\Xi\pbc$ for $\phiscal=0$ increases, while its functional form does not change significantly.

In \cref{fig_Xi_pbc_x0}, the scaling functions of the \CCF are shown as functions of the scaled mean OP $\phiscal$ for $\tscal=0$.
In the thin-film limit $\rho=0$ [\cref{fig_Xi_pbc_x0}(a)], in which the perturbative results at this order in $\epsilon$ are reliable in the whole domain of $\phiscal$, the only difference between $\ring\Xi\pbc$ and $\Xi\pbc$ is due to \cref{eq_XiGC_C_sameTheta}.
We conclude that, in contrast to $\delta \ring\Xi$ [\cref{eq_Casi_constrcorr_scal_perNeu}], the constraint-induced effect expressed in \cref{eq_dPhi_dL} increases the value of $\ring\Xi$ compared to the one of $\Xi$.
For nonzero $\rho$ [Figs.\ \ref{fig_Xi_pbc_x0}(b) and \ref{fig_Xi_pbc_x0}(c)], the OP constraint decreases the value of the canonical CCF relative to the grand canonical one by the amount given in \cref{eq_Casi_constrcorr_pbc_x0}.
Accordingly, for nonzero aspect ratios $\rho$ and in the limit $\phiscal\to\infty$, the canonical \CCF approaches a negative value \footnote{A similar result has been obtained in Ref.\ \cite{gross_critical_2016} for $(++)$ boundary conditions.}.
Figure \ref{fig_Xi_pbc_x0}(d) illustrates in more detail the dependence of the canonical scaling function $\ring\Xi\pbc$ on $\phiscal$ for $\tscal=0$ upon changing the aspect ratio. In passing, we mention that in the limit $\phiscal\to\infty$ the CCF defined under the condition of constant volume (see \cref{app_CCF_vol}) vanishes in both ensembles.

\begin{figure*}[t]\centering
  \subfigure[]{\includegraphics[width=0.42\linewidth]{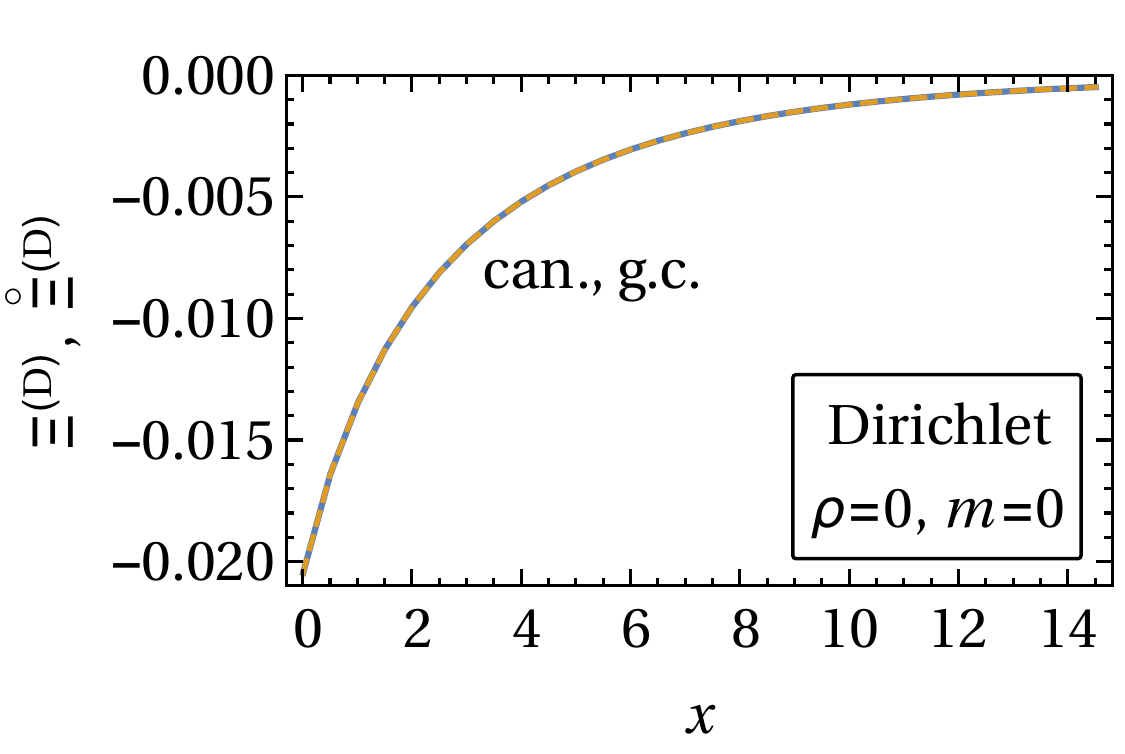}}\qquad
  \subfigure[]{\includegraphics[width=0.4\linewidth]{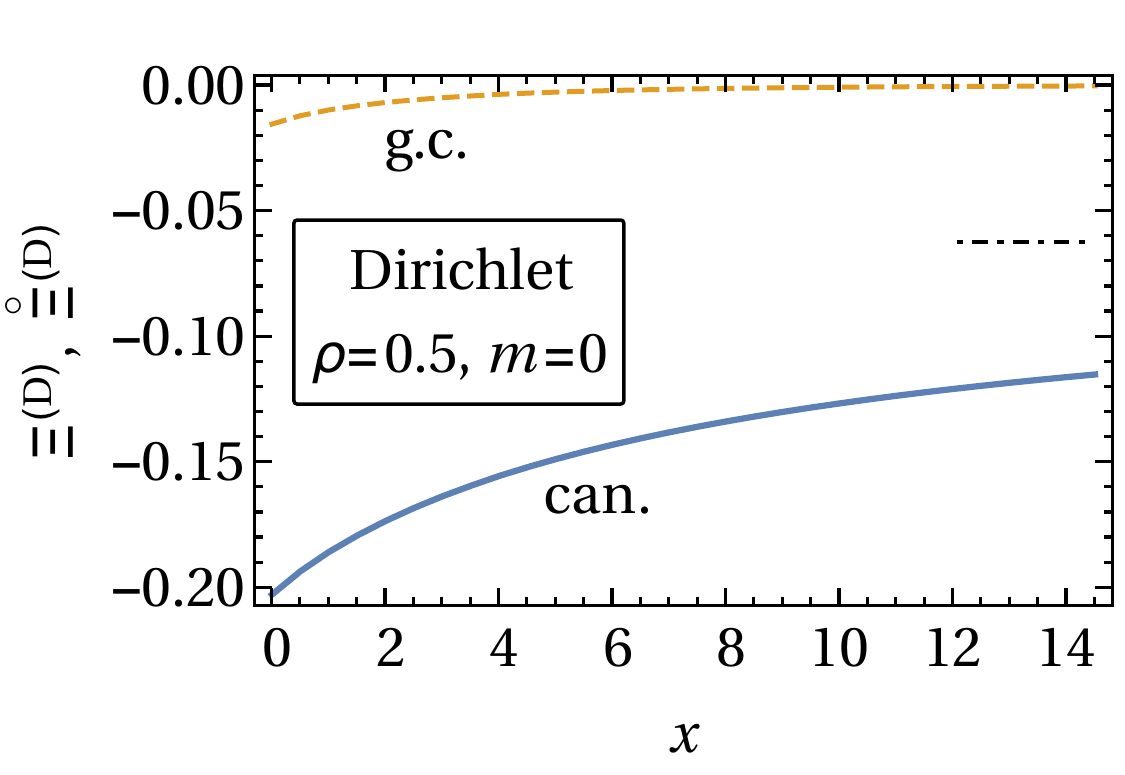}}
  \subfigure[]{\includegraphics[width=0.4\linewidth]{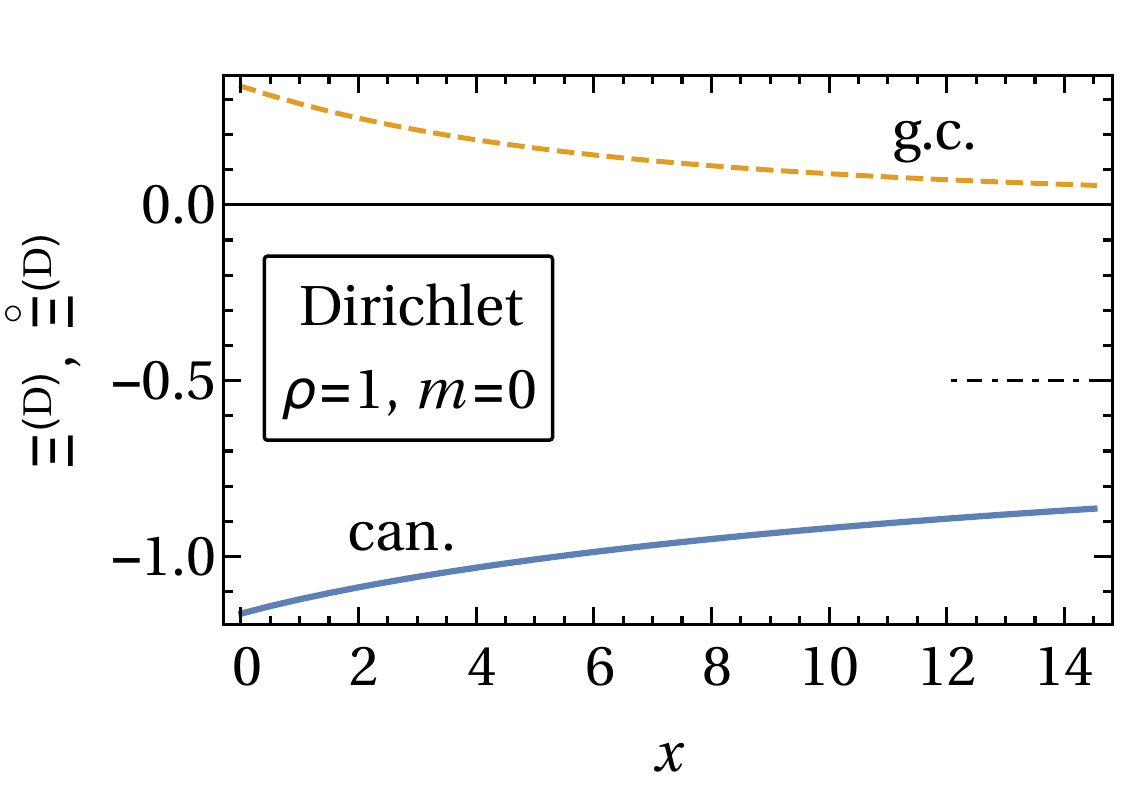}}\qquad
  \subfigure[]{\includegraphics[width=0.4\linewidth]{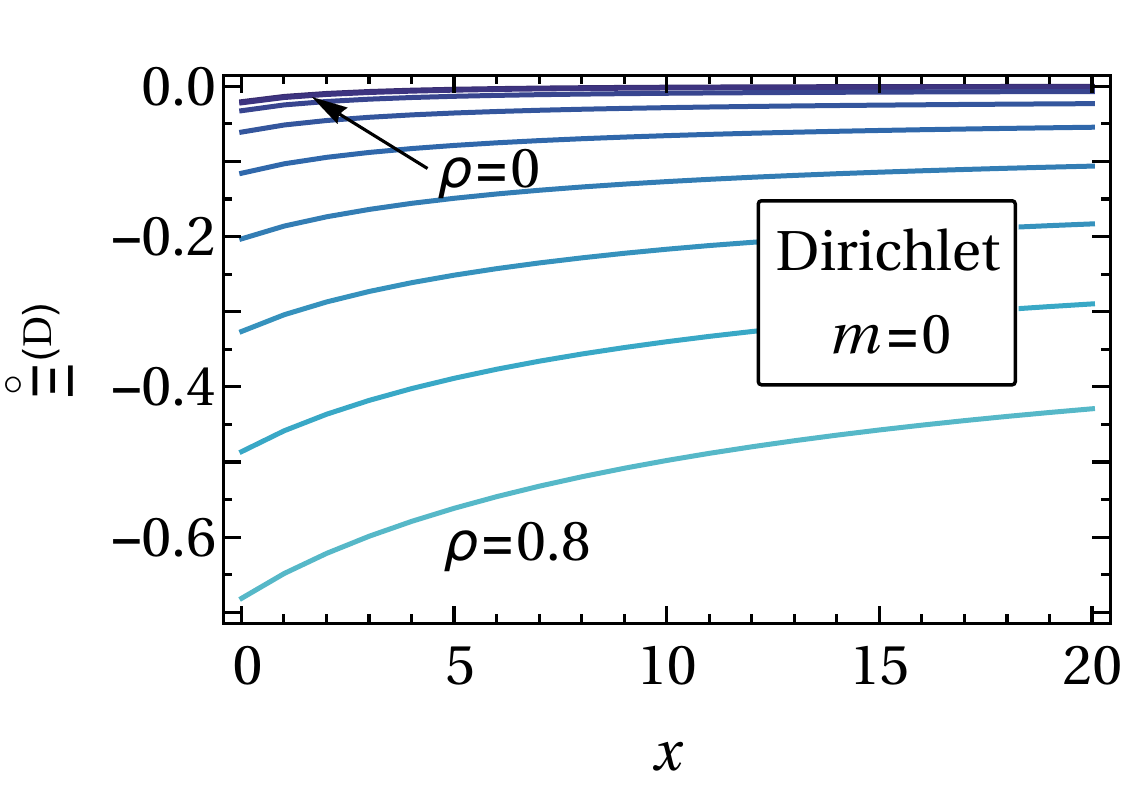}}
\caption{(a)--(c) Scaling functions of the \CCF [\cref{eq_CCF_def_Fres}] at $\Ocal(\epsilon^0)$ for Dirichlet \bcs in the canonical and the grand canonical ensemble [\cref{eq_Casi_force_c_scalf,eq_Casi_force_gc_scalf_m}, respectively] as functions of the scaled temperature $\tscal$ for $\phiscal=0$ and three aspect ratios $\rho$. For $\phiscal=0$, $\ring\Xi$ and $\Xi$ differ only by the constraint-induced term given in \cref{eq_Casi_constrcorr_scal_Dir}. As a consequence, for $\rho>0$ the canonical CCF attains a nonzero value in the limit $\tscal\to \infty$ [see \cref{eq_Casi_constrcorr_Dir_lim} and the short dash-dotted lines in (b) and (c)]. In the case of Dirichlet \bcs, this asymptotic value is approached slower than for periodic ones [see \cref{fig_Xi_pbc_m0}]. (d) Dependence of $\ring\Xi\Dbc(\tscal,\phiscal=0,\rho)$ on $\tscal$ for various values of the aspect ratio $\rho$, increasing from 0 to 0.8 in steps of 0.1 from the top to the bottom curve (with distinct blue shading).}
  \label{fig_Xi_Dir}
\end{figure*}

\subsection{Dirichlet \bcs}

\subsubsection{Residual finite-size free energy}
For Dirichlet \bcs we consider only the case $\phiscal=\hscal=0$; hence the scaling functions of the \resFE [\cref{eq_Fres_c_scalfunc,eq_Fres_gc_scalfunc}] depend solely on $\tscal$. 
The only difference between the residual finite-size free energies in the two ensembles is provided by the constraint-induced term $\delta F\Dbc$ [\cref{eq_constrcorr_Dir}], which contributes to $\ring\Theta\Dbc$ with the expression [\cref{eq_constrcorr_s_Dir}]
\begin{multline} \delta\ring\Theta\Dbc\st{s}(\tscal,\rho) \equiv \rho^{d-1}\delta F\Dbc\st{s}(\tscal,\rho) \\ =  \onehalf \rho^{d-1} \ln \left(\left[\frac{1}{\tscal} - \frac{2}{\tscal^{3/2}} \tanh\left(\sqrt{\tscal}/2\right) \right] 2\pi \rho^{-d+1} \right).
\label{eq_constrcorr_Dir_scalf}\end{multline} 
In the thin-film limit $(\rho\to 0$), $\delta\ring\Theta\st{s}\Dbc(\tscal,\rho)$ vanishes, so that in this case the canonical and grand canonical scaling functions are identical.
In Fig.\ \ref{fig_Fres_Dir} the canonical ($\ring\Theta\Dbc$) and grand canonical ($\Theta\Dbc$) scaling functions are shown for Dirichlet \bcs, for $\phiscal=0$, and for various aspect ratios $\rho$.
Due to \cref{eq_constrcorr_Dir_scalf}, $\ring\Theta\Dbc$ significantly differs from $\Theta\Dbc$ upon increasing the aspect ratio $\rho$.
In particular, while $\Theta\Dbc$ vanishes exponentially for $\tscal\to \infty$, $\ring\Theta\Dbc$ diverges logarithmically in the same limit; this latter behavior is similar to the one discussed above for periodic \bcs [see \cref{sec_pbc_resFE}] and is due to the constraint-induced contribution [see \cref{eq_Fres_largeTau}].

\subsubsection{Critical Casimir force}

Since here we are considering $\phiscal=0$, according to \cref{eq_Casi_force_gc_c_rel_Oeps} the constraint-induced term $\delta\ring\Xi\Dbc$ in \cref{eq_Casi_constrcorr_scal_Dir} provides the only difference between the canonical and grand canonical \CCFs.
Therefore, in the thin-film limit ($\rho\to 0$) the \CCFs for Dirichlet \bcs and $\phiscal=0$ are identical in the two ensembles, as are the corresponding residual finite-size free energies.
The quantity $\delta\ring\Xi\Dbc$ attains two distinct $\rho$-dependent values for $\tscal\to 0$ and $\tscal\to \infty$:
\beq \delta\ring\Xi\Dbc = \begin{cases}\displaystyle
    -\frac{3}{2}\rho^{d-1}\qquad &\text{for }\tscal\to 0,\\[10pt] \displaystyle
    -\onehalf \rho^{d-1}\qquad &\text{for }\tscal\to\infty,
                           \end{cases}
\label{eq_Casi_constrcorr_Dir_lim}\eeq 
which coincide with the corresponding limits of $\delta \ring\Xi\pbc$ for periodic \bcs [see \cref{eq_Casi_constrcorr_pbc_lim}].
Accordingly, while $\Xi\Dbc$ vanishes in the limit $\tscal\to \infty$, the scaling function $\ring\Xi\Dbc$ of the canonical \CCF does not.
This means that the effect of the OP constraint [\cref{eq_constr}] on the fluctuations manifests itself in the form of an attractive contribution to the CCF even for arbitrarily thick films.
The scaling functions $\Xi\Dbc$ and $\ring\Xi\Dbc$ of the \CCF for Dirichlet \bcs are illustrated in Figs.~\ref{fig_Xi_Dir}(a)$-$(c) for various aspect ratios.
In general, the canonical \CCF turns out to be attractive for all aspect ratios considered here and its strength is found to be significantly larger than that of the grand canonical \CCF. We remark that a similar constraint-induced effect is present also for the CCF defined under the constraint of a fixed volume and for Dirichlet \bcs (see \cref{app_CCF_vol}). 
As \cref{fig_Xi_Dir}(d) shows, the strength of the canonical CCF significantly grows upon increasing the aspect ratio from thin-film geometry towards a cubical system. In contrast to the canonical CCF, the grand canonical CCF changes its character from attractive to repulsive upon increasing the aspect ratio $\rho$ [see \cref{fig_Xi_Dir}(c)].

\begin{figure*}[t]\centering
  \subfigure[]{\includegraphics[width=0.38\linewidth]{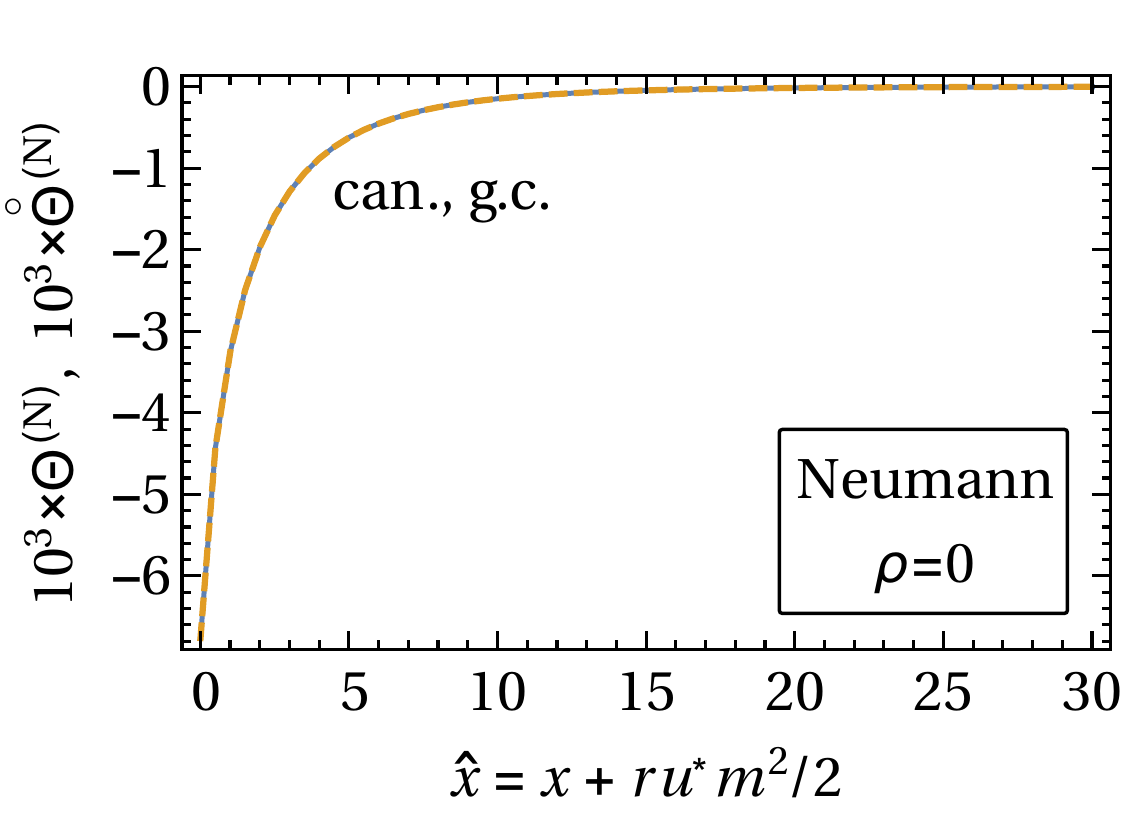} \label{fig_Fres_Neu_film}}\qquad
  \subfigure[]{\includegraphics[width=0.4\linewidth]{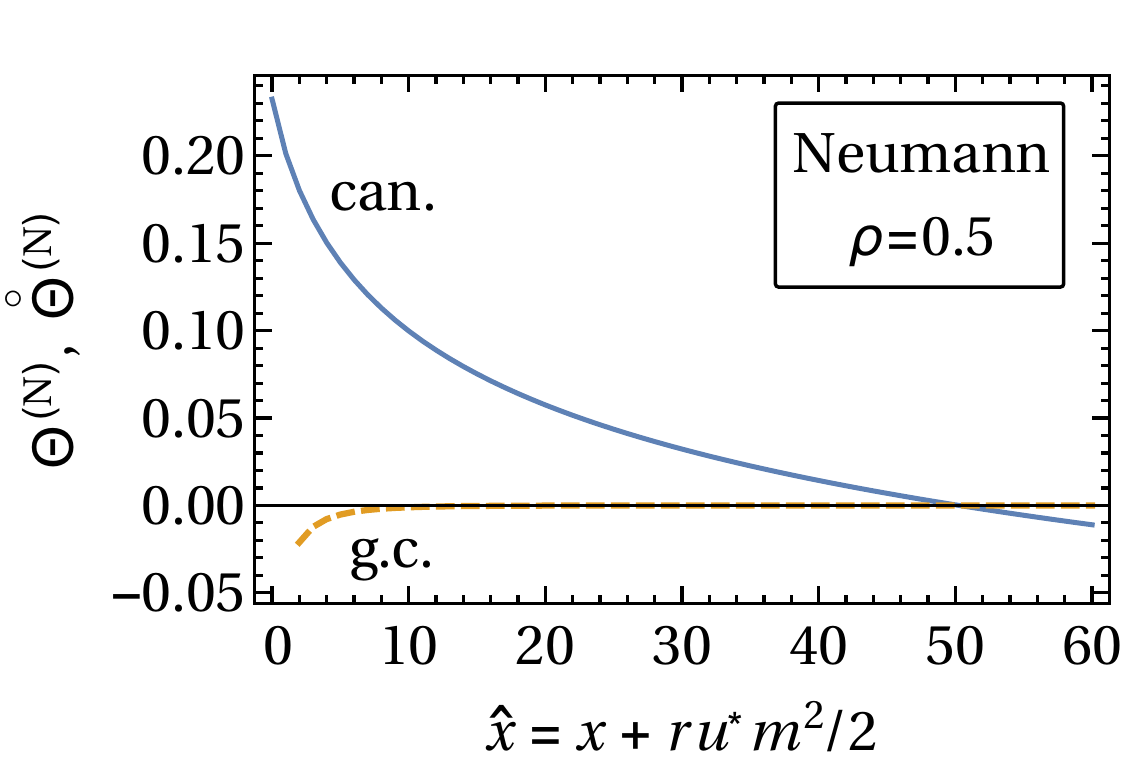}}
  \subfigure[]{\includegraphics[width=0.39\linewidth]{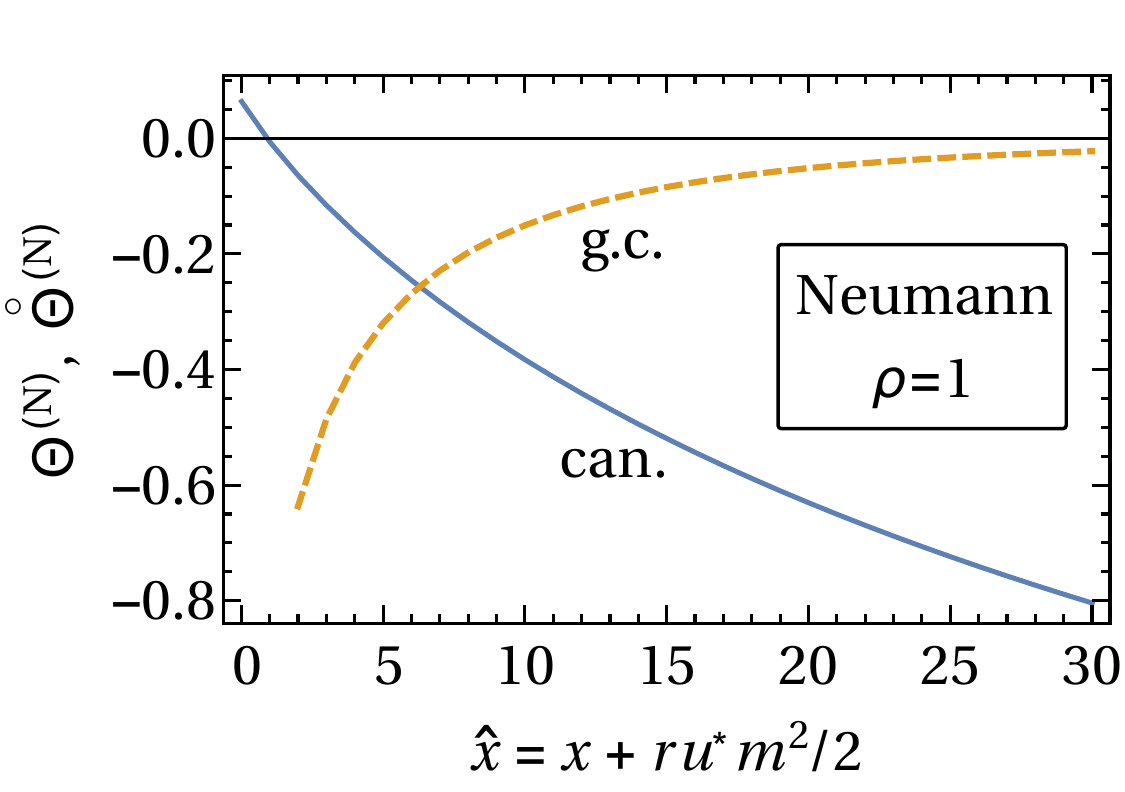}}\qquad
  \subfigure[]{\includegraphics[width=0.4\linewidth]{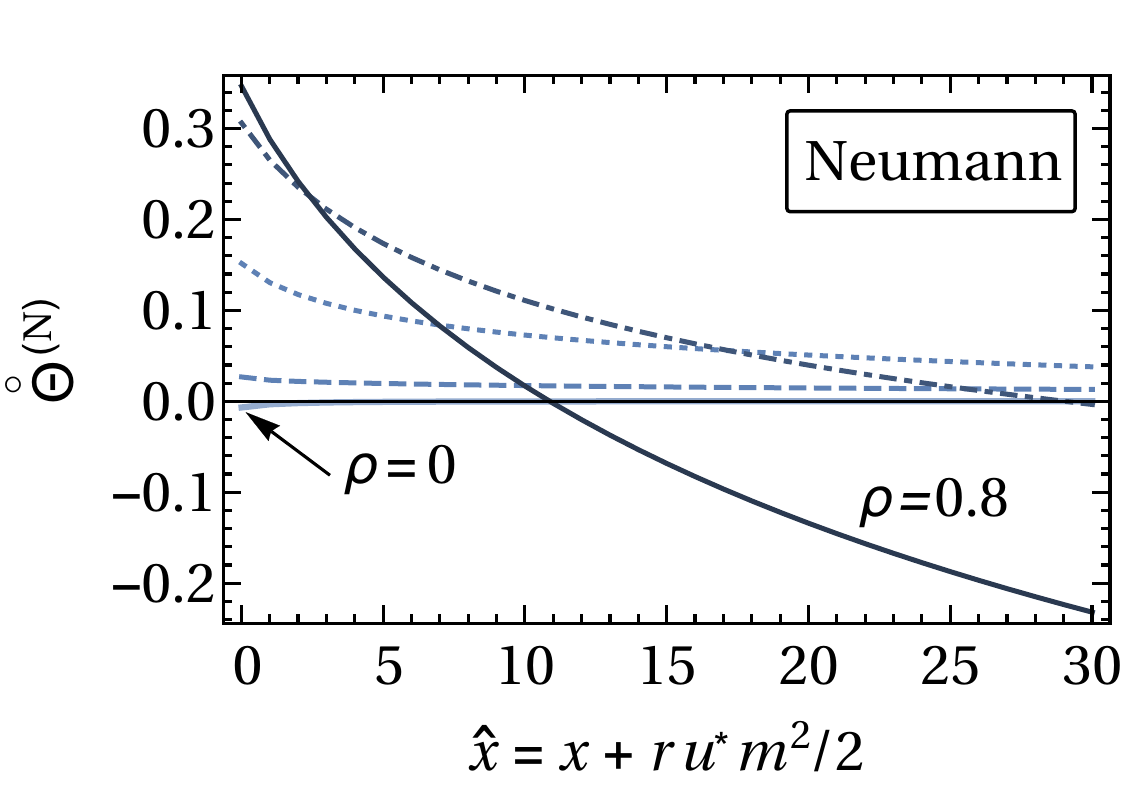}}
  \caption{(a)--(c) Scaling functions at $\Ocal(\epsilon^0)$ of the \resFE for Neumann \bcs in the canonical [\cref{eq_Fres_c_scalfunc,eq_S_Nbc}, solid line] and the grand canonical [\cref{eq_Fres_gc_scalfunc,eq_S_Nbc}, dashed line] ensemble for various aspect ratios $\rho$. In both ensembles, the scaling functions depend on the scaled temperature $\tscal$ and on the scaled mean OP $\phiscal$ via the quantity $\mtscal$ [see \cref{eq_tscal_mod}]. In the grand canonical ensemble, $\phiscal$ is related to the scaled bulk field $\hscal$ according to \cref{eq_EOS_MFT_scalf}. For $\rho>0$, the perturbative expression of $\Theta$ reported in \cref{eq_Fres_gc_scalfunc} is reliable only for $\mtscal\gtrsim 1$. For $\rho=0$ [panel (a)], the canonical and the grand canonical scaling functions are identical. For $\rho>0$, the difference between $\ring\Theta\Nbc$ and $\Theta\Nbc$ stems solely from the constraint-induced correction term [given in \cref{eq_constrcorr_s_perNeu}]. For $\rho\neq 0$ and $\mtscal\to\infty$ the scaling function $\ring\Theta\Nbc$ diverges $\propto -\rho^{d-1}\ln \tscal$. Panel (d) illustrates how the dependence on $\mtscal$ of the canonical scaling function $\ring\Theta\Nbc$ changes upon varying the aspect ratio $\rho$. The unlabeled dashed, dotted, and dash-dotted curves (with distinct blue shading) correspond to $\rho=0.2,0.4$, and 0.6, respectively.}
  \label{fig_Fres_Neu}
\end{figure*}

\subsection{Neumann \bcs}

\subsubsection{Residual finite-size free energy}

The scaling functions $\ring\Theta\Nbc$ and $\Theta\Nbc$ of the canonical and the grand canonical \resFE for Neumann \bcs [\cref{eq_Fres_c_scalfunc,eq_Fres_gc_scalfunc}, respectively] are shown in Fig.~\ref{fig_Fres_Neu} as functions of the scaling variable $\mtscal$ [\cref{eq_tscal_mod}] for various values of the aspect ratio $\rho$. 
Due to the presence of the constraint-induced term $\delta F\st{s}\Nbc$ [\cref{eq_constrcorr_s_perNeu}], $\ring\Theta$ and $\Theta$ are equal only for $\rho=0$, while they increasingly differ for larger values of $\rho$. 
The qualitative behavior of $\Theta\Nbc$ is similar to that of the scaling function $\Theta\pbc$ for periodic \bcs [see Fig.~\ref{fig_Fres_pbc}].
However, for $\rho=0$, $\Theta\Nbc$ is about 50 times smaller in strength than $\Theta\pbc$; the strengths become comparable only for $\rho\simeq 1$.
As discussed in \cref{sec_F_gc}, in the grand canonical ensemble and for $\rho>0$, the perturbative expressions for the \resFE and the CCF are reliable only for $\mtscal\gtrsim 1$.

\subsubsection{Critical Casimir force}

\begin{figure*}[t!]\centering
  \subfigure[]{\includegraphics[width=0.43\linewidth]{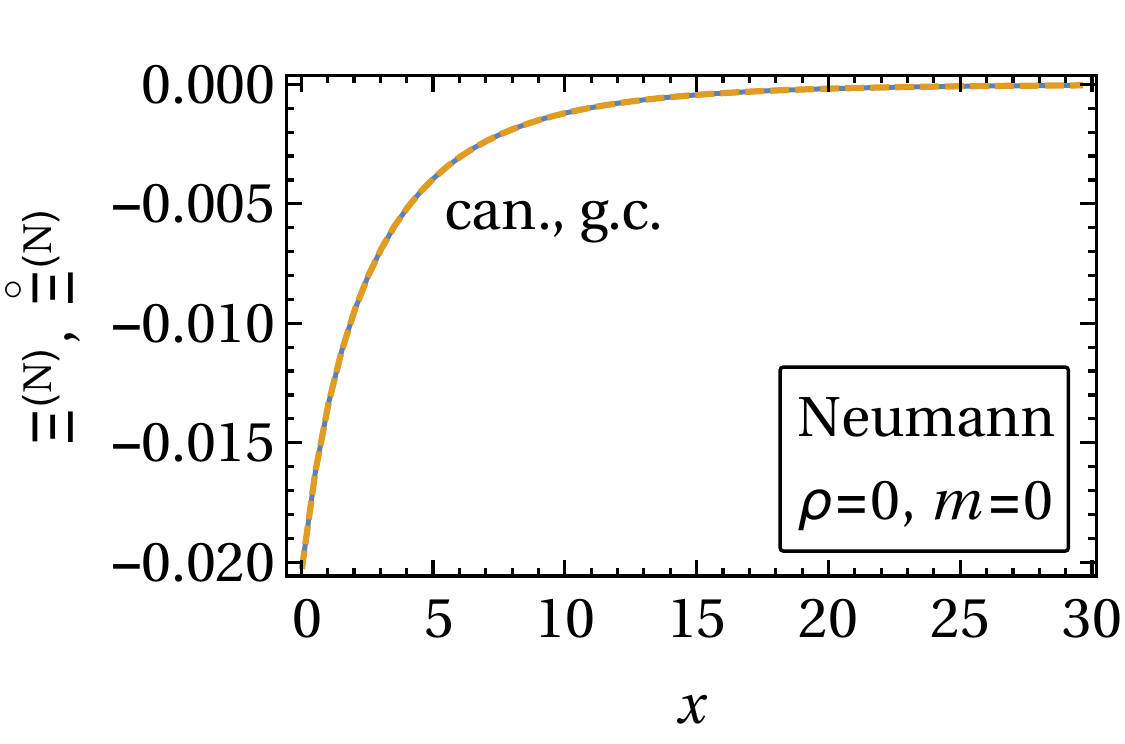}}\qquad
  \subfigure[]{\includegraphics[width=0.4\linewidth]{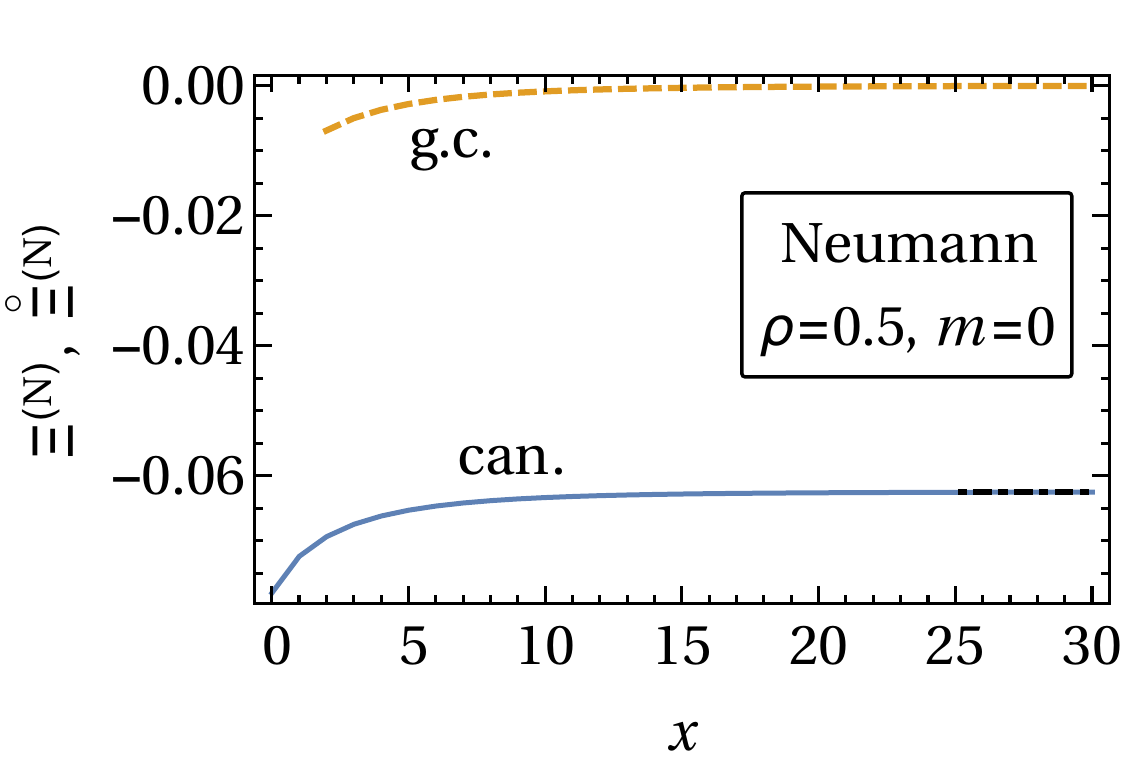}}
  \subfigure[]{\includegraphics[width=0.4\linewidth]{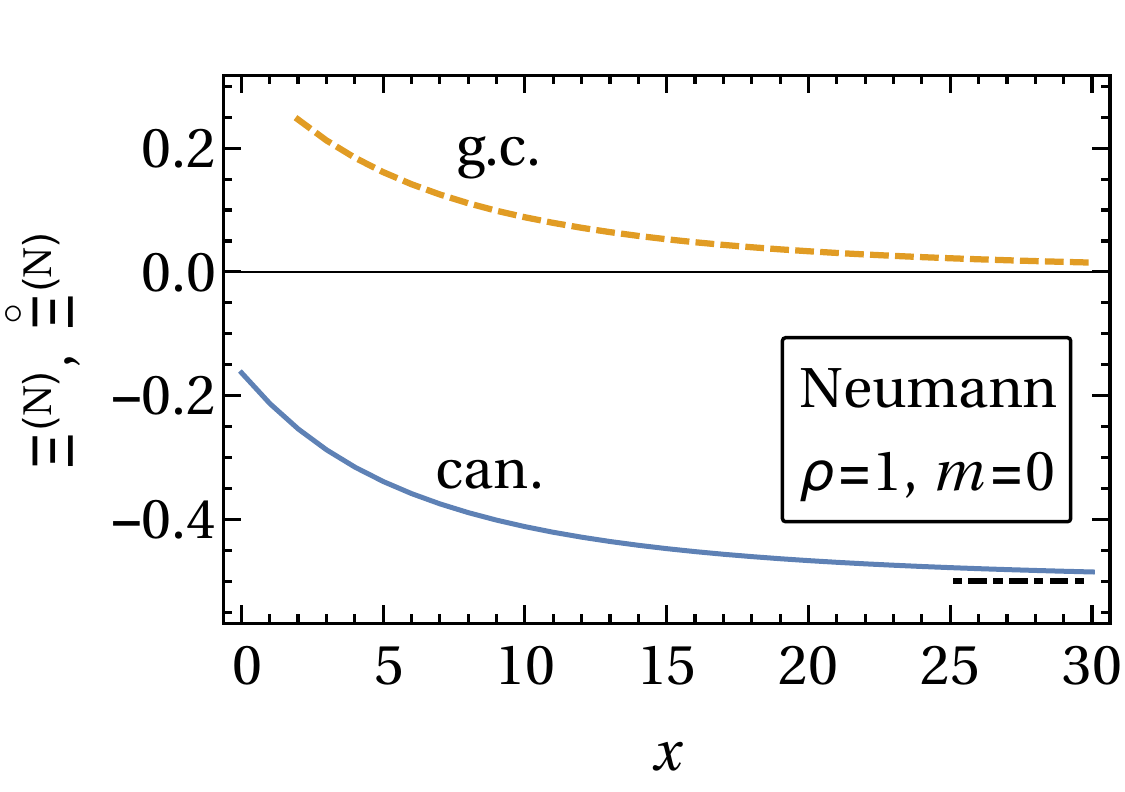}}\qquad
  \subfigure[]{\includegraphics[width=0.4\linewidth]{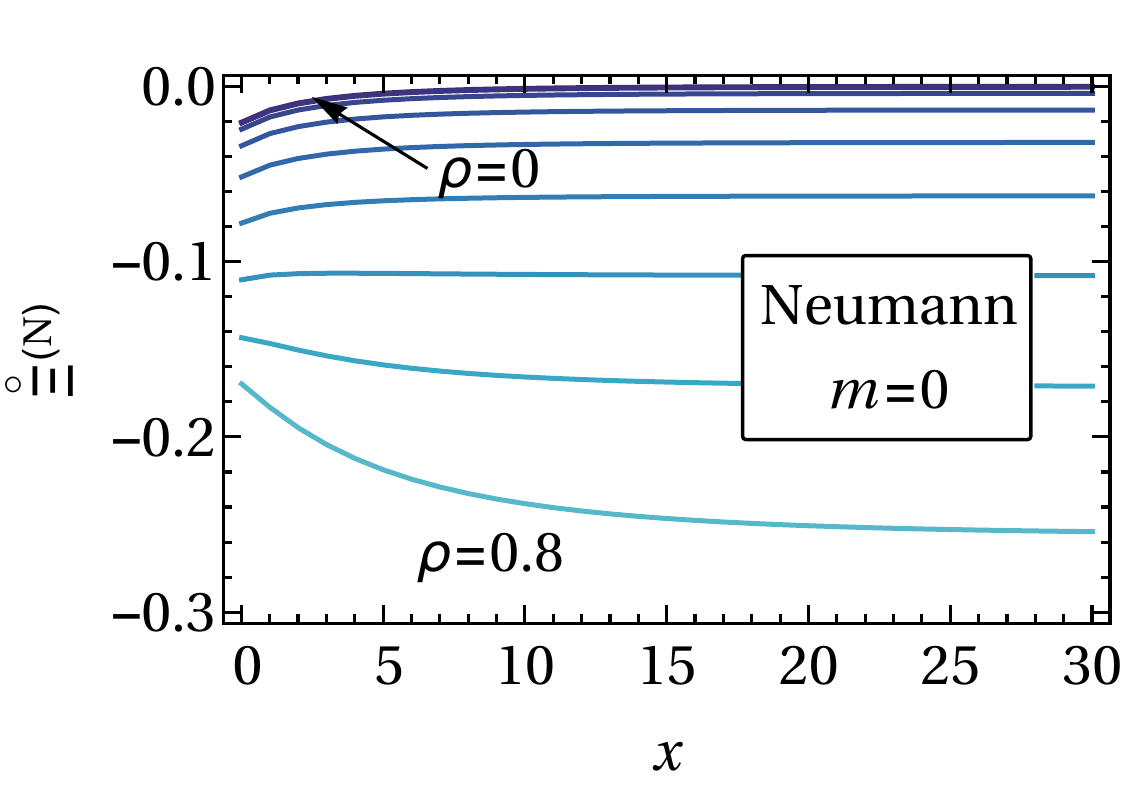}}
  \caption{(a)--(c) Scaling functions of the \CCF [\cref{eq_CCF_def_Fres}] at $\Ocal(\epsilon^0)$ for Neumann \bcs in the canonical and the grand canonical ensembles [\cref{eq_Casi_force_c_scalf,eq_Casi_force_gc_scalf_m}, respectively] as functions of the scaled temperature $\tscal$ for $\phiscal=0$ and three aspect ratios $\rho$. While the grand canonical CCF vanishes both for $\tscal\to\infty$ and for $\phiscal\to\infty$, in these limits the canonical CCF approaches the values given by \cref{eq_Casi_constrcorr_Nbc_lim} [short dash-dotted lines in (b) and (c)]. (d) Dependence of $\ring\Xi\Nbc(\tscal,\phiscal=0,\rho)$ on $\tscal$ for various values of the aspect ratio $\rho$, increasing from 0 to 0.8 in steps of 0.1 from the top to the bottom curve (with distinct blue shading).}
  \label{fig_Xi_Neu_m0}
\end{figure*}

\begin{figure*}[t!]\centering
  \subfigure[]{\includegraphics[width=0.42\linewidth]{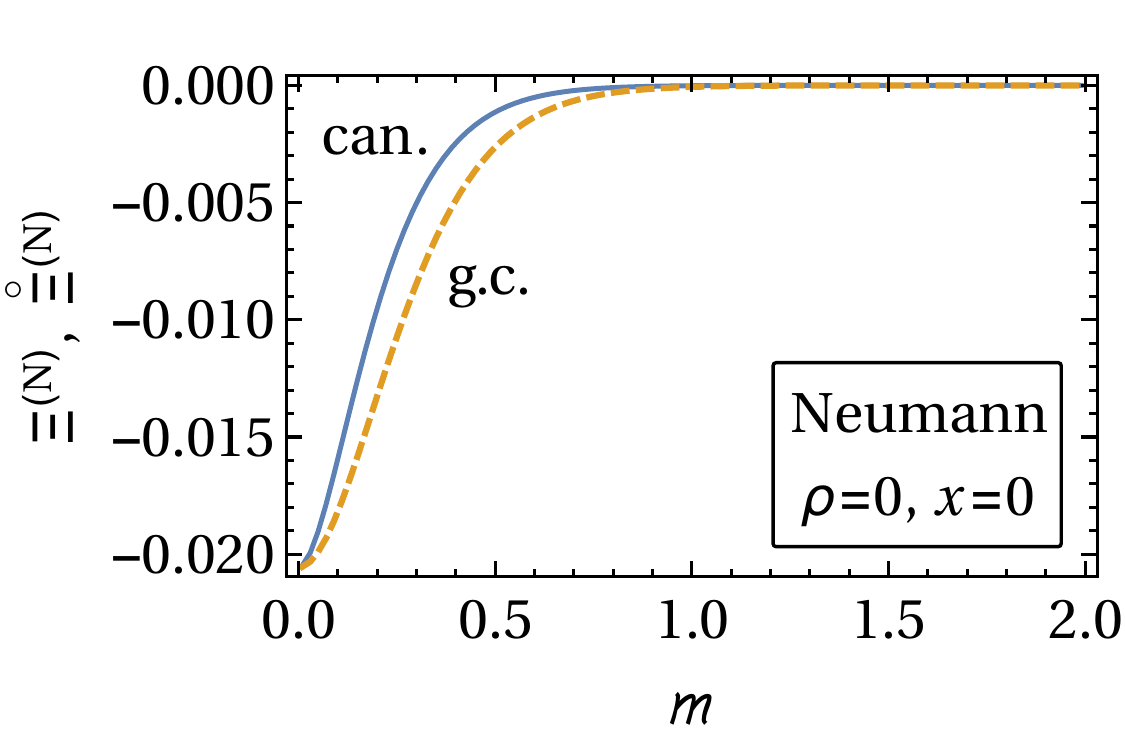}}\qquad
  \subfigure[]{\includegraphics[width=0.4\linewidth]{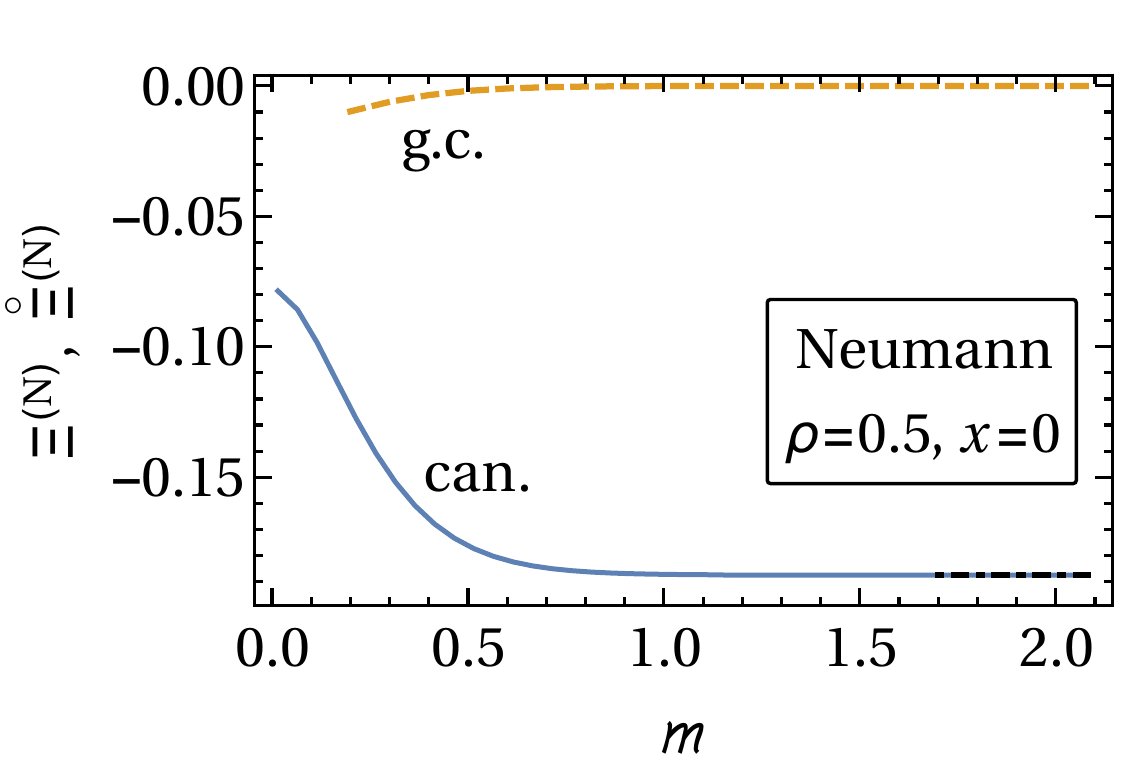}}
  \subfigure[]{\includegraphics[width=0.4\linewidth]{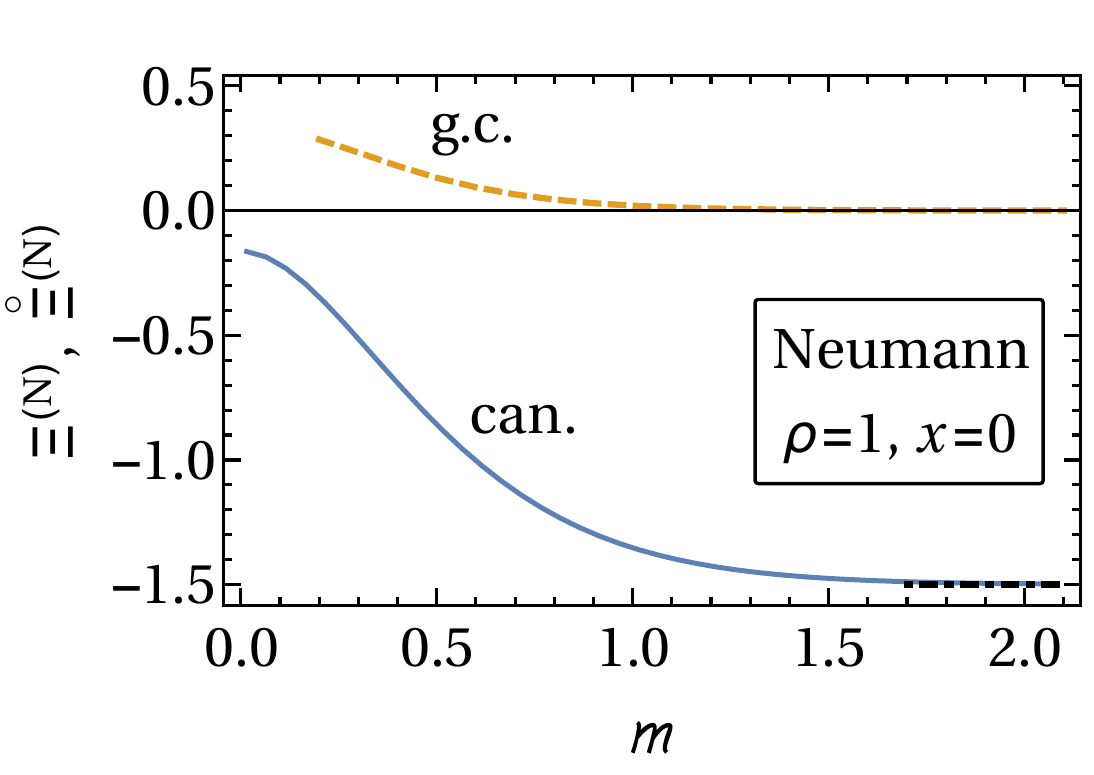}}\qquad 
  \subfigure[]{\includegraphics[width=0.4\linewidth]{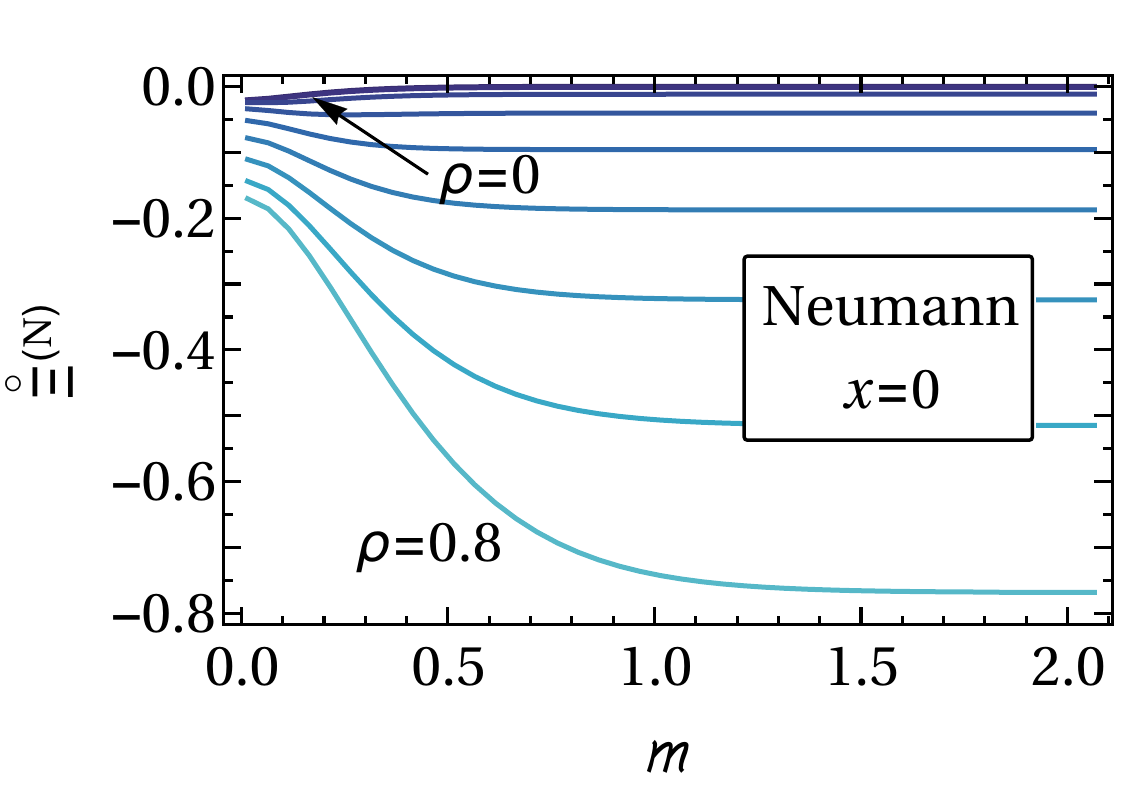}}
 \caption{(a)--(c) Scaling functions of the \CCF [\cref{eq_CCF_def_Fres}] at $\Ocal(\epsilon^0)$ for Neumann \bcs in the canonical and the grand canonical ensembles [\cref{eq_Casi_force_c_scalf,eq_Casi_force_gc_scalf_m}, respectively] as functions of the scaled magnetization $\phiscal$ for $\tscal=0$ and three aspect ratios $\rho$. While the grand canonical CCF vanishes both for $\tscal\to\infty$ and for $\phiscal\to\infty$, in these limits the canonical CCF approaches the values given by \cref{eq_Casi_constrcorr_Nbc_lim} [short dash-dotted lines in (b) and (c)]. (d) Dependence of $\ring\Xi\Nbc(\tscal=0,\phiscal,\rho)$ on $\phiscal$ for various values of the aspect ratio $\rho$, increasing from 0 to 0.8 in steps of 0.1 from the top to the bottom curve (with distinct blue shading).}
  \label{fig_Xi_Neu_x0}
\end{figure*}

The scaling functions $\ring\Xi\Nbc$ and $\Xi\Nbc$ of the canonical and the grand canonical CCF are shown in \cref{fig_Xi_Neu_m0} as functions of the scaled temperature $\tscal$ (for $\phiscal=0$) and in \cref{fig_Xi_Neu_x0} as functions of the scaled mean OP $\phiscal$ (for $\tscal=0$) for various aspect ratios $\rho$.
For $\rho=0$, the contribution $\delta\ring\Xi\Nbc$ in \cref{eq_Casi_constrcorr_scal_perNeu} vanishes, such that, according to \cref{eq_Casi_force_gc_c_rel_Oeps,eq_tscal_mod}, the scaling functions of the canonical and the grand canonical CCF are related as follows:
\begin{multline} 
\Xi\Nbc(\tscal, \phiscal, \rho=0) \\ = \ring\Xi\Nbc(\tscal,\phiscal, \rho=0) - \phiscal\pd_\phiscal \ring\Theta\Nbc(\mtscal(\tscal,\phiscal), \rho=0) .
\label{eq_XiGC_C_sameTheta_Nbc}\end{multline} 
Hence, for $\phiscal=0$ and $\rho=0$, the CCFs in the two ensembles are identical, as illustrated in \cref{fig_Xi_Neu_m0}(a). For nonzero mean OP $\phiscal\neq 0$ and $\rho=0$, the difference between the CCFs stems from the last term in \cref{eq_XiGC_C_sameTheta_Nbc}, the presence of which is a direct consequence of \cref{eq_dPhi_dL}. As shown in \cref{fig_Xi_Neu_x0}(a), similarly to the case with periodic \bcs, this contribution causes the canonical CCF to be less attractive than the grand canonical one.
In contrast, for nonzero aspect ratios $\rho>0$, the contribution $\delta\ring\Xi\Nbc$ [\cref{eq_Casi_constrcorr_scal_perNeu}] dominates in \cref{eq_Casi_force_gc_c_rel_Oeps} and typically leads to a more attractive canonical CCF compared to the grand canonical one. This is illustrated by the panels (b) and (c) of \cref{fig_Xi_Neu_m0,fig_Xi_Neu_x0}. 
For $\rho>0$, $\ring\Xi\Nbc$ approaches a nonzero constant in the limit $\mtscal\to\infty$, whereas $\Xi\Nbc$ vanishes.
Specifically, \cref{eq_Casi_constrcorr_scal_perNeu} implies
\begin{subequations}\begin{align}
\delta\ring\Xi\Nbc(\tscal,\phiscal=0,\rho) = \delta\ring\Xi\Nbc(\tscal\to\infty,\phiscal,\rho) &= -\onehalf \rho^{d-1} \label{eq_Casi_constrcorr_Nbc_m0}
\intertext{and}
\delta\ring\Xi\Nbc(\tscal=0,\phiscal,\rho) = \delta\ring\Xi\Nbc(\tscal,\phiscal\to\infty,\rho) &=- \frac{3}{2} \rho^{d-1}, \label{eq_Casi_constrcorr_Nbc_x0}
\end{align}\label{eq_Casi_constrcorr_Nbc_lim}\end{subequations}
as in the case of periodic and Dirichlet \bcs.
The canonical CCF remains attractive for all aspect ratios $\rho$ considered here and its strength grows significantly upon increasing $\rho$ [see Figs.\ \ref{fig_Xi_Neu_m0}(d) and \ref{fig_Xi_Neu_x0}(d)]. In contrast, the grand canonical CCF changes its character from attractive to repulsive upon increasing $\rho$ [see Figs.\ \ref{fig_Xi_Neu_m0}(c) and \ref{fig_Xi_Neu_x0}(c)].
Under the constraint of a fixed total volume, the associated CCF for Neumann \bcs (see \cref{app_CCF_vol}) turns out to be identical in the two ensembles for all values of $\tscal$, $\phiscal$, and $\rho$ considered here.

\section{Summary and outlook}
\label{sec_summary}
In this study, we have investigated the implications of a global constraint on a scalar OP within a field theoretical approach.
Generic results, which are independent of the specific form of the field-theoretic action describing the confined system are summarized in \cref{sec_fieldtheory_summary} and will not be repeated here.
We have subsequently applied this formalism to a Landau-Ginzburg model for a finite cubical volume $V=AL$ in the supercritical regime ($T\geq T_{c}$, where $T_c$ is the bulk critical temperature). We have considered periodic, Dirichlet, and Neumann \bcs along the transverse direction of extent $L$ and periodic \bcs along the remaining lateral directions [see \cref{fig_setup}].
Our approach is expected to be applicable for films, i.e., for systems with aspect ratios $\rho$ [\cref{eq_aspectratio} and \cref{fig_setup}] fulfilling $0\leq \rho\lesssim 1$. 
Perturbative expressions for the \resFE and the CCF have been obtained within an $\epsilon$-expansion to $\Ocal(\epsilon^0)$, corresponding to the Gaussian approximation of the free energy and to the mean-field approximation of the equation of state.
While most of the present study focuses on the CCF emerging under the condition of fixed transverse area $A$ [see \cref{eq_CCF_def_Fres}], \cref{app_CCF_vol} considers the alternative condition of having a fixed volume $V$, which is also briefly summarized below.
Within the Gaussian approximation, the OP constraint can be implemented exactly and one obtains the expressions of the canonical free energy presented in \cref{eq_F_can_pert}.
Non-Gaussian contributions to the field-theoretic action can be accounted for perturbatively, based on a suitably defined Green function [\cref{eq_G_redef}]. Apart from the contribution of the $\phi^4$-term to the effective temperature variable $\mtau$ [\cref{eq_mtau}], non-Gaussian effects have not been considered. 

The consequences of the constraint on the \resFE and on the CCF are summarized as follows:
\begin{enumerate}
 \item The canonical \resFE $\fcalc\res$ (per volume $AL$ and per $k_B T_c$, see \cref{eq_Fres_def}) differs at $\Ocal(\epsilon^0)$ from the grand canonical $\fcal\res$ by an extra contribution $\delta F(\mtau,L,A)$ [\cref{eq_constrcorr}].
 This term is induced by fluctuations only. Its presence can be most easily understood for periodic \bcs: in this case, in fact, it simply removes the zero mode from the mode sum in the free energy [see \cref{eq_free_en_per}]. $\delta F$ can be decomposed into a scaling and a non-scaling contribution as in \cref{eq_constrcorr_def_sns}, where the latter explicitly depends on the film thickness $L$. For periodic and Neumann \bcs, the constraint-induced contribution cancels the divergence of the grand canonical \resFE at criticality caused by the zero-mode \cite{brezin_finite_1985,rudnick_finite-size_1985}. In the limit $\tscal\to\infty$ or $\phiscal\to\infty$, $\delta F$ gives rise to a logarithmic divergence of the scaling function of the \resFE in the canonical ensemble, $\ring\Theta\propto \rho^{d-1}\ln \mtscal$, independently of the choice of the \bcs [see \cref{eq_Fres_largeTau} and panels (b), (c), and (d) of Figs.\ \ref{fig_Fres_pbc}, \ref{fig_Fres_Dir}, and \ref{fig_Fres_Neu}.]. 
 
 \item For a vanishing aspect ratio ($\rho=0$), the constraint-induced contribution to the \resFE vanishes [see \cref{eq_Fres_c,eq_Fres_c_scalfunc}]. This holds for all \bcs considered here and to all orders in perturbation theory (see the discussion in \cref{sec_fieldtheory_summary}). As a consequence, at $\Ocal(\epsilon^0)$ the canonical and the grand canonical residual finite-size free energies are identical [see panel (a) in Figs.\ \ref{fig_Fres_pbc}, \ref{fig_Fres_Dir}, and \ref{fig_Fres_Neu}] and reduce (for $\phiscal=0$) to the one-loop results of Ref.\ \cite{krech_free_1992}. For the boundary conditions considered here, the equation of state acquires finite-size corrections only beyond the leading order in the $\epsilon$-expansion \footnote{For nonzero symmetry breaking surface fields $h_1$, the equation of state acquires finite-size contributions already at the mean-field level \cite{gross_critical_2016}. This is expected to hold also for Dirichlet \bcs for a nonzero mean OP.}. 

 \item The CCF depends on whether it is defined under the constraint of a constant transverse area $A$ [\cref{sec_renorm_CCF}] or a constant total volume $V=AL$ [\cref{app_CCF_vol}]. In the latter case, at $\Ocal(\epsilon^0)$ the OP constraint has \emph{no} effect on the scaling functions for periodic and Neumann \bcs  (see \cref{fig_XiV}), i.e., the canonical and the grand canonical CCF coincide. For Dirichlet \bcs, instead, ensemble differences are present for both definitions of the CCF. They vanish, however, in the thin-film limit ($\rho\to 0$).
 
 \item If the CCF is defined under the condition of a fixed transverse area $A$, the OP constraint is reflected by two distinct contributions to the canonical CCF: first, the fluctuation-induced term $\delta F$ in the \resFE yields, via \cref{eq_CCF_def_Fres}, a contribution $\delta \ring\Xi$ to the scaling function of the CCF [see \cref{eq_Casi_constrcorr}]. Within the Gaussian approximation, the expressions of $\delta F$ and hence $\delta \ring\Xi$ coincide for periodic and Neumann \bcs [see \cref{eq_Casi_constrcorr_scal_perNeu}; $\delta\ring\Xi\Dbc$ for Dirichlet \bcs is reported in \cref{eq_Casi_constrcorr_scal_Dir}]. These contributions vanish in the thin-film limit ($\rho\to 0$). A second difference between the canonical and the grand canonical CCFs arises from the fact that the mean OP $\avOP$, which enters into the definition of the OP scaling variable $\phiscal$ [see \cref{eq_Casi_force_c}], is affected, via \cref{eq_dPhi_dL}, by the constraint of having a fixed total OP $\intOP$ and a fixed area $A$. This effect occurs also within MFT \cite{gross_critical_2016}. 
 
 \item For nonzero aspect ratios ($\rho>0$), the canonical CCF is typically more attractive than the grand canonical CCF (see Figs.\ \ref{fig_Xi_pbc_m0}, \ref{fig_Xi_pbc_x0}, \ref{fig_Xi_Dir}, \ref{fig_Xi_Neu_m0}, and \ref{fig_Xi_Neu_x0}). Indeed, a restriction on the OP [see \cref{eq_fluct_constr}] is expected to reduce the fluctuation contribution to the pressure of the confined system and therefore leads to an additional, attractive contribution to the CCF.
 However, this effect is absent for periodic and Neumann \bcs if the CCF is defined with a fixed total volume $V$ (see \cref{app_CCF_vol}).
 In the limit $\tscal\to\infty$ or $\phiscal\to\infty$, the scaling function $\ring\Xi$ of the canonical CCF for fixed transverse area approaches a negative constant $\propto \rho^{d-1}$ [see \cref{eq_Casi_constrcorr_pbc_lim}]. This asymptotic value is the same for all \bcs studied here. In contrast, the CCF defined with constant volume vanishes in the limit $\tscal\to\infty$ or $\phiscal\to\infty$ for all \bcs considered.
 
 \item For $\rho=0$ and vanishing mean OP $\phiscal=0$, the canonical and the grand canonical CCF defined for fixed transverse areas coincide [see panels (a) of \cref{fig_Xi_pbc_m0,fig_Xi_Dir,fig_Xi_Neu_m0}] and reduce to the expressions reported in Ref.\ \cite{krech_free_1992}. In contrast, for $\rho=0$ and nonzero mean OPs $\phiscal\neq 0$, the OP constraint yields, via \cref{eq_dPhi_dL}, a repulsive contribution to the canonical CCF [see Figs.\ \ref{fig_Xi_pbc_x0}(a) and \ref{fig_Xi_Neu_x0}(a)], although in that case the corresponding residual finite-size free energies are identical.
 In general, this repulsive contribution is absent for the CCF defined for constant volume, because in that case $\d\avOP/\d L=0$.
 For $\rho=0$, in general the canonical CCF vanishes in the limits $\tscal\to\infty$ or $\phiscal\to\infty$.
 
\end{enumerate}

We mention that the perturbative results obtained here for the grand canonical CCF agree qualitatively with corresponding Monte Carlo simulation data \cite{hucht_aspect-ratio_2011, hasenbusch_thermodynamic_2011, vasilyev_universal_2009}.
If one aims at improving the analytical predictions in the grand canonical ensemble, in particular the issues associated with the presence of a  zero mode must be dealt with appropriately (see, e.g., Refs.\ \cite{esser_field_1995, dohm_critical_2011, dohm_pronounced_2014, gruneberg_thermodynamic_2008, diehl_fluctuation-induced_2006}).
The purpose of the present study is, however, not to present quantitatively accurate expressions for the grand canonical CCF, but to provide a self-contained treatment of ensemble differences due to fluctuations in a near-critical, confined fluid.
We remark that, for the Ising model at very high temperatures [see Eqs.~(131) and (132) in Ref.\ \cite{gross_critical_2016}], the constraint induced contribution to the free energy and CCF reduces (up to irrelevant constants) to the expressions given in \cref{eq_Fres_largeTau,eq_Casi_constrcorr_pbc_lim}.

The present study can be considered as a sequel to Ref.\ \cite{gross_critical_2016}, where we have investigated the ensemble differences within MFT for $(++)$ and $(+-)$ \bcs, i.e., for a confined, near-critical fluid which exhibits strong adsorption at the container walls in the transverse direction. 
For periodic and Neumann \bcs, as well as for the disordered phase with Dirichlet \bcs, a mean-field contribution to the \resFE is in general absent. 
Periodic \bcs, although not experimentally relevant (see, however, Ref.\ \cite{ranc_con_critical_2016}), are arguably the simplest case for which the influence of a constraint on the OP fluctuations can be studied analytically. 
Dirichlet \bcs apply at the RG fixed point of the so-called ordinary surface universality class \cite{diehl_field-theoretical_1986}. Generically, confining surfaces exhibit a preference for one of the two species of a binary liquid mixture, which gives rise to a symmetry breaking surface field and therefore to ($++$) or ($+-$) \bcs. If the surface is endowed with a periodically striped pattern of alternating surface fields, for thick films such surfaces behave effectively as if there is a Dirichlet boundary condition (see Sec.\ IIIB in Ref.\ \cite{sprenger_forces_2006}). This way, Dirichlet \bcs can be realized even for classical binary liquid mixtures.
Neumann \bcs apply at the fixed-point of the so-called special surface universality class and correspond to weak adsorption.
Our predictions lend themselves to be tested by Monte Carlo \cite{vasilyev_monte_2015, gross_critical_2016} or molecular dynamics \cite{puosi_direct_2016, roy_structure_2016} simulations. 
In future studies, the theory developed here could be extended to the sub-critical region, where, so far, predictions for the canonical CCF are not available.

\acknowledgements{We thank O. Vasilyev for informing us about preliminary simulation results and D.\ Dantchev and M.\ Kr\"uger for useful discussions.}


\appendix

\section{Gaussian lattice field theory and dimensional considerations}
\label{app_lattice}
Here, we consider an uncorrelated Gaussian random field on a $d$-dimensional hypercubic lattice of volume $V = N a^d$, where $N$ is the number of lattice points and $a$ the lattice constant.
This is arguably the simplest system for which the effect of the constraint on the OP field [\cref{eq_totMass}] can be studied exactly. In addition, the finite lattice constant of the model provides a natural regularization.
Before proceeding, we recall that the OP field $\phi$, the reduced temperature $\tau$, and the bulk field $h$ appearing in the action [see \cref{eq_landau_fe}] must have the engineering dimensions 
\beq [\phi] = a^{1-d/2},\quad [\tau] = a^{-2}, \quad [h]=a^{-1-d/2},
\eeq 
in order to render $\Hcal$ in \cref{eq_Z_unconstr}, and thus also the free energy $\Fcal$ in units of $k_B T_c$, dimensionless \cite{amit_field_2005}.  

\subsection{Grand canonical ensemble}
We consider a dimensionless Hamiltonian of the form 
\beq \Hcal(\{\phi_i\},h) = \frac{\tau}{2} a^d\sum_{i=1}^N \phi_i^2 - a^d h \sum_{i=1}^N \phi_i,
\label{eq_lattice_Hgc}\eeq
corresponding to a Gaussian ensemble of uncorrelated random variables $\phi_i$.
Using the lattice constant $a$ to render the integration measure dimensionless, the lattice partition function corresponding to \cref{eq_Z_unconstr} is given by
\beq \begin{split} 
\Zcal(h) &= \prod^N_{i=1} \int_{-\infty}^\infty \frac{\d\phi_i}{a^{1-d/2}} \exp\left(-\frac{\tau}{2} a^d \sum_{i=1}^N  \phi_i^2 + a^d h\sum_{i=1}^N \phi_i \right) \\
&= \left(\frac{2\pi}{\tau a^2}\right)^{N/2} \exp\left( \frac{N a^d h^2}{2\tau} \right)\\
& = \Zcal(0) \exp\left( \frac{V h^2}{2\tau} \right).
\end{split}\label{eq_lattice_Zgc}\eeq 
We point out that the contribution which diverges in the \emph{continuum limit} $a\to 0$ (with fixed volume $V$) is contained completely in $\Zcal(0)$. It is therefore convenient to define the actual grand canonical partition function by dividing $\Zcal(h)$ by $\Zcal(0)$ \footnote{This is, in fact, a commonly adopted definition of a path integral \cite{kleinert_path_2009}.}. However, since this term does not interfere with others, we carry it along in our calculations.
The bulk field $h$ can be related to the average $\bra\Phi\ket$ of the total OP 
\beq \Phi \equiv a^d\sum_{i=1}^N \phi_i
\label{eq_lattice_totalOP}\eeq 
by noting that, according to \cref{eq_lattice_Zgc}, $\left\bra \Phi \right\ket = \sfrac{\pd \ln\Zcal}{\pd h} =Vh/\tau$ and thus
\beq h = \tau \frac{\bra \Phi\ket}{V}.
\eeq 
Accordingly, the free energy in units of $k_B T_c$ follows as
\beq \begin{split}\Fcal(h) &= -\ln \Zcal(h) \\ &= -\frac{N}{2} \ln 2\pi + \frac{N}{2} \ln \left[ \tau L^2 \left(\frac{a}{L}\right)^2\right] -\onehalf \tau V\left(\frac{\bra\Phi\ket}{V}\right)^2 \\
&= -\ln\Zcal(0) -\onehalf \tau V\left(\frac{\bra\Phi\ket}{V}\right)^2. \end{split}
\label{eq_lattice_Fgc}\eeq 
Below, we compare these results are compared with the corresponding expressions in the canonical ensemble.

\subsection{Canonical ensemble}
The counterpart of \cref{eq_lattice_Hgc} in the canonical ensemble is given by
\beq \Hcal(\{\phi_i\}) = a^d\sum_{i=1}^N \frac{\tau}{2} \phi_i^2,
\eeq
subject to a constraint of the form
\beq a^d \sum_{i=1}^N \wt_i \phi_i = \Phi,
\eeq 
which is imposed on the field $\{\phi_i\}$ such that \cref{eq_lattice_totalOP} is recovered for $\wt_i=1$. Here, we keep the general expression involving $\wt_i$ in order to be able to track the influence of the constraint.
Using the lattice constant $a$ in order to render the integration measures and the argument of the $\delta$ function dimensionless (note that $[\Phi]= a^{1+d/2}$), the \emph{constrained} lattice partition function corresponding to \cref{eq_Z_constr} is given by
\begin{widetext}
\beq \begin{split} 
\Zcalc(\Phi) &= \left( \prod^N_{j=1} \int_{-\infty}^\infty \frac{\d\phi_j}{a^{1-d/2}} \right) \exp\left(-\frac{\tau}{2} a^d \sum_{j=1}^N  \phi_j^2 \right) \delta\left(\left(a^d \sum_{j=1}^N \wt_j \phi_j - \Phi\right) a^{-(1+d/2)}\right) \\
&= \frac{1}{2\pi}\int_{-\infty}^\infty \frac{\d J}{a^{-1-d/2}} \left( \prod_{j=1}^N \int \frac{\d\phi_j}{a^{1-d/2}} \right) \exp\left(-\frac{\tau}{2} a^d \sum_{j=1}^N \phi_j^2 +  \im  J a^d \sum_{j=1}^N \wt_j\phi_j - \im J \Phi \right) .
\end{split}\label{eq_lattice_Zcan}\eeq 
Because we require the weights $\wt_i$ to be dimensionless, the dimension of the auxiliary integration variable $ J$ is $[J]= a^{-(1+d/2)}$.
Accordingly, $\Zcalc$ in \cref{eq_lattice_Zcan} is dimensionless.
In \cref{eq_lattice_Zcan}, first performing the Gaussian integrals over $\{\phi_j\}$ and then the remaining one over $ J$, we obtain
\beq \begin{split} 
\Zcalc(\Phi) &= \frac{1}{2\pi}\int_{-\infty}^\infty \frac{\d J}{a^{-(1+d/2)}} \left(\frac{2\pi}{\tau a^2}\right)^{N/2} \exp\left[-\onehalf  J^2 \frac{a^d \sum_j \wt_j^2}{\tau} - \im J \Phi \right]\\
&= \left(\frac{2\pi}{\tau a^2}\right)^{N/2} \left( \frac{\tau a^{2+d}}{2\pi a^d \sum_j \wt_j^2} \right)^{1/2} \exp\left[ - \frac{\tau \Phi^2}{2 a^d \sum_j \wt_j^2}  \right] \\
&\overset{\wt_j=1}{=} \Zcal(0) \left(\frac{\tau a^{2+d}}{2\pi V}\right)^{1/2} \exp\left(-\frac{\tau \Phi^2}{2V}\right).
\end{split}\eeq 
This result can also be obtained directly from \cref{eq_lattice_Zgc,eq_lattice_Zcan} by noting that
\beq \Zcalc(\Phi) = \frac{1}{2\pi}\int_{-\infty}^\infty \frac{\d J}{a^{-(1+d/2)}} \exp(-\im J\Phi) \Zcal(\im J).
\eeq 
Accordingly, the free energy of the constrained system, in units of $k_B T_c$, is given by
\beq\begin{split}
\Fcalc(\Phi) = -\ln \Zcalc(\Phi) &= -\ln\Zcal(0) + \onehalf \ln\left( \frac{2\pi}{\tau a^{2}} \sum_j \wt_j^2 \right) + \frac{\tau}{2} \frac{\Phi^2}{a^d\sum_j\wt_j^2} \\
&\overset{\wt_j=1}{=}  -\ln \Zcal(0) + \onehalf \ln \left(\frac{2\pi \rho^{-d+1} }{\tau L^2} \frac{L^{2+d}}{a^{2+d}}\right) +  \frac{\tau}{2} V \left(\frac{\Phi}{V}\right)^2.\end{split}\label{eq_lattice_freeE}
\eeq 
\end{widetext}
In the last step of \cref{eq_lattice_freeE}, we have introduced the aspect ratio $\rho=L/A^{1/(d-1)}$, assuming a lattice of cubical geometry and volume $V=AL$.
Introducing the lattice correlation function $\Gcal_{ij}$ for this model of an uncorrelated random field $\{\phi_i\}$ in the form $\Gcal_{ij} = \delta_{i,j}/(a^d \tau)$ (which has an engineering dimension of $a^{2-d}$), the second term on the r.h.s.\ of the second equation in \cref{eq_lattice_freeE} can be written alternatively as $ \onehalf \ln \left(2\pi a^{d-2} \sum_{i,j} \wt_i \Gcal_{ij} \wt_j\right)$.
In the continuum limit ($a\to 0$ with fixed $V$), this expression reduces, upon recalling that $\sum_i a^d \to \int \d V$ and by neglecting a divergent factor $a^{-d-2}$, to the term $\onehalf \ln \left(2\pi(\wt,\Gcal,\wt)\right)$ in the corresponding expression for the free energy in \cref{eq_freeE_Gauss}.
The second term on the r.h.s. of the last equation in \cref{eq_lattice_freeE} is a constraint-induced contribution which appears in the same form also for periodic and Neumann \bcs within the corresponding continuum field theory (see \cref{eq_constrcorr_per,eq_constrcorr_Neu}, respectively).
The last term in \cref{eq_lattice_freeE} is the usual canonical bulk free energy. Its sign is opposite to that of the analogous term in \cref{eq_lattice_Fgc} because the bulk contributions to $\Fcal(h)$ and $\Fcalc(\Phi)$ are related via a Legendre transform.

\section{Fourier transforms}
\label{app_Fourier}

We consider a function $f(\rv)$ which is periodic with respect to a $D$-dimensional macroblock of volume $A=\Pi_{\alpha=1}^D L_\alpha$ ($D\leq d$) with edges of length $L_\alpha$:
\begin{multline} f(\rv)=f(\rv+\bv{T}_\bv{m}),\quad \bv{T}_\bv{m}=(m_1 L_1, m_2 L_2,\ldots, m_D L_D),\\ \bv{m}\in \mathbb{Z}^D.
\label{eq_Fourier_periodic}\end{multline} 
This function can be expressed in terms of a Fourier series 
\beq 
f(\rv) = \frac{1}{A} \sum_\kv \exp(\im \kv \cdot \rv) \hat f(\kv) .\label{eq_Fourier_inv}
\eeq
\Cref{eq_Fourier_periodic} implies that $\kv$ is discrete, i.e., $\kv= 2\pi \left(n_1/L_1,\ldots, n_D/L_D\right)$ with $\bv{n}\in \mathbb{Z}^D$.
Forming \mbox{$\int_A \d^D r\, \exp(-\im \kv\cdot\rv) f(\rv)$} and inserting \cref{eq_Fourier_inv} yields the inverse Fourier transform
\beq 
\hat f(\kv) = \int_A \d^{D}r\, \exp(-\im\kv\cdot \rv) f(\rv) \label{eq_Fourier}
\eeq 
by using 
\beq \int_A \d^D r \exp[\im (\kv-\kv')\cdot\rv] = A \delta_{\kv,\kv'}.
\label{eq_expIK_int}\eeq 
The Fourier transform $\hat F(\kv,\kv')$ of a function $F(\rv,\rv')= f(\rv-\rv')$, where $f$ is periodic and $k_\alpha$ and $k_\alpha'$ are discrete (as above), is given by
\begin{widetext}
\beq\begin{split} \hat F(\kv,\kv') &=\int_A \d^{D} r\int_A \d^{D} r'\, \exp(-\im \kv\cdot \rv-\im\kv'\cdot\rv') F(\rv,\rv') \\
&= \frac{1}{A} \sum_\pv \int_A \d^{D} r\int_A \d^{D} r'\, \exp[-\im (\kv-\pv)\cdot \rv - \im ( \kv'+\pv)\cdot\rv'] \hat f(\pv) \\
&= \frac{1}{A} \sum_\pv \hat f(\pv) \int_A \d^D r \exp[-\im (\kv-\pv)\cdot\rv] \int_A \d^D r' \exp[-\im(\kv'+\pv)\cdot\rv'] \\
&= \frac{1}{A} \sum_\pv \hat f(\pv) A \delta_{\kv,\pv} A \delta_{\kv',-\pv} = A \hat f(\kv) \delta_{\kv,-\kv'},
\end{split}\label{eq_Fourier_translinv}\eeq
where \cref{eq_expIK_int} has been used.

\section{Canonical finite-size free energy}
\label{app_calc_fcan}

\subsubsection{Periodic \bcs} 
\label{sec_Fres_per}
In order to determine the regularized finite-size free energy for a system with periodic \bcs in all spatial directions and arbitrary aspect ratio, we follow the approach as taken in Refs.\ \cite{dohm_diversity_2008, dohm_critical_2011}. In these studies, only the case $\avOP=0$ was considered. Within the present theory, the generalization of the free energy to nonzero $\avOP$ amounts to replacing the temperature parameter $\tau$ by $\mtau$ [see \cref{eq_mtau}].
In order to extract the finite-size part of the mode sum [see \cref{eq_freeE_inhom,eq_free_en_per}]
\beq 
\msumBare\pbc_{d}(\mtau, L,A) \equiv \sum_\kv \ln(\kv^2+\mtau) ,
\label{eq_modesum_per0}\eeq 
we introduce as a regularization the subtraction of the corresponding bulk expression:
\beq 
\msumBare\pbc_{d,\reg}(\mtau, L,A) \equiv \sum_\kv \ln(\kv^2+\mtau) - AL \int \frac{\d^{d}k}{(2\pi)^{d}} \ln(\kv^2 + \mtau).
\label{eq_modesum_per1}\eeq 
As shown in Refs.\ \cite{dohm_diversity_2008, dohm_critical_2011}, this expression can be simplified to
\beq 
\msumBare\pbc_{d,\reg}(\mtau,L, A) = A L^{-d+1} \modesum\pbc_{d,\reg}(\mtau L^2,\rho)
\label{eq_modesum_per2}\eeq 
with
\beq \modesum\pbc_{d,\reg}(\mtscal,\rho) = \int_0^\infty \d y\, y^{-1} \exp\left(-\frac{\mtscal y}{4\pi^2}\right) \left\{\left(\frac{\pi}{y}\right)^{d/2} - \left[\rho \vartheta(\rho^2 y)\right]^{d-1} \vartheta(y) \right\},
\label{eq_modesum_per}\eeq 
where
\beq \vartheta(y) \equiv \theta_3(0|e^{-y}) = \sum_{n=-\infty}^\infty e^{-y n^2}
\label{eq_thetafunc}\eeq
is the elliptic Jacobi theta function  $\theta_3(z|q)$ \cite{olver_nist_2010}.
Due to the presence of the theta function, $\modesum\pbc_{d,\reg}$ is not a homogeneous function of its first argument, i.e., there is \emph{no} value of $\kappa$ for which, with arbitrary $\mtscal$ and $b$, one has $\modesum\pbc_{d,\reg}(b \mtscal,\rho)\, =\, b^\kappa \modesum\pbc_{d,\reg}(\mtscal,\rho)$. 
In order to facilitate the analysis of the scaling behavior (see Sec.\ \ref{sec_scaling}), $\msumBare\pbc_{d,\reg}$ in \cref{eq_modesum_per2} has been brought directly into a suitable scaling form.

It is useful to note the limiting behaviors $\vartheta(y\to \infty)=1$ and $\vartheta(y\to 0) \simeq (\pi/y)^{1/2}[1+2\exp(-\pi^2/y)]$.
Accordingly, the integrand in \cref{eq_modesum_per} vanishes in the limit $y\to 0$ for all $\rho$ and $\mtscal$ \footnote{For $\rho<1$, $y\to 0$, and all $\mtscal$, the integrand in \cref{eq_modesum_per} behaves in leading order as $2\pi^{d/2}y^{-d/2-1} \exp(-\pi^2/y)$.}, and decays exponentially as a function of $y$ for $y\to\infty$ and $\mtscal \neq 0$.
Thus $\modesum\pbc_{d,\reg}$ is finite for all $d$ and $\mtscal \neq 0$.
In contrast, for $\mtscal\to 0$ and nonzero $\rho$, $\modesum\pbc_{d,\reg}$ diverges asymptotically as 
\beq \modesum\pbc_{d,\reg}(\mtscal\to 0,\rho\neq 0)\simeq \rho^{d-1}\ln \mtscal,
\label{eq_modesum_per_div} \eeq 
due to the leading behavior of the integrand in \cref{eq_modesum_per} at the upper limit of integration \footnote{For a sufficiently large constant $A$ ensuring $\vartheta(A)\simeq 1$ and $B\gg A$, the r.h.s.\ of \cref{eq_modesum_per} can be estimated as $C(\mtscal) - D(\mtscal)$, where $D(\mtscal)\equiv \rho^{d-1}\int_A^B \d y y^{-1}\exp(-\mtscal y/(4\pi^2))$ and $C(\mtscal)$ captures the remaining contributions of the integrand for small $y$. According to the preceding discussion in the main text, $C(\mtscal)$ is thus finite for all $\mtscal$. In fact, the dominant contribution for $\mtscal\to 0$ stems from the term $D(\mtscal)$. This follows from noting that, for small $\mtscal$, the integrand in $D(\mtscal)$ contributes only if $y\ll 4\pi^2/\mtscal$, for which $\exp(-\mtscal y/(4\pi^2))\simeq 1$. Accordingly, one obtains the estimates $D(\mtscal)\sim -\rho^{d-1}\int_A^{4\pi^2/\mtscal} \d y y^{-1} \sim -\rho^{d-1} \ln(4\pi^2/(\mtscal A))$ up to $\mtscal$-independent terms, which gives the asymptotic result in \cref{eq_modesum_per_div}.}.
Below this property will be discussed further [see \cref{eq_Fres_scalf_pbc}].
Since $\lim_{\rho\to 0} \rho\vartheta(\rho^2 y) = \sqrt{\pi/y}$, \cref{eq_modesum_per} reduces in the thin-film limit $\rho\to 0$ to
\beq \modesum\pbc_{d,\reg}(\mtscal,\rho= 0) = \frac{1}{\pi} \int_0^\infty \d y \, \exp\left(-\frac{\mtscal y}{4\pi^2}\right) \left(\frac{\pi}{y}\right)^{\sfrac{(d+1)}{2}} \left[\left(\frac{\pi}{y}\right)^{1/2} - \vartheta(y) \right],
\label{eq_modesum_per_film}\eeq
which can be shown \cite{kastening_finite-size_2010} to be identical to the expression derived in Ref.\ \cite{krech_free_1992}:
\beq \modesum\pbc_{d,\reg}(\mtscal,\rho= 0) = -\frac{2^{2-d} \pi^{(1-d)/2} \mtscal^{d/2}}{\Gamma((d-1)/2)} \gfunc_{(d-1)/2}\left(\sqrt{\mtscal}/2\right),
\label{eq_modesum_per_film_KD}\eeq 
where $\Gamma$ is the Gamma function and
\beq \gfunc_a(x) = \frac{1}{a} \int_1^\infty \d t\, \frac{(t^2-1)^a}{\exp(2xt)-1}.
\label{eq_gfunc_KD}\eeq 
As implied by \cref{eq_modesum_per_div}, $\modesum\pbc_{d,\reg}(\tscal,\rho=0)$ [\cref{eq_modesum_per_film}] is finite for all $\mtscal\geq 0$. 

In order to evaluate the bulk expression appearing in the subtraction in \cref{eq_modesum_per1}, we note that, for an arbitrary constant $a>0$, one has in dimensional regularization \cite{rudnick_finite-size_1985, amit_field_2005}:
\beq\begin{split} 
\int \frac{\d^{d}k}{(2\pi)^{d}} \ln(\kv^2 + a) -  \int \frac{\d^{d}k}{(2\pi)^{d}} \ln(\kv^2) &= \int_0^a \d s \int \frac{\d^{d}k}{(2\pi)^{d}} \frac{1}{\kv^2+s} = -A_d \int_0^a \d s\, s^{d/2-1} = -\frac{2 A_d}{d}a^{d/2},
\end{split}\label{eq_log_int_eval}\eeq 
with 
\beqn A_d \equiv -(4\pi)^{-d/2}\Gamma\left(1-d/2\right).
\eeqn
In summary, for periodic \bcs and finite aspect ratio $\rho$, the total free energy defined in \cref{eq_freeE_inhom} takes the form
\beq \Fcalc\pbc =A L \left(\Lcal_b(\avOP)-\frac{A_d}{d} \mtau^{d/2} \right) + \onehalf AL \int \frac{\d^{d}k}{(2\pi)^{d}} \ln(\kv^2 ) + \onehalf AL^{-d+1} \modesum\pbc_{d,\reg}(\mtau L^2,\rho) + \delta F\pbc(\mtau, A, L),
\label{eq_finiteAR_F_pbc}\eeq
where $\Lcal_b$ is defined in \cref{eq_landau_fe} and the constraint-induced contribution $\delta F\pbc$ is reported in \cref{eq_constrcorr_per}.
The contribution in \cref{eq_finiteAR_F_pbc} involving the term $\int \d^dk \ln(\kv^2)$ formally vanishes in dimensional regularization \cite{zinn-justin_quantum_2002} and will be disregarded henceforth. (This term would be canceled also by additive renormalization of the total free energy \cite{krech_free_1992}.)

Following \cref{eq_Fres_def}, we extract from \cref{eq_finiteAR_F_pbc} the residual finite-size free energy per volume,
\beq \ring\fcal\pbc\res(\mtau,\rho,L) = \onehalf L^{-d} \modesum\pbc_{d,\reg}(\mtau L^2,\rho) + \frac{1}{AL}\delta F\pbc(\mtau L^2) =  L^{-d} \left[ \ring\Theta\pbc(\mtau L^2,\rho) + \rho^{d-1} \delta F\st{ns}(L)\right],
\label{eq_Fres_pbc}\eeq
with the scaling function
\beq \ring\Theta\pbc(\mtscal,\rho) \equiv \onehalf \modesum\pbc_{d,\reg}(\mtscal,\rho) + \rho^{d-1} \delta F\pbc\st{s}(\mtscal,\rho) ,
\label{eq_Fres_scalf_pbc}\eeq 
where $\delta F\pbc\st{s} = -\frac{1}{2} \ln (\sfrac{\rho^{d-1} \mtscal}{(2\pi)})$ [\cref{eq_constrcorr_s_perNeu}] is the scaling contribution to the constraint-induced term and $\delta F\st{ns}$ is the non-scaling contribution [\cref{eq_constrcorr_nsterm}].
The divergence of $\modesum\pbc_{d,\reg}$ expressed in \cref{eq_modesum_per_div} is canceled by $\delta F\pbc\st{s}$ in $\ring\Theta\pbc$, which therefore remains finite for all $\mtscal \geq 0$ and all aspect ratios $\rho$.
In contrast, for an unconstrained system with $\rho\neq 0$, the corresponding \resFE diverges for $\mtscal\to 0$ as in \cref{eq_modesum_per_div}.
This divergence originates from the contribution of the mode with $\kv=\bv0$ in the mode sum in \cref{eq_modesum_per0} \footnote{We remark that, also in the thin-film limit $\rho\to 0$, a problematic infrared divergence of the \resFE occurs at higher orders in perturbation theory \cite{diehl_fluctuation-induced_2006, gruneberg_thermodynamic_2008}, despite the finiteness of $\fcal\res$ at the Gaussian level [see \cref{eq_modesum_per_film_KD}]. However, this fact is immaterial for the present discussion.}.
Since $\modesum\pbc_{d,\reg}(\mtscal\to \infty)\to 0$, the presence of the constraint-induced term $\delta F\pbc\st{s}$ leads to a logarithmic divergence of $\ring\Theta\pbc$ for $\mtscal\gg 1$: 
\beq \ring\Theta\pbc(\mtscal\gg 1,\rho) \simeq -\onehalf\rho^{d-1} \ln \frac{\mtscal \rho^{d-1}}{2\pi}.
\label{eq0_Fres_pbc_largeTau}\eeq 

\subsubsection{Dirichlet \bcs}
\label{sec_F_Dir}
We consider a $d$-dimensional box with periodic \bcs in the $d-1$ lateral directions and Dirichlet \bcs at $z=0,L$ (see \cref{sec_Dir_gen}).
In the basic expression for the free energy in \cref{eq_freeE_inhom}, we have $\psi=\avOP=0$; $E_n$ is defined in \cref{eq_Dirichlet_eigenval} and the quantity $(1/2) \ln(\wt,\Gcal,\wt)$ is reported in \cref{eq_constrcorr_Dir}. 
The expression of the corresponding mode sum can be obtained from the one for periodic \bcs by noting that, due to Eqs.~\eqref{eq_freeE_inhom} and \eqref{eq_Dirichlet_eigenval}, one has
\beq \begin{split}
\msumBare\Dbc_d(\tau,L,A) \equiv \sum_{n=1}^\infty \sum_{\kv_\parallel} \ln\left[\kv_\parallel^2 + \left(\frac{\pi n}{L}\right)^2 + \tau \right] 
&= \onehalf \sum_{\substack{n=-\infty\\ n\neq 0}}^\infty \sum_{\kv_\parallel} \ln\left[ \kv_\parallel^2 + \left(\frac{2\pi n}{L'}\right)^2 + \tau\right] \\
&= \onehalf \sum_{p'} \sum_{\kv_\parallel} \ln\left[ \kv_\parallel^2 + p'^2 + \tau\right] - \onehalf \sum_{\kv_\parallel} \ln(\kv_\parallel^2 +\tau) \\
&= \onehalf \msumBare\pbc_d(\tau, 2L, L_\parallel^{d-1}) - \onehalf \msumBare\pbc_{d-1}(\tau,L_\parallel,L_\parallel^{d-2}),
\end{split}\label{eq_Dirichlet_modesum_rewr}\eeq
with $L'\equiv 2L$ and where we introduced the wavenumber $p'\equiv 2\pi n/L'$ with $n\in \mathbb{Z}$.
In the last line in \cref{eq_Dirichlet_modesum_rewr}, we have identified the first term as (half of) the mode sum of a $d$-dimensional system with periodic \bcs and aspect ratio $L'/L_\parallel$, and the second term as (half of) the mode sum of a $(d-1)$-dimensional cubic system of volume $L_\parallel^{d-1}$ with periodic \bcs (see  \cref{eq_modesum_per1}).
Using \cref{eq_modesum_per2}, we can thus express the regularized mode sum for Dirichlet \bcs as
\beq \msumBare\Dbc_{d,\reg}(\tau L^2, L, L_\parallel) = A L^{-d+1} \modesum\Dbc_{d,\reg}(\tau L^2, \rho) ,
\label{eq_modesum_D_reg}\eeq 
with
\beq \modesum\Dbc_{d,\reg}(\tscal, \rho) \equiv 2^{-d} \modesum\pbc_{d,\reg}(4\tscal, 2\rho) - \onehalf \rho^{d-1} \modesum\pbc_{d-1,\reg}(\tscal/\rho^2, 1).
\label{eq_modesum_D_reg_scalf}\eeq 
Since, for $\tscal\to\infty$, $\modesum\pbc_{d-1,\reg}(x,1)$ vanishes exponentially as a function of $x$, upon using \cref{eq_modesum_per_film_KD} one recovers in the thin-film limit ($\rho\to 0$) the expression contained in Ref.\ \cite{krech_free_1992}:
\beq 
\modesum\Dbc_{d,\reg}(\tscal,\rho=0) = \frac{2^{2-d} \pi^{(1-d)/2} x^{d/2}}{\Gamma((d-1)/2)} \gfunc_{(d-1)/2}\left(\sqrt{x}\right).
\label{eq_modesum_D_film}\eeq 
Furthermore, due to \cref{eq_modesum_per_div}, the divergences for $\tscal\to 0$ of the two separate terms in \cref{eq_modesum_D_reg_scalf} cancel, rendering $\modesum\Dbc_{d,\reg}$ finite for $\tscal= 0$ and all aspect ratios.
Taking into account \cref{eq_modesum_per1,eq_log_int_eval}, the free energy of the constrained system [\cref{eq_freeE_inhom}] for Dirichlet \bcs with vanishing mean OP follows as
\beq\begin{split}
\Fcalc\Dbc &=  \onehalf AL \int \frac{\d^{d}k}{(2\pi)^{d}} \ln(\kv^2 + \tau) - \onequarter A \int \frac{\d^{d-1}k_\parallel}{(2\pi)^{d}} \ln(\kv_\parallel^2 + \tau) + \onehalf \msumBare\Dbc_{d,\reg}(\tau L^2,\rho) + \delta F\Dbc(\tau L^2)\\
&= - A L \frac{A_d}{d} \tau^{d/2} + \onehalf AL \int \frac{\d^{d}k}{(2\pi)^{d}} \ln(\kv^2 ) + \frac{A}{2} \frac{A_{d-1}}{d-1} \tau^{(d-1)/2} - \onequarter A \int \frac{\d^{d-1}k_\parallel}{(2\pi)^{d-1}} \ln(\kv_\parallel^2 )\\
&\qquad + \onehalf A L^{-d+1} \modesum\Dbc_{d,\reg}(\tau L^2,\rho) + \delta F\Dbc(\tau, A, L).
\end{split}\label{eq_finiteAR_F_Dir}\eeq
The constraint-induced term $\delta F\Dbc$ is reported in \cref{eq_constrcorr_Dir}.
From \cref{eq_Fres_def}, one obtains the residual finite-size free energy per volume:
\beq \ring \fcal\Dbc\res(\tau,\rho,L) =  L^{-d}\left[\ring\Theta\Dbc(\tau L^2,\rho) +  \rho^{d-1} \delta F\st{ns}(L) \right]
\label{eq_Fres_Dir}\eeq 
with the scaling function
\beq \ring\Theta\Dbc(\tscal,\rho) \equiv \onehalf  \modesum\Dbc_{d,\reg}(\tscal,\rho) + \rho^{d-1} \delta F\st{s}\Dbc(\tscal,\rho),
\label{eq_Fres_scalf_Dir}\eeq 
where $\delta F\st{s}$ and $\delta F\st{ns}$ are given in \cref{eq_constrcorr_s_Dir,eq_constrcorr_nsterm}, respectively. 
For $\tscal\gg 1$, $\modesum\Dbc_{d,\reg}(\tscal,\rho)$ vanishes exponentially so that the asymptotic behavior of $\delta F\st{s}\Dbc$ dominates, resulting in a logarithmic divergence of $\ring\Theta\Dbc$:
\beq \ring \Theta\Dbc(\tscal\gg 1) \simeq -\onehalf\rho^{d-1} \ln \frac{\tscal \rho^{d-1}}{2\pi},
\label{eq_Fres_Dbc_largeTau}\eeq 
analogously to $\Theta\pbc$ [\cref{eq0_Fres_pbc_largeTau}].
Since both $\modesum\Dbc_{d,\reg}$ and $\delta F\st{s}\Dbc$ are finite for $\tscal\to 0$ [see \cref{eq_Dirichl_wGw_limits}], also $\ring\Theta\Dbc$ remains finite in that limit.

\subsubsection{Neumann \bcs}
\label{sec_Fres_Neu}

The mode sum for Neumann \bcs can be related to the one for Dirichlet \bcs [\cref{eq_Dirichlet_modesum_rewr}] by writing 
\beq \begin{split}
\msumBare\Nbc_d(\mtau,L,A) \equiv \sum_{n=0}^\infty \sum_{\kv_\parallel} \ln\left[\kv_\parallel^2 + \left(\frac{\pi n}{L}\right)^2 + \mtau \right] = \msumBare\Dbc_d(\mtau,L,A) + \sum_{\kv_\parallel} \ln(\kv_\parallel^2 + \mtau).
\end{split}\label{eq_Neumann_modesum_rewr}\eeq
From \cref{eq_modesum_D_reg,eq_modesum_D_reg_scalf} we thus obtain the regularized mode sum 
\beq \msumBare\Nbc_{d,\reg}(\mtau L^2, L, L_\parallel) = A L^{-d+1} \modesum\Nbc_{d,\reg}(\mtau L^2, \rho) ,
\eeq 
with
\beq \modesum\Nbc_{d,\reg}(\mtscal, \rho) \equiv 2^{-d} \modesum\pbc_{d,\reg}(4\mtscal, 2\rho) + \onehalf \rho^{d-1} \modesum\pbc_{d-1,\reg}(\mtscal/\rho^2, 1),
\label{eq_modesum_N_reg}\eeq 
where $\msumBare\pbc_{d,\reg}$ is given by \cref{eq_modesum_per}.
In the thin-film limit ($\rho\to 0$) the second term in \cref{eq_modesum_N_reg} vanishes so that
\beq \modesum\Nbc_{d,\reg}(\mtscal,\rho=0) = \modesum\Dbc_{d,\reg}(\mtscal,\rho=0),
\label{eq_modesum_N_D_id}\eeq
with $\modesum\Dbc_{d,\reg}(\mtscal,\rho=0)$ given by \cref{eq_modesum_D_film}, in agreement with explicit calculations for $\rho=0$ reported in Ref.\ \cite{krech_free_1992}.
However, in contrast to the case of Dirichlet \bcs, $\modesum\Nbc_{d,\reg}$ diverges for $\mtscal\to 0$ and nonzero $\rho$:
\beq \modesum\Nbc_{d,\reg}(\mtscal\to 0,\rho\neq 0) \simeq  \rho^{d-1} \ln \mtscal,
\label{eq_modesum_N_div}\eeq 
due to \cref{eq_modesum_per_div}.
The free energy of the constrained system [\cref{eq_freeE_inhom}] for Neumann \bcs is given by
\beq\begin{split}
\Fcalc\Nbc &= A L\left(\Lcal_b[\avOP] - \frac{A_d}{d} \mtau^{d/2}\right) + \onehalf AL \int \frac{\d^{d}k}{(2\pi)^{d}} \ln(\kv^2 ) - \frac{A}{2} \frac{A_{d-1}}{d-1} \mtau^{(d-1)/2} + \onequarter A \int \frac{\d^{d-1}k_\parallel}{(2\pi)^{d-1}} \ln(\kv_\parallel^2 )\\
&\qquad + \onehalf A L^{-d+1} \modesum\Nbc_{d,\reg}(\mtau L^2,\rho) + \delta F\Nbc(\mtau, A, L),
\end{split}\label{eq_finiteAR_F_Neu}\eeq
where $\Lcal_b$ is defined in \cref{eq_landau_fe} and the constraint-induced contribution $\delta F\Nbc$ is reported in \cref{eq_constrcorr_Neu}.
We remark that, for $\avOP=0$, the grand canonical $\Fcal\Nbc$ and $\Fcal\Dbc$ are identical [\cref{eq_finiteAR_F_Dir}], except that the sign of the surface contribution is reversed.
The residual finite-size free energy per volume follows from \cref{eq_Fres_def} as
\beq \ring \fcal\Nbc\res(\mtau,\rho,L) =  L^{-d}\left[\ring\Theta\Nbc(\mtau L^2,\rho) + \rho^{d-1} \delta F\st{ns}(L) \right]
\label{eq_Fres_Neu}\eeq 
with the scaling function
\beq \ring\Theta\Nbc(\mtscal,\rho) \equiv \onehalf  \modesum\Nbc_{d,\reg}(\mtscal,\rho) + \rho^{d-1} \delta F\st{s}\Nbc(\mtscal,\rho),
\label{eq_Fres_scalf_Neu}\eeq 
where $\delta F\st{s}$ and $\delta F\st{ns}$ are given in \cref{eq_constrcorr_s_perNeu,eq_constrcorr_nsterm}, respectively.
In the thin-film limit ($\rho\to 0$) and for a vanishing mean OP $\avOP$ (implying $\mtscal\to \tscal$, see \cref{eq_tscal_mod}), the scaling functions for Dirichlet and Neumann \bcs are identical, $\ring\Theta\Dbc=\ring\Theta\Nbc$, as a consequence of \cref{eq_modesum_N_D_id}.
Since $\modesum\Nbc_{d,\reg}(\mtscal,\rho)\to 0$ for $x\to\infty$, it follows from the presence of $\delta F\st{s}\Nbc$ that $\ring\Theta\Nbc$ diverges logarithmically for $\mtscal\gg 1$:
\beq \ring \Theta\Nbc(\mtscal\gg 1) \simeq -\onehalf\rho^{d-1} \ln \frac{\mtscal \rho^{d-1}}{2\pi}.
\label{eq_Fres_Nbc_largeTau}\eeq 
The divergences for $\mtscal\to 0$ of $\modesum\Nbc_{d,\reg}$ [\cref{eq_modesum_N_div}] and $\delta F\st{s}\Nbc$ [\cref{eq_constrcorr_s_perNeu}] cancel in $\ring\Theta\Nbc$, rendering the \resFE in the canonical case and at bulk criticality finite for all aspect ratios.
In contrast, in the grand canonical case, \cref{eq_modesum_N_div} implies a divergent \resFE for $\mtscal\to 0$ and nonzero $\rho$.
This divergence is due to a zero mode in the fluctuation spectrum, as it is also the case for periodic \bcs.
\end{widetext}

\section{Critical Casimir forces obtained from pressure differences}
\label{app_CFF_pressure}
Alternatively to the definition based on the \resFE [\cref{eq_CCF_def_Fres}], the \CCF $\Kcal$ can be defined as the difference between the pressure $p$ in the confined system and the pressure $p_b$ in the surrounding bulk medium:
\beq \Kcal = p - p_b.
\label{eq_CCF_def_pdiff}\eeq 
For fixed area $A=V/L$, these pressures follow from the corresponding free energy densities $\fcal$ and $\fcal_b$:
\begin{subequations}\begin{align} 
p &= -\frac{\d (L \fcal)}{\d L},\label{eq_pf_def}\\
p_{b} &= -\frac{\d (L \fcal_b)}{\d L}. \label{eq_pb_def}
\end{align}\end{subequations}
The same relations apply also in the canonical ensemble.
The bulk pressure can be obtained from the thermodynamic limit:
\beq p_b = \lim_{\substack{L\to\infty,\\ A\to\infty}} p,
\label{eq_pb_lim}\eeq 
which is to be performed by keeping a fixed mean OP $\avOP$ in the canonical ensemble \cite{gross_critical_2016} and a fixed bulk field $h$ in the grand canonical ensemble.

Turning first to the canonical ensemble, we employ the decomposition property in \cref{eq_Fcan_decomp} to formally write the pressure $\ring p$ as consisting of bulk, surface, and residual finite-size contributions:
\beq \ring p = \ring p_b + \ring p_s + \ring p\res,
\label{eq_p_decomp}\eeq 
where 
\beq \ring p_s \equiv -\frac{\d \fcalc_s}{\d L}
\label{eq_p_surf_def}\eeq 
is a ``surface'' pressure and 
\beq \ring p\res \equiv -\frac{\d (L\fcalc\res)}{\d L}
\eeq 
is the excess contribution.

In the following, we focus on \emph{Neumann} \bcs, because only in this case the CCF derived from \cref{eq_CCF_def_pdiff} differs from the one obtained on the basis of \cref{eq_CCF_def_Fres}.
For simplicity, we analyze the regularized (but not yet renormalized) expressions of the free energy, as given in \cref{sec_Fres_calc,sec_F_gc}. 
Renormalization produces (via additive counterterms) contributions to the bulk free energy \cite{eisenriegler_helmholtz_1987}, but does not change the conclusions of this section regarding the CCF.
According to \cref{eq_F_can_pert_Neu} the \emph{bulk} free energy density is given by
\beq 
\fcalc_b = \Lcal_b(\avOP) - \frac{A_d}{d} \mtau^{d/2} ,
\label{eq_FR_bulk}\eeq 
where $\Lcal_b$ is defined in \cref{eq_landau_fe}. 
Inserting $\fcalc_{b}$ [\cref{eq_FR_bulk}] into \cref{eq_pb_def} yields the bulk pressure 
\beq \ring p_b = -\frac{\d (L\fcalc_{b})}{\d L} = -\left[ \fcalc_{b} - \avOP \frac{\pd\fcalc_{b}}{\pd\avOP}\right],
\label{eq_CCFcan_bulk_contrib}\eeq 
where we made use of \cref{eq_dPhi_dL}.
As a manifestation of ensemble equivalence in the thermodynamic limit, the \emph{grand canonical} bulk free energy density $\fcal_{b}$ can be obtained from $\fcalc_{b}$ [\cref{eq_FR_bulk}] via a Legendre transform: 
\beq \fcal_{b}(\tau,h,\rho,L) = \fcalc_{b}(\tau,\avOP(h)) - h  \avOP(h),
\label{eq_FR_bulk_gc}\eeq 
with $\avOP=\avOP(h)$ determined from the implicit equation
\beq h = \frac{\pd \fcalc_{b} }{ \pd\avOP}.
\label{eq_EOS_Lbulk}\eeq
In the grand canonical case, $h$ is an external field and \cref{eq_EOS_Lbulk} does not introduce any dependence on $L$. 
Therefore, by using the above equations, the grand canonical film pressure follows as
\beq p_{b} = -\frac{\d (L\fcal_{b})}{\d L} = -\fcal_{b} = \ring p_{b}.
\label{eq_CCFgc_bulk_contrib}\eeq 
As expected, the canonical and the grand canonical bulk pressures are identical.

From \cref{eq_F_can_pert_Neu} one infers the canonical surface free energy per area for Neumann \bcs:
\beq 
\fcalc\Nbc_{s} = - \frac{1}{2} \frac{A_{d-1}}{d-1}\mtau^{(d-1)/2} .
\label{eq_FR_surf}
\eeq 
Upon using \cref{eq_dPhi_dL}, the ``surface'' pressure [\cref{eq_p_surf_def}] follows as
\beq \ring p_s\Nbc = -\frac{\pd \fcalc_s\Nbc}{\pd \avOP} \frac{\d \avOP}{\d L} = \frac{\pd \fcalc_s\Nbc}{\pd \avOP} \frac{\avOP}{L} = -\frac{1}{2 L}A_{d-1}  \mtau^{(d-3)/2} u \avOP^2.
\label{eq_p_surf_Nbc_can}\eeq 
Due to the OP constraint [\cref{eq_dPhi_dL}], the surface pressure is nonzero in the canonical ensemble.
This result can be compared with the corresponding one in the grand canonical ensemble, in which, according to \cref{eq_Fgc_surf_Neu}, the surface free energy $\fcal_s\Nbc$ has the same formal expression as $\fcalc_s\Nbc$ [\cref{eq_FR_surf}], except that $\avOP=\avOP_b(\tau,h)$ is a function of the external field $h$ via the \emph{bulk} equation of state. Since the latter is independent of $L$, we immediately infer, analogously to \cref{eq_p_surf_Nbc_can}, that
\beq p_s\Nbc =  -\frac{\pd \fcal_s\Nbc}{\pd \avOP}\Big|_{\avOP_b} \frac{\d \avOP_b}{\d L} = 0,
\eeq
as expected.
As a direct consequence of \cref{eq_p_surf_Nbc_can}, the canonical \CCF $\ring\Kcal$ defined by \cref{eq_CCF_def_pdiff} is in general different from the \CCF defined by \cref{eq_CCF_def_Fres}, because the latter simply coincides with $\ring p\res$. 
Following \cref{sec_scaling}, \cref{eq_p_surf_Nbc_can} can be cast into scaling form:
\beq\begin{split} 
\ring p_{s}=  L^{-d} \ring\Theta\st{s}\left(\left(\frac{L}{\amplXip}\right)^{1/\nu} \tau, \left(\frac{L}{\amplXip}\right)^{\beta/\nu} \avOP\right),
\end{split}\label{eq_p_surf_scal}\eeq 
with the scaling function
\beq \ring\Theta\st{s}(\tscal,\phiscal) = - \frac{A_{d-1}}{2} r u^* \phiscal^2 \left[\tscal + \onehalf r u^* \phiscal^2 \right]^{(d-3)/2},
\eeq 
which is to be evaluated for $d=4$.
In \cref{eq_p_surf_scal} $\ring p_s$, is to be understood as per $k_B T_c$, so that $\ring\Theta\st{s}$ is dimensionless.

\section{Critical Casimir force for constant volume}
\label{app_CCF_vol}

In \cref{eq_CCF_def_Fres} we have defined the CCF under the condition of a fixed transverse area $A$, implying a change of the volume of the film upon its action and thereby of the mean OP [see \cref{eq_dPhi_dL}].
In the case of a binary liquid mixture, the near-incompressibility of the liquid (close to demixing) strongly opposes changes of volume and, therefore, of the distance between the plates realizing the confinement. 
In the grand canonical ensemble the change of volume of the film occurs (easily) via exchange with the reservoir, but not due to compression. 
Alternatively, one may thus consider the CCF (per area) under the constraint of \emph{constant} volume $V=AL$,
\beq \Kcal_V \equiv -\frac{1}{A} \frac{\d (V \fcal\res)}{\d L}\Big|_{V=\,\text{const}}.
\label{eq_CCF_def_Fres_V}\eeq 
An analogous definition applies to the corresponding canonical CCF $\ring{\Kcal}_V$, where, as before, additionally to $V$ also the total OP $\intOP$ is held constant.
We further note that, for constant volume, \cref{eq_aspectratio} implies $\d\rho/\d L = \rho L d/(d-1)$ and
\beq \frac{\d \avOP}{\d L}\Big|_{V=\,\text{const}} = 0,
\eeq 
instead of \cref{eq_dPhi_dL}.
Using \cref{eq_CCF_def_Fres_V,eq_Fres_c_scal_L}, the canonical CCF $\ring\Kcal_V$ can be shown to fulfill \cref{eq_Casi_force_c} with the scaling function
\begin{widetext}
\beq \ring\Xi_V(\tscal,\phiscal,\rho) = d \ring{\hat \Theta}(\tscal,\phiscal,\rho) - \frac{1}{\nu}\tscal \pd_\tscal \ring{\hat \Theta}(\tscal,\phiscal,\rho) - \frac{\beta}{\nu}\phiscal\pd_\phiscal \ring{\hat \Theta}(\tscal,\phiscal,\rho) - \frac{d}{d-1}\rho\pd_\rho \ring{\hat \Theta}(\tscal,\phiscal,\rho) + \delta\ring\Xi\st{ns}(\rho),
\label{eq_Casi_force_c_scalf_V}\eeq 
instead of $\ring\Xi$. The expression for $\delta \ring\Xi\st{ns}$ is the same as in \cref{eq_CCF_can_nscal}, and the expressions of the scaling functions $\ring\Theta(\mtscal(\tscal,\phiscal),\rho)=\ring{\hat\Theta}(\tscal,\phiscal,\rho)$ are reported in \cref{eq_Fres_c_scalfunc,eq_constrcorr_scal,eq_Sreg_coll} for the various \bcs. 
Using, analogously, \cref{eq_CCF_def_Fres_V,eq_Fres_gc_scal_L}, the grand canonical CCF $\Kcal_V$ fulfills \cref{eq_Casi_force_gc} with the scaling function
\beq \tilde\Xi_V(\tscal, \hscal, \rho) =  d \tilde\Theta(\tscal, \hscal, \rho) - \frac{1}{\nu} \tscal \pd_{\tscal}\tilde\Theta(\tscal, \hscal, \rho) - \frac{\beta\delta}{\nu} \hscal \pd_{\hscal}\tilde\Theta (\tscal, \hscal, \rho)- \frac{d}{d-1} \rho \pd_{\rho}\tilde\Theta(\tscal, \hscal, \rho)
\label{eq_Casi_force_gc_scalf_V}\eeq 
instead of $\tilde\Xi$.
Analogously to \cref{eq_Casi_force_gc_scalf_m}, expressing $\tilde\Theta(\tscal,\hscal,\rho)$ in terms of $\hat\Theta(\tscal,\phiscal(\tscal,\hscal,\rho),\rho)$ and using the scaling form of the equation of state in \cref{eq_phi_bulk_scalfunc} (which is valid in the bulk limit as well as at $\Ocal(\epsilon^0)$), yields
\beq 
\Xi_V(\tscal, \phiscal, \rho) =  d \hat\Theta(\tscal, \phiscal, \rho) - \frac{1}{\nu} \tscal \pd_{\tscal}\hat\Theta(\tscal, \phiscal, \rho) - \frac{\beta}{\nu} \phiscal \pd_{\phiscal}\hat\Theta(\tscal, \phiscal, \rho) - \frac{d}{d-1}\rho \left(\frac{\pd \phiscal}{\pd\rho}\pd_\phiscal+ \pd_{\rho}\right)\hat\Theta(\tscal, \phiscal, \rho).
\label{eq_Casi_force_gc_scalf_m_V}\eeq 
Since, at $\Ocal(\epsilon^0)$, $\pd \phiscal/\pd\rho=0$ for the considered \bcs, the scaling functions $\ring\Xi_V$ and $\Xi_V$ in \cref{eq_Casi_force_c_scalf_V,eq_Casi_force_gc_scalf_m_V} have formally identical expressions in terms of the corresponding scaling functions $\ring{\hat\Theta}$ and $\hat\Theta$.

In order to asses the actual difference between the two ensembles, we must take into account that, in the canonical ensemble, the constraint-induced term $\delta F$ [\cref{eq_constrcorr}] contributes to $\ring{\hat\Theta}$ with a term which is given in \cref{eq_constrcorr_scal}. 
According to \cref{eq_CCF_def_Fres_V}, the total constraint-induced contribution to the CCF $\ring\Kcal_V$ [including the term $\delta \ring\Xi\st{ns}$ in \cref{eq_CCF_can_nscal}] is given by
\beq \delta \ring\Kcal_V(\tren,A,L) = -\frac{1}{A} \frac{\d\, \delta F}{\d L}\Big|_{V=\const} = L^{-d} \delta \ring\Xi_V\left(\tscal= \left(\frac{L}{\amplXip}\right)^{1/\nu} \tren,\rho\right)
\label{eq_Casi_constrcorr_V}\eeq
with 
\beq \delta\ring \Xi_V(\tscal,\rho)= \begin{cases} 
	0, \qquad &\text{periodic and Neumann,}\\
	\displaystyle \onehalf \rho^{d-1}\frac{\sqrt{x}-\sinh\sqrt{x} }{\cosh^2(\sqrt{x}/2)[\sqrt{x}-2\tanh(\sqrt{x}/2)]},  &\text{Dirichlet.}
\end{cases}\label{eq_Casi_constrcorr_V_scal}
\eeq 
These expressions can be contrasted to the corresponding ones for $\delta\ring\Xi$ reported in \cref{eq_Casi_constrcorr_scal}.
Notably, the constraint-induced contribution to the CCF defined with fixed volume $V$ vanishes for periodic and Neumann \bcs. 
At $\Ocal(\epsilon^0)$, upon using \cref{eq_tscal_mod} and $\beta=1/2$, we can express  \cref{eq_Casi_force_gc_scalf_m_V} in terms of the scaling function $\Theta(\mtscal(\tscal,\phiscal),\rho) = \hat\Theta(\tscal,\phiscal,\rho)$ as
\beq \Xi_V(\tscal, \phiscal, \rho) =  d \Theta(\mtscal, \rho) - \frac{1}{\nu} \mtscal \pd_{\mtscal} \Theta(\mtscal, \rho) - \frac{d}{d-1}\rho \pd_{\rho} \Theta(\mtscal, \rho) + \Ocal(\epsilon).
\label{eq_Casi_force_gc_scalf_V_simpl}\eeq 
Since \cref{eq_Casi_constrcorr_V_scal} contains the contributions to the CCF from both the scaling and non-scaling terms in the \resFE, \cref{eq_Casi_force_c_scalf_V} can, owing to \cref{eq_Fres_c_scalfunc,eq_Fres_gc_scalfunc}, be expressed analogously in terms of $\Theta(\mtscal,\rho)$ as
\beq \ring\Xi_V(\tscal, \phiscal, \rho) =  d \hat\Theta(\mtscal, \rho) - \frac{1}{\nu} \mtscal \pd_{\mtscal} \hat\Theta(\mtscal, \rho) - \frac{d}{d-1}\rho \pd_{\rho} \hat\Theta(\mtscal, \rho) + \delta\ring\Xi_V(\tscal,\rho) + \Ocal(\epsilon).
\label{eq_Casi_force_c_scalf_V_simpl}\eeq 
Accordingly, at $\Ocal(\epsilon^0)$, the canonical and the grand-canonical CCFs are identical for periodic and Neumann \bcs:
\beq \ring\Xi_V\ut{(p,N)}(\tscal,\phiscal,\rho) = \Xi_V\ut{(p,N)}(\tscal,\phiscal,\rho). 
\label{eq_Casi_force_c_gc_V_ident}\eeq 
For Dirichlet \bcs with $\phiscal=0$ we have, instead,
\beq \ring\Xi_V\Dbc(\tscal,\rho) = \Xi_V\Dbc(\tscal,\rho) + \delta\ring\Xi_V\Dbc(\tscal,\rho),
\label{eq_Casi_force_c_gc_V_Dir}\eeq 
where $\delta\ring\Xi_V$ is negative for all $\tscal$ and vanishes in the limit $\tscal\to\infty$.
We finally note that, for a fully isotropic cube, the CCF for conserved volume is expected to vanish by symmetry.
Indeed, using \cref{eq_tscal_mod,eq_Casi_force_gc_scalf_V_simpl,eq_modesum_per}, for periodic \bcs with $\rho=1$ and $d=4$ one finds
\beq 
\ring{\Xi}_V\pbc(\tscal,\phiscal,\rho=1) = \Xi_V\pbc(\tscal,\phiscal,\rho=1) = \int_0^\infty \d y\, \pd_y \left\{ \exp\left(-\frac{\mtscal y}{4\pi^2}\right) \left[ \vartheta^d(y) - \left(\frac{\pi}{y}\right)^{d/2} \right] \right\} = 0,
\label{eq_Casi_froce_V_pbc_rho1}\eeq 
where the last step follows from the asymptotic behavior of the theta function  $\vartheta(y)$ [see \cref{eq_thetafunc} and the associated comments].
 
\begin{figure}[t!]\centering
  \subfigure[]{\includegraphics[width=0.4\linewidth]{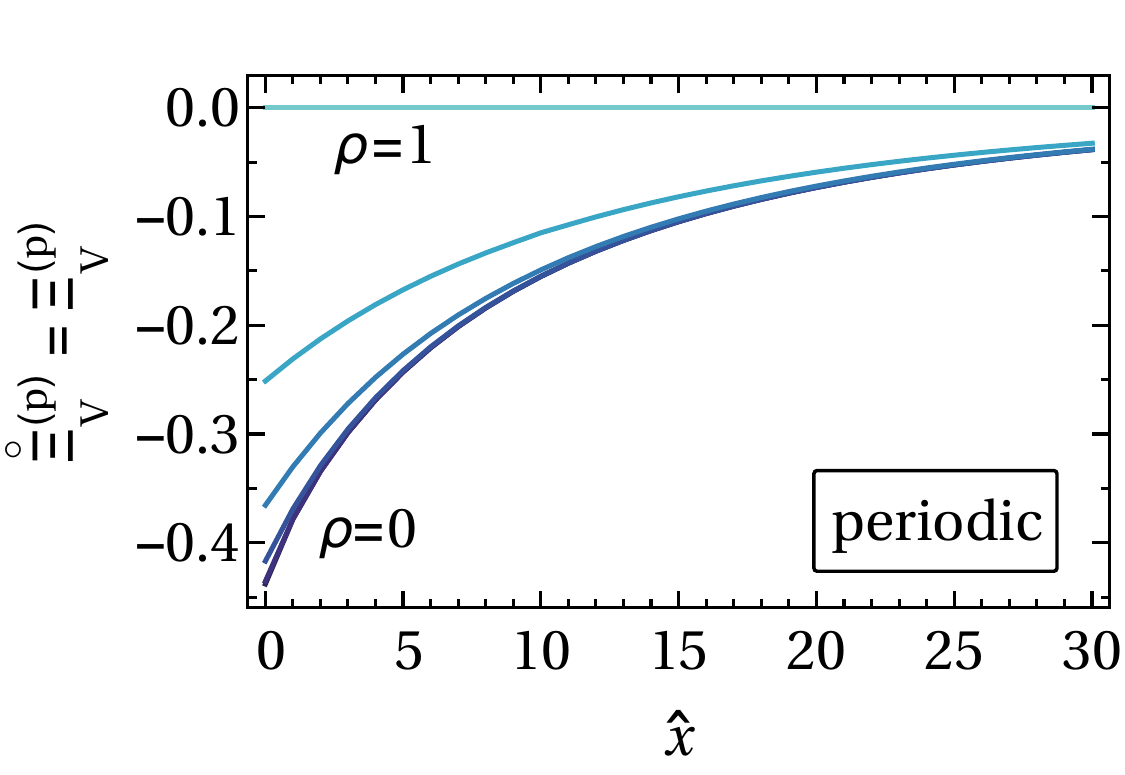}}\qquad
  \subfigure[]{\includegraphics[width=0.4\linewidth]{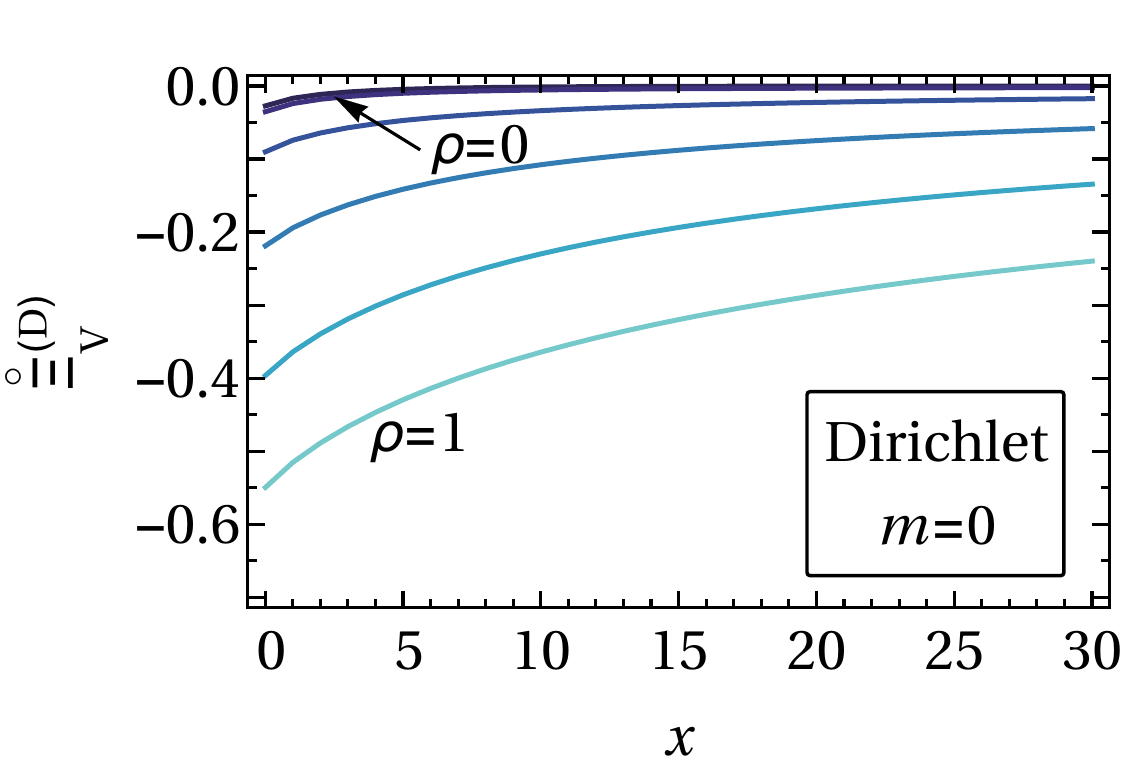}}\qquad
  \subfigure[]{\includegraphics[width=0.41\linewidth]{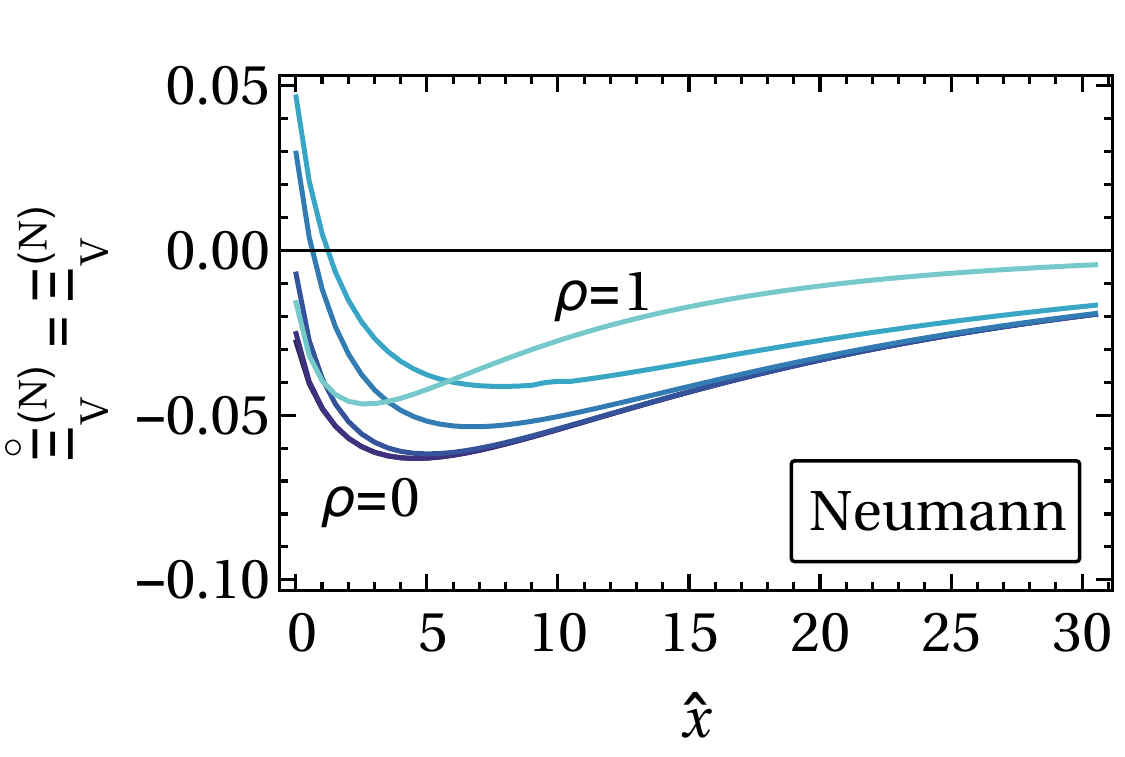}}
  \caption{Scaling functions $\ring\Xi_V(\tscal,\phiscal,\rho)$ [\cref{eq_Casi_force_c_scalf_V_simpl}] of the canonical CCF defined according to \cref{eq_CCF_def_Fres_V} under the condition of fixed volume for various aspect ratios $\rho$. These scaling functions depend on $\tscal$ and $\phiscal$ via $\mtscal$ [\cref{eq_tscal_mod}] only (noting that $\mtscal(\tscal,\phiscal=0)=\tscal$ and $\delta\ring \Xi_V(\tscal,\rho)=0$ for periodic and Neumann \bcs). For periodic (p) and Neumann (N) \bcs, the canonical and the grand canonical scaling functions coincide, i.e., $\ring\Xi_V=\Xi_V$ at $\Ocal(\epsilon^0)$ [see \cref{eq_Casi_force_c_gc_V_ident}]. For Dirichlet (D) \bcs, we consider a mean OP $\phiscal=0$, for which the scaling functions depend on $\tscal$ and are related according to \cref{eq_Casi_force_c_gc_V_Dir}. The aspect ratio $\rho$ increases in steps of $0.2$ between the unlabeled curves as indicated by the distinct blue shading of the curves [in panel (a), the curves for $\rho=0$ and $0.2$ are not distinguishable].}
  \label{fig_XiV}
\end{figure}

Figure \ref{fig_XiV} shows the numerically evaluated scaling functions $\ring\Xi_V\ut{(p,D,N)}$ of the canonical CCF for conserved volume. Note that, according to \cref{eq_Casi_constrcorr_V_scal,eq_Casi_force_gc_scalf_V_simpl,eq_Casi_force_c_scalf_V_simpl}, $\Xi_V$ and $\ring\Xi_V$ can be considered as functions of the combined scaling variable $\mtscal$ [\cref{eq_tscal_mod}].
For periodic \bcs [\cref{fig_XiV}(a)], the CCF at constant volume is identical in the two ensembles and its absolute strength decreases upon increasing the aspect ratio $\rho$. 
This trend is opposite to the behavior of the CCF at constant transverse area displayed in \cref{fig_Xi_pbc_m0,fig_Xi_pbc_x0}. 
For Dirichlet \bcs, the CCF at constant volume [\cref{fig_XiV}(b)] is qualitatively similar to that at constant transverse area [\cref{fig_Xi_Dir}], except that, in the latter case, $\ring\Xi\Dbc$ attains a nonzero value for $\tscal\to\infty$, whereas $\ring\Xi_V\Dbc$ vanishes in that limit.
In the case of Neumann \bcs [\cref{fig_XiV}(c)], the scaling function $\ring\Xi_V\Nbc=\Xi_V\Nbc$ of the CCF at constant volume shows a behavior distinct from that of $\ring\Xi\Nbc$ [see \cref{fig_Xi_Neu_m0,fig_Xi_Neu_x0}], as it depends non-monotonically on the effective scaled temperature $\mtscal$ and exhibits a pronounced minimum at intermediate values of $\mtscal$.

\end{widetext}

%


\end{document}